\documentclass[twocolumn,twocolappendix]{aastex63}
\pdfoutput=1
\usepackage{amsmath,amstext}
\usepackage[T1]{fontenc}
\usepackage{times}
\usepackage{apjfonts} 
\usepackage[figure,figure*]{hypcap}
\usepackage{natbib}
\usepackage{hyperref}
\usepackage{booktabs}
\usepackage{graphicx}
\usepackage{multirow}
\usepackage[utf8x]{inputenc} 
\usepackage[caption=false]{subfig}

\received{2022 August 1}
\revised{2023 January 5}
\accepted{2023 January 23}
\submitjournal{the Astrophysical Journal}
\shorttitle{Centers and Outskirts of $z\sim2.3$ Galaxies}
\shortauthors{Cutler et al.}
\begin{document}
\title{The Differential Assembly History of the Centers and Outskirts of Main Sequence Galaxies at $z\sim2.3$}

\correspondingauthor{Sam E. Cutler}
\email{secutler@umass.edu}

\author[0000-0002-7031-2865]{Sam E. Cutler}
\affiliation{Department of Astronomy, University of Massachusetts, Amherst, MA 01003, USA}
\author[0000-0002-7831-8751]{Mauro Giavalisco}
\affiliation{Department of Astronomy, University of Massachusetts, Amherst, MA 01003, USA}
\author[0000-0001-7673-2257]{Zhiyuan Ji}
\affiliation{Department of Astronomy, University of Massachusetts, Amherst, MA 01003, USA}
\author[0000-0001-8551-071X]{Yingjie Cheng}
\affiliation{Department of Astronomy, University of Massachusetts, Amherst, MA 01003, USA}

\begin{abstract}
    We present a study of spatially-resolved star formation histories (SFHs) for 60 $z\sim2.3$ main-sequence, star-forming galaxies selected from the MOSDEF spectroscopic survey in the GOODS-N field, with median stellar mass $\log(M_\star/M_\odot)=9.75$ and  spanning the range $8.6<\log(M_\star/M_\odot)<11.5$. Photometry is decomposed into a central and outer spatial component using observed $z_\mathrm{F850LP}-H_\mathrm{F160W}$ colors. The \textsc{Prospector} code is used to model spectral energy distributions for the center, outskirt, and integrated galaxy using HST/ACS and WFC3, \textit{Spitzer}/IRAC, and ground-based photometry, with additional constraints on gas-phase metallicity and spectroscopic redshift from MOSDEF spectroscopy. For the low-resolution bands, spatially-resolved photometry is determined with an iterative approach. The reconstructed SFHs indicate that the majority of galaxies with $\log(M_\star/M_\odot)<10.5$ are observed while their central regions undergo relatively recent ($<100$ Myr) bursts of star formation, while the outskirts have a smooth, quasi-steady SFH that gently increases towards the redshift of observation. The enhanced star formation activity of the central parts is broadly consistent with the idea that it is produced by highly dissipative gas compaction and accretion. The wide range of central densities and sizes observed in the sample suggests that for the selected galaxies such a process has started but is still far from being completed. The implication would be that selecting star-forming galaxies at cosmic noon frequently includes systems in an ``evolved'' evolutionary phase where the centers have recently started a burst of star formation activity that will likely initiate inside-out quenching in the next several hundred million years.
\end{abstract}


\section{Introduction}
A fundamental question in galaxy evolution is the formation history of the dense stellar cores associated with galactic bulges. In the canonical picture, galaxy growth and quenching is an inside-out process. Massive galaxies ($>10^{11}~M_\odot$) build their outer regions (i.e., increase in size) at lower redshift \citep{vanDokkum10,Whitney19,Mosleh20,Cutler22,Ji22} while the central parts of these galaxies are in place as early as $z\sim2$ \citep{Carrasco10}. Similar studies have shown this inside-out growth is prominent in galaxies across the main sequence and even below the main sequence \citep[e.g.,][]{vanderWel14b,Nelson16,Dimauro22}. The prominence of galactic bulges is correlated both with the stellar mass of the galaxy and scale length of the disk \citep{Shen03,vanderWel14b}, indicating that the growth of the disk is tied to the formation and structure of the bulge. Similarly, the growth of the central supermassive black hole in galaxies is also correlated with the bulge strength \citep{Haring04,Kormendy13}, which suggests that the bulge may play a role in active galactic nuclei (AGN) strength and corresponding quenching processes \citep{Chen20}. Bulges have also been tied to slow, inside-out quenching processes, in which massive bulges stabilize the gas in the disk and prevent it from collapsing \citep[so called ``morphological quenching'',][]{Martig09}. Further studies suggest that the bending of the star-forming main sequence (SFMS) to lower specific star formation rates ($\mathrm{sSFR}\equiv\mathrm{SFR}/M_\star$) at higher stellar mass is evidence of the presence of old bulges in massive galaxies \citep{Abramson14}, though this is disputed \citep[e.g.,][]{Guo15,Schreiber16,Dimauro22}.

Dense stellar cores have been directly observed in the optical and near-infrared (NIR) for the most massive galaxies as early as $z\sim2.5$ \citep[e.g.,][]{Carrasco10,vanDokkum14}. Analysis of the mass evolution in these optically-detected galaxies suggests it predominantly occurs in the outer stellar envelope, while these central regions maintain a roughly constant mass \citep[e.g.,][]{vanDokkum14}. The forming dense stellar cores of massive, main-sequence galaxies have also been probed using observations of obscured star formation with the Atacama Large Millimeter/submillimeter Array (ALMA). These studies found dense, starbursting cores that likely formed at the centers of extended disks due to dissipative collapse \citep[e.g.,][]{Tadaki17,Tadaki20}, suggesting an outside-in formation process counter to that suggested by \cite{vanDokkum14}. Alternatively, mergers could play a significant role in forming compact starburst regions by driving gas into the galactic center \citep{Puglisi19,Puglisi21}. These centrally-concentrated starbursts may reflect the final stage of star-formation before quiescence \citep{Elbaz18} and have very short depletion times, which become even shorter for more compact systems \citep{Franco20,Gomez-Guijarro22}.

On the other hand, the low dark matter fractions measured in these galaxies are similar to those of $z=0$ massive, fast-rotating ellipticals, suggesting these star forming galaxies may be the progenitors of today's massive ellipticals \citep{Genzel17,Genzel20}. Moreover, there isn't clear observational evidence for the direct detection of distinct centers in less massive, potential Milky Way progenitors at higher redshifts. In contrast to the results of massive galaxy studies, some observations suggest that cores are not fully formed in Milky Way progenitors at high redshift and significant mass evolution occurs at all radii \citep{vanDokkum13}. In this scenario, bulges likely form alongside disks through migration of star forming clumps \citep{Dekel09a,Dekel09}. This is driven by the accretion of cold gas filaments into the disk \citep[as seen in simulations][]{Keres05}. The massive amount of gas entering the disk subsequently fragments into dense clumps through violent disk instabilities and these clumps become sites of intense bursts of star-formation. Dynamical friction then causes these clumps to migrate to the center of the galaxy and then merge to form a bulge \citep{Dekel09a,Dekel09,Ceverino15,Mandelker17,Renzini20}. Observations support this formation pathway, as clumps closer to galactic centers tend to be older and less active \citep{Guo12}, though other secular evolution processes \citep[e.g. bar instabilities,][]{Kormendy04} may also be play a role.

In another scenario, a compaction event causes the dissipative collapse of cold gas into the galactic center \citep{Dekel14,Zolotov15,Tacchella18,Nelson19}. The so-called wet disk contraction triggers a burst of star formation in the center of the galaxy resulting in a so-called ``blue nugget''. Compaction events can be caused by a number of physical mechanisms, including violent disk instabilities triggered by cold gas accretion \citep{Dekel14}, mergers \citep[which can also cause disk instabilities, see][]{Zolotov15,Inoue16}, and collisions of counter-rotating streams \citep{Dekel19}. The removal of gas from the disk, as well as other internal (star-formation and AGN feedback) or external (halo heating of CGM gas) processes, then leads to the galaxy quenching. However, the galaxies may only stay quenched permanently if they have reached a certain cutoff in halo mass and the CGM is heated to the point where cold mode accretion stops. If this threshold is not reached, these galaxies may be able to regrow their disks \citep{Tacchella16}. This could explain the discrete bimodal distribution seen in chemical diagrams of the Milky Way center \citep{Queiroz20}: $\alpha$-rich, metal-poor populations (the bulge and chemical thick disk) are formed in an early burst caused by gas contraction which briefly causes the galaxy to quench and eject gas out of the galactic center, then cold mode accretion resumes forming an $\alpha$-poor, metal-enhanced population in the thin disk that mixes into the bulge/thick disk over time.

The progenitors of today's Milky Way/$L^*$ galaxies are less massive than the galaxies studied in \cite{Genzel17} or \cite{Genzel20}. Due to their location in the middle of the SFMS, lower-mass ($\log(M_\star/M_\odot)<11$ UV/optically-selected galaxies have been identified as potential candidates for progenitors of Milky-Way-like galaxies \citep{Giavalisco96,Steidel96,Papovich01,Shapley01,Giavalisco02,Williams14,Steidel16}. 
Moreover, chemical and kinematic studies of the bulge of the Milky Way in \cite{Queiroz21} suggest that the bulge is most likely an old, pressure-supported component, which formed around $z\sim2-3$. As such, observations of these lower-mass, UV/optically-selected galaxies galaxies between $2<z<3$ could reveal forming bulges and dense centers and help constrain the physical processes behind bulge formation in lower-mass galaxies. In this paper, we search for these objects in a sample of UV-bright, low-obscuration, star-forming galaxies at $z\sim2.3$ in the Great Observatories Origins Deep Survey-North (GOODS-N). 

\begin{figure*}[ht!]
    \centering
    \includegraphics[width=\linewidth]{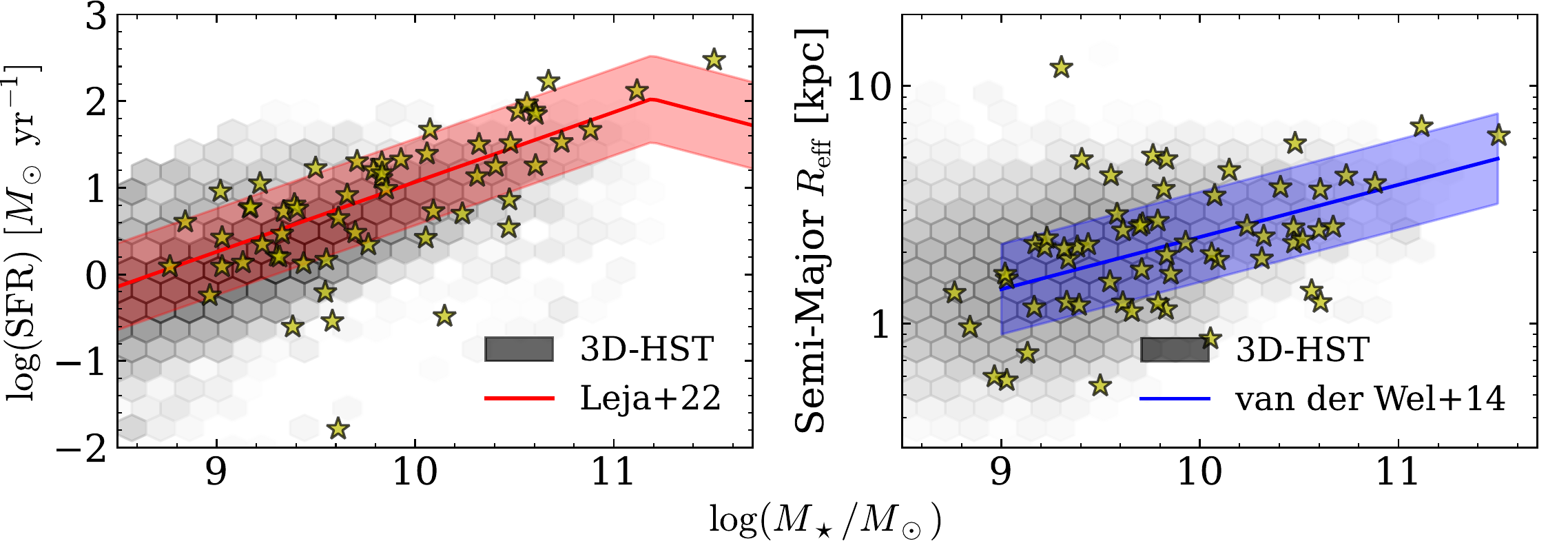}
    \caption{Location of sample galaxies (yellow stars) on the SFMS (left) and the star-forming galaxy size-mass relation(right). Galaxies in this sample occupy the middle of the SFMS at $z\sim2.3$ and fall within the scatter of the size-mass relation. Grey hexbins indicate an underlying sample of galaxies from the 3D-HST morphological catalog \citep{vanderWel14}. The SFMS from \cite{Leja21} is shown with a 0.5 dex scatter in red and the size-mass relation and intrinsic size scatter of 0.19 dex are from \cite{vanderWel14}. This sample contains an overabundance of compact galaxies ($>0.19$ dex below the size-mass relation).}
    \label{fig:sample}
\end{figure*}

It is important to note that throughout this paper the term centers or cores is used in lieu of bulges to refer to the compact and dense central regions of the galaxies under consideration, as ``bulge'' may invoke the stricter definition of a classical bulge (or pseudobulge) used in connection to local galaxies. In the Milky Way this central component includes the chemical thick disk, which is also believed to form around $2<z<3$ and is confined to the inner few kpcs of the galaxy \citep{Queiroz21, Miglio21}. Thus, the centers in this paper refer to the central regions of the galaxies with no consideration, in defining the term, to their star formation activity, age, metallicity, light profile or dynamical state, and likely includes significant contributions from a forming bulge and thick disk. In fact, the primary goal of this paper is to investigate if the central regions of star-forming galaxies at the cosmic noon that are plausible candidates for today's MW-like galaxies are already characterized by different star formation history than the outer regions. Similarly, the term outskirts is used instead of disk, as this refers to the outer regions of the galaxy which likely covers the thin disk with some contamination from the outer thick disk.

We reconstruct the star formation history of our target galaxies using the fully-Bayesian \textsc{Prospector} code \citep{Johnson21}, which we use to fit the sensitive CANDELS \citep{Grogin11,Koekemoer11} panchromatic spectral energy distributions (SEDs) of the two resolved, color-selected subcomponents, the center and the outskirts, to spectral populations synsthesis evolution models. To achieve as accurate age measures as possible, we also use spectroscopic redshift and gas-phase metallicity measures for our galaxies from the MOSFIRE Deep Evolution Field (MOSDEF) spectroscopic survey \citep{Kriek15} as strong priors in the fits. MOSDEF is also biased towards UV-selected galaxies (due to lower spectroscopic success rate for red galaxies), which provides ideal targets for this study. Metallicity measurements, along with coverage in the observed infrared (IR) from \textit{Spitzer}/IRAC can help break the age-metallicity degeneracy, while the rest-frame optical spectroscopy also provides  accurate redshifts for SED fitting. 

The primary goal of this study is to study the star-formation history of the central regions and outskirts of our targets to help provide empirical constraints the mechanisms of their assembly. If the central regions of these galaxies have been following substantially different evolutionary path than the outskirts, we should see this in the resulting star formation histories (SFHs): centers could form earlier than or coevally with the outskirts but most models expect the formation to be bursty, occur later on, and decline faster than the outer region \citep{Dekel09,Guo12,Franco20,Dimauro22} potentially in an outside-in fashion, i.e. centers exhibit a declining and sharply peaked SFH when compared to the outer components. 

In Section \ref{sec:data}, we describe the photometric and spectroscopic data used in SED fitting and core decomposition. Section \ref{sec:analysis} explains the techniques used to decompose galaxies into resolved central and outer components and deal with unresolved photometry ($K$-band/IRAC), as well as the \textsc{Prospector} model we use to fit SEDs. We discuss the resulting SFHs and their impact on our understanding of galaxy formation in Section \ref{sec:sfhs}. A summary of the conclusions is presented in Section \ref{sec:summary}. Throughout the paper we assume a flat $\Lambda$CDM cosmology with $\Omega_m=0.3$, $\Omega_\Lambda=0.7$, and $\mathrm{H}_0=70\mathrm{~km~s}^{-1}~\mathrm{Mpc}^{-1}$, as well as a Kroupa initial mass function \citep{Kroupa01} for stellar masses. All magnitudes are in the AB system.

\section{Data and Sample Selection}\label{sec:data}
Our sample consists of 60 $z\sim2.3$ galaxies from the GOODS-N field. GOODS-N is chosen because has both rest-frame optical spectroscopy from MOSDEF \citep{Kriek15} and photometric measurements ranging from the rest-frame ultraviolet (UV) to mid-IR (MIR). The sample of galaxies is chosen by matching sources without photometric contamination by other adjacent sources (\texttt{FLAGS=0}) from the the multiwavelength \textit{Hubble Space Telescope} (HST) $H_\mathrm{F160W}$-selected CANDELS/SHARDS catalogs of \cite{Barro19} to a sample of MOSDEF galaxies from \cite{Sanders18}, which has robust spectroscopic redshifts and metallicities measured at $z\sim2.3$. Roughly 5\% of the \cite{Barro19} catalog is removed by applying the contamination flags selection. Of this non-contaminated sample, only 464 galaxies (1.3\%) have MOSDEF coverage and 63 of those are in the metallicity sample from \cite{Sanders18}. The \cite{Sanders18} sample also removes potential AGN using mid-IR (MIR) \textit{Spitzer}/IRAC selections presented in \cite{Coil15}, as well as X-ray and emission line selections. Figure \ref{fig:sample} shows the final sample of galaxies relative to the SFMS (left) and the galaxy size-mass relation (right). These galaxies have an average stellar mass of $\log(M_\star/M_\odot)=9.75\pm0.29$ and an average size of $R_{\mathrm{eff}}=2.16\pm1.86$ kpc. The median SFR is $\dot{M}=39.2\pm26.5~M_\odot~\mathrm{yr}^{-1}$ on a 100 Myr timescale and $\dot{M}=7.5\pm6.1~M_\odot~\mathrm{yr}^{-1}$ on a 10 Myr timescale. Stellar masses and SFRs are measured from the global \textsc{Prospector} fits discussed in Section \ref{sec:analysis} and sizes are taken from the 3D-HST morphological catalogs \citep{vanderWel14}. In general, the galaxies span a wide range of masses and SFRs, covering most of the SFMS at $z\sim2.3$. These galaxies also generally fall within the intrinsic size scatter (0.19 dex) of the $z=2.25$ star-forming galaxy size-mass relation from \cite{vanderWel14}, though 17\% of the sample lies $>0.19$ dex below the relation, suggesting a slight overabundance of compact galaxies.

\subsection{Photometric Data}\label{sec:phot}
Photometric data is taken directly from the CANDELS/SHARDS catalogs \citep{Barro19} for each cross matched galaxy. In the UV, we include Kitt Peak North \textit{U}-band photometry from the Hawaii Hubble Deep Field North survey \citep{Capak04}. The optical photometry is composed of HST/ACS observations in the F435W, F606W, F775W, and F850LP filters from GOODS \citep{Giavalisco04}, as well as F814W from CANDELS \citep{Grogin11,Koekemoer11}. CANDELS data is also used in the near-IR (NIR) with the HST/WFC3 F105W, F125W, and F160W filters, with additional NIR measurements coming from HST/WFC3 F140W in the AGHAST survey GO: 11600 (PI: B. Weiner) and Subaru/MOIRCS \textit{K}-band \citep{Kajisawa11}. The photometry is rounded out by MIR \textit{Spitzer}/IRAC observations in the 3.6, 4.5, 5.6 and 8 $\mu$m filters \citep{Ashby13,Dickinson03}. Stellar masses from the catalog are also included as initial estimates for the stellar mass of the galaxy. These masses are derived with the FAST code \citep{Kriek09}.

\begin{figure*}
    \centering
    \includegraphics[width=\linewidth]{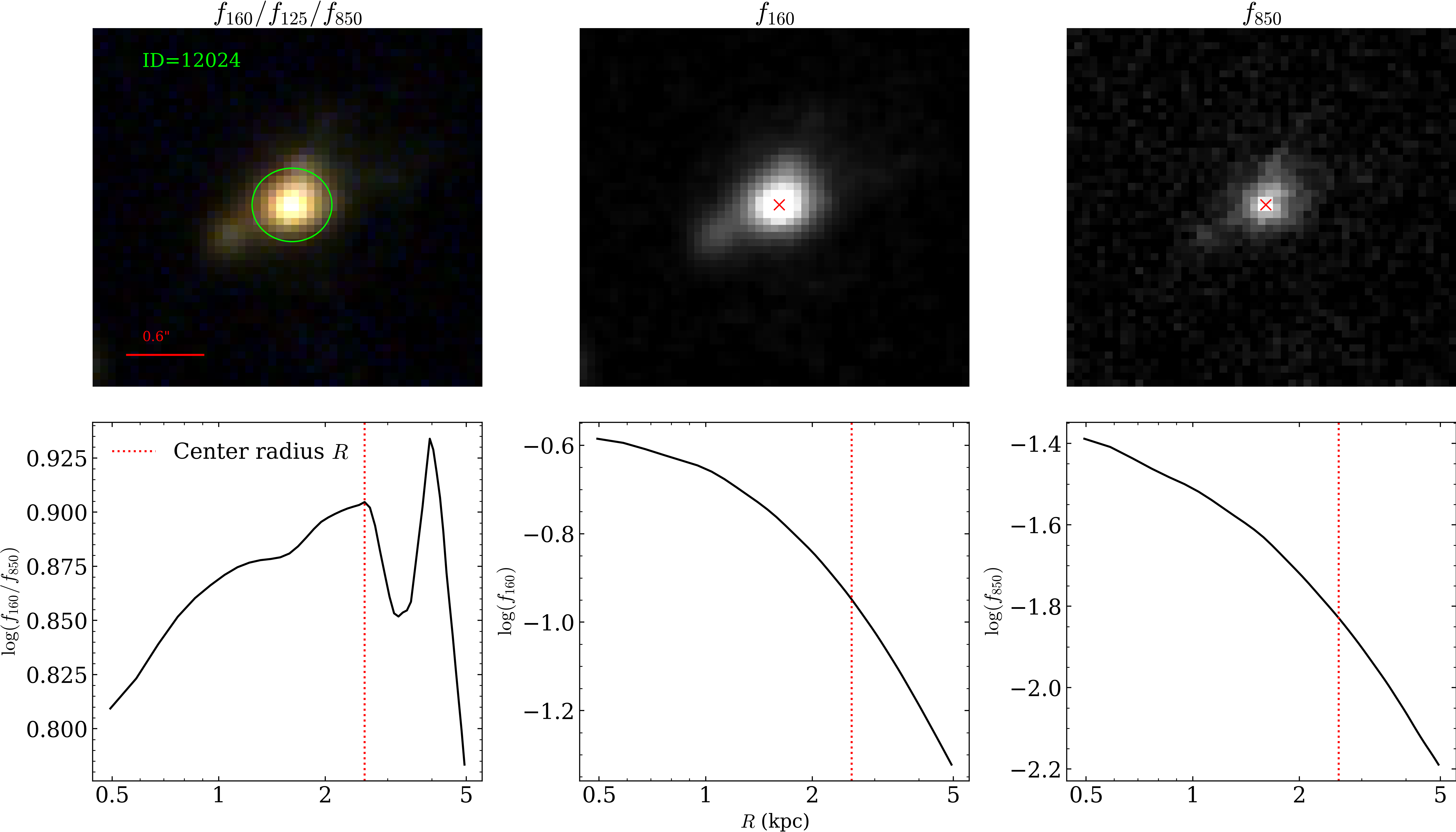}
    \caption{Example central-outer decomposition for a galaxy in our sample. The left column shows the target galaxy in a 3-color image: red=F160W, green=F125W, blue=F850LP. The bottom panel shows the growth curve of the $z_\mathrm{F850LP}-H_\mathrm{F160W}$ color (equivalently $\log(f_\mathrm{F160W}/f_\mathrm{F850LP})$) with increasing aperture radius $R$. Apertures are centered on the F160W centroid (red x in middle panel). The selected center aperture is indicated with a green circle in the 3-color image and with a dotted red line in the growth curve. The middle and right columns show the F160W and F850LP images of the target galaxy, with corresponding light curves below. Centroids for both filters are shown with red x's. The secondary peak in the $z_\mathrm{F850LP}-H_\mathrm{F160W}$ color is due to the depth of the F850LP image: since the F850LP imaging is shallower than F160W, enclosing more of the faint outskirts of the galaxy makes the growth curve tend redder, even though the intrinsic color might not be changing.}
    \label{fig:decomp}
\end{figure*}

\subsection{Spectroscopic and Metallicity Data}
\textit{H}-, \textit{J}-, and \textit{K}-band (rest-frame optical) spectroscopic measurements are incorporated into the sample of galaxies from GOODS-N using MOSDEF \citep{Kriek15}. MOSDEF is ideal for a sample of low-mass, UV/optically-selected galaxies, as the survey is biased towards galaxies in the middle of the SFMS in particular \citep[see Fig. 16 in][]{Kriek15}. However, due to the low S/N continuum of the MOSDEF spectra, this spectroscopy proved unsuitable for direct use in our SED fitting methods. Instead, we utilize spectroscopic redshifts and metallicities measured from MOSDEF spectra by \cite{Sanders18}. This sample contains galaxies with robustly measured redshifts at $2\leq z\leq2.7$, $\log(M_\star/M_\odot)\geq9$, and high S/N H$\alpha$ and H$\beta$ measurements. AGN are also excluded from this sample with a combination of IR, X-ray, and emission line diagnostics. Multiple indicators are used to measure the oxygen abundance of each galaxy. Cross-matching via R.A. and Dec. to the \cite{Sanders18} catalog returns our final sample size of 60 galaxies. Oxygen abundances are converted to metallicities using a solar oxygen abundance of $12+\log(\mathrm{O/H})_\odot=8.69$ \citep{Asplund09}.

\section{Analysis}\label{sec:analysis}
In this section we discuss the methods used to measure the SEDs of resolved subcomponents of galaxies. Galaxies are decomposed into central and outer components with color selections in Section \ref{sec:decomp}. In Section \ref{sec:prospect}, we discuss the \textsc{Prospector} code \citep{Johnson21} and the various settings and templates used. Subsections \ref{sec:met} and \ref{sec:dust} discuss specific priors and assumptions used in the SED modeling, namely the metallicities and dust-obscured star formation. Lastly, Section \ref{sec:iter} includes a description of the iterative method used to estimate the unresolved IRAC photometry of the central and outer components.

\subsection{Center and Outskirts Decomposition}\label{sec:decomp}
In the present universe, the Milky Way bulge and thick disk is known to be older and redder than the surrounding regions of the galaxy \citep[e.g.,][]{Bensby17,Barbuy18,Queiroz21,Miglio21}, possibly forming at $2<z<3$. In particular, \cite{Queiroz21} detect a population of counter-rotating stars in the Milky Way bar and bulge, which could potentially be remnants of the instability-induced clumpy star formation that helped form the bulge and is observed in star-forming disks at $2<z<3$ \citep{Elmegreen08,Huertas-Company20}. As such, a forming stellar core can potentially be identified via its rest-frame optical colors. Using resolved, observed-frame $z_\mathrm{F850LP}-H_\mathrm{F160W}$ colors, we decompose each galaxy in our sample into separate central and outer components. This specific color is chosen because it spans the 4000\AA{} break, which is a known age indicator and can help identify older stellar populations \citep[e.g.,][]{Kauffmann03,Wild09}. Moreover, the central regions of galaxies can often experience significant dust attenuation in their centers \citep[e.g.,][]{Tadaki17,Tadaki20,Elbaz18,Tacchella18,Puglisi19,Puglisi21,Franco20,Gomez-Guijarro22}, which in turn would allow us to select dense galactic cores by looking for significantly red regions of galaxies. Thus, we should be able to detect central regions that form in a number of ways, including mechanisms such as accretion of star forming clumps and central starbursts from mergers or gas accretion.

Image cutouts of each galaxy in both filters and segmentation map cutouts are made with Montage v6.0. Nearby objects are masked and the centroid in F160W is measured via the center of mass. A series of 50 circular apertures with radii ranging from 1 to 10 pixels (0.06\arcsec{} to 0.6\arcsec{}) are placed at the centroid and aperture photometries for both F850LP and F160W are measured. Fixed circular apertures are used because any dense central core at this redshift would likely be small enough to fall within the point spread function (PSF) FWHM. We then choose the center aperture that maximizes the color (i.e. contains the reddest flux within). To do this, we impose a condition that chooses a local maximum if the following local minimum (at a larger aperture radius) is less than 0.01 times the local maximum color, and the global maximum otherwise. This prevents overestimation of the center aperture by secondary peaks driven by background flux or contamination from dusty star-forming clumps in the outskirts. An example decomposition is shown in Figure \ref{fig:decomp} in a color image (F160W/F125W/F850LP for the red, green, and blue filters, respectively), as well as F160W and F850LP individually. The central aperture is shown with a green circle and highlighted by a red point in the inset $z_\mathrm{F850LP}-H_\mathrm{F160W}$ growth curve. The color images and apertures of all 60 galaxies are shown in Appendix \ref{sec:montage}.

The central apertures determined by this method are then confirmed visually, and in cases where a distinct center is not visually identifiable we rely on the identification from color selection. As another check, we examine the residuals of best-fit exponential disk models for a subsample of 10 galaxies. Subtracting the exponential disk model from the galaxy should leave behind light from the dense stellar core (or other clumpy regions of star-formation, which we discuss later), making the presence of this central feature apparent.  This test ensures the color-selected central components correspond to distinct morphological regions of the galaxies. The exponential disk fits are not used to subtract the outskirts' contributions from the centers. Exponential disks are fit to the F160W cutouts of the subsample using GALFIT \citep{Peng02,Peng10} and the best-fitting model is subtracted from the image cutout. For all galaxies in the subsample, the color-selected center aperture encloses the majority of the residual flux, suggesting that these selections are capturing the dense stellar core. Centers selected for these galaxies have an average radius of $R=2.46\pm0.47$ kpc, compared to an average half-light radius of $R_\mathrm{eff}=2.16\pm1.86$ kpc. 58\% of the galaxies in the sample have a central size greater than their half-light radius, and the median ratio of the two sizes is $R/R_\mathrm{eff}=1.11$, suggesting the defined centers tend to contain roughly half of the light in a given galaxy.

The center fluxes are then computed via aperture photometry for each HST band. Photometric uncertainties are computed by summing in quadrature the sigma image ($1/\sqrt{w}$, where $w$ is the weight image). The corresponding outskirts fluxes are computed by subtracting the center from the total catalog fluxes for each filter. Outskirts uncertainties are computed by subtracting the center photometric error from the catalog flux error in quadrature. 

\begin{figure*}
    \centering
    \includegraphics[width=0.78\linewidth,trim={1.5cm 1cm 1cm 0.5cm},clip]{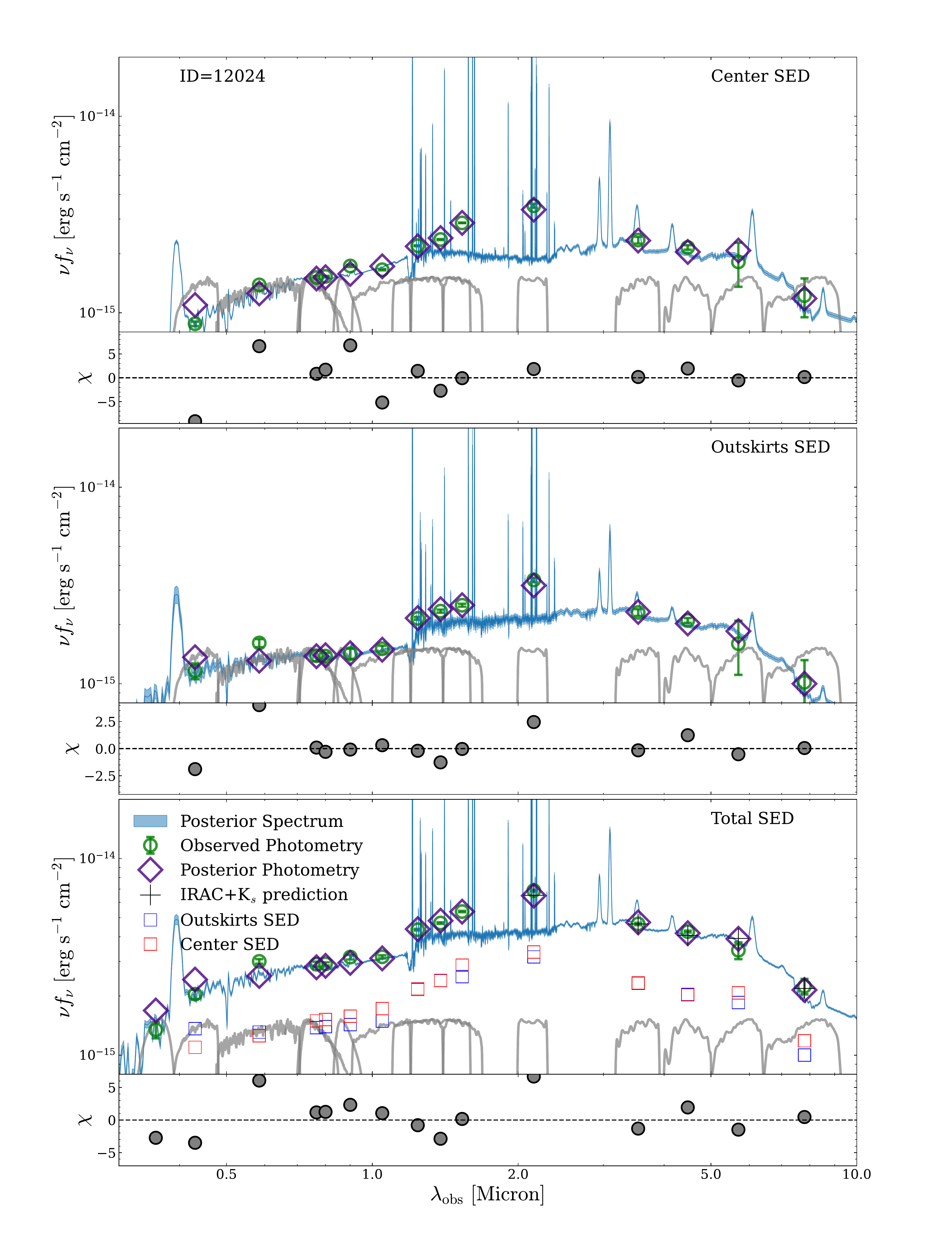}
    \caption{Example SED fits to a galaxy from our sample. The 3 figures show the SEDs of the center, outskirts, and total galaxy from top to bottom. The top panel in each figure shows the SED and the bottom shows the uncertainty-normalized residual ($\chi=(f_\mathrm{obs}-f_\mathrm{model})/\sigma_{f,\mathrm{obs}}$) between the observed and predicted photometry as a function of wavelength. The x-axis of each figure is the observed wavelengths in microns. The blue line is the best-fit spectrum from \textsc{Prospector} with the 16$-$84 percentile range indicated by the shaded region. Best-fit photometry is shown with purple diamonds, and error bars indicate the 16$-$84 percentile range. Observed photometry and uncertainties (including converged IRAC and $K$-band data for the center and outskirts) are shown with black and green circles. For comparison, the center and outskirts best-fit photometries are shown with red and blue squares in the bottom figure, and the best-fit IRAC and $K$-band photometry from both components is added and indicated with black crosses. Grey curves show the transmission for all included filters.}
    \label{fig:SED_ex}
\end{figure*}

As mentioned earlier, these centers also likely include significant contribution from the chemical thick disk. However, we are interested in the formation and evolution of the galactic center as a whole, so investigating the evolution of both these components (bulge and thick disk) is useful in understanding the galaxy formation histories as a whole. Moreover, most quenched galaxies at $z\sim2$ tend to be fast rotators with large velocity dispersions \citep{Newman18}, and thus are most likely dominated by chemical thick disks, so understanding the formation of this component as well as the rest of the galactic center might help explain this behavior.

\subsubsection{Impact of Dust Attenutation and Obscured Star Formation}\label{sec:dust}
One potential contaminant in our central SFHs is dust-obscured star formation. Dust attenuation is known to play a large role in the centers of massive star-forming galaxies and starbursts \citep[e.g.][]{Tadaki17,Tadaki20,Elbaz18,Tacchella18,Puglisi19,Puglisi21,Franco20,Gomez-Guijarro22}, and could potentially contribute significant obscured star formation to the centers of these less-massive galaxies as well. Since our photometry only samples out to the rest frame NIR ($<2~\mu m$), the SFHs measured from these SEDs will likely not capture these effects. 

To constrain the impact of dust-obscured star formation, we check for far-IR (FIR) counterparts to our galaxies in the GOODS-N ``super-deblended'' catalog presented in \cite{Liu18}. We find only 3 of our sources have significant (S/N$>1$) flux at $850~\mu m$ when crossmatching to FIR sources within 1\arcsec{}. These FIR measurements are not robust detections (S/N$<2$), but will allow us to compute upper limits on the obscured star formation in these galaxies. Using the $S_{850\mu m}-\mathrm{SFR}$ conversion for Lyman-break galaxies \citep[see Eqn. 10 in][]{Webb03}, we measure an upper limit of $<40~M_\odot~\mathrm{yr}^{-1}$ on the three FIR counterparts, much less than the peak SFRs measured from the integrated galaxy SEDs ($>100~M_\odot~\mathrm{yr}^{-1}$). Moreover, the number of detections is a small fraction of the total sample (5\%) and is limited to some of the most massive galaxies in the sample ($\log(M/M_\odot)>10.8$), which are already known to have more significant contributions from obscured star formation. 

The marginal detections (if detections at all) for a small fraction of sample galaxies, the higher mass of these detections with relation to the whole sample, and the relatively low dust-obscured SFRs suggest that optically-thick dust obscuration of star formation (i.e., star formation that does not appreciably contribute to the UV/optical SED) does not seem to be a common and substantial contribution to the total SFR budget in the sample considered here.

\subsection{\textsc{Prospector} Inputs}\label{sec:prospect}
SED fitting is done using the fully-Bayesian \textsc{Prospector} code \citep{Johnson21}. \textsc{Prospector} forward models observed data (both spectra and photometry) of composite stellar populations (CSPs) given a set of parameters describing the CSP (e.g. mass formed, metallicity, dust extinction, etc.) and other observational effects (e.g. filters, redshift, etc.), which can either be fixed or varied. Using these models, likelihood and posterior probabilities are computed via comparison to observed data and noise properties. CSPs are generated with Flexible Population Stellar Synthesis \citep[\textsc{FSPS},][]{Conroy09,Conroy10a,Conroy10b} code using \textsc{Python-FSPS} \citep{ForemanMackey14}. 

Priors are also important ingredients to \textsc{Prospector}, given the Bayesian forward-modeling approach this code takes. Since significant parts of SED parameter space can be highly degenerate (e.g. the age-dust-metallicity degeneracy), priors can help shape the posterior distribution and capture more accurate stellar population properties. Moreover, choice of prior is crucial in determining nonparametric (piecewise step function) SFHs with lower quality data. In a non-Bayesian framework (using regularization), \cite{Ocvirk06} determine that only 8 episodes of star formation can be recovered with high quality spectroscopy ($R=10,000$, S/N$=100$). However, \cite{Leja19} find that with \textsc{Prospector}, nonparametric SFHs can be useful with photometry or low-quality spectra if the SFH prior is well tuned.

In this work, the redshift is fixed at the spectroscopic redshift measured from MOSDEF. The metallicity ($\log Z_\star$) is handled with Gaussian priors for SED fits to the total galaxy and both subcomponents. These priors are centered on the median MOSDEF metallicities from \cite{Sanders18} with widths equal to the 16-84th percentile range from the \cite{Sanders18} metallicity sample. The Gaussian priors are only applied to the outer component because we assume that the metallicity of the galaxy should be dominated by the metallicity from the outskirts. This is supported by the fact that the Milky Way bulge consists of an older, metal poor population \citep[e.g.,][]{Queiroz21} and thus centers of star forming galaxies at $z\sim2.3$ may be more metal poor than the rest of the light from the galaxies. The impact of this metallicity prior on the different components, as well as the robustness of the integrated galaxy SED fit with and without a metallicity prior, are discussed in Section \ref{sec:met} and Appendix \ref{sec:sfhcomp}, respectively. Dust and nebular emission are incorporated with the $V$-band optical depth ($\tau_V=1.086A_V$) and ionization parameter ($U_\mathrm{neb}$) allowed to vary.

\begin{figure}[t!]
    \centering
    \includegraphics[width=\linewidth,trim={0 0 0 0cm},clip]{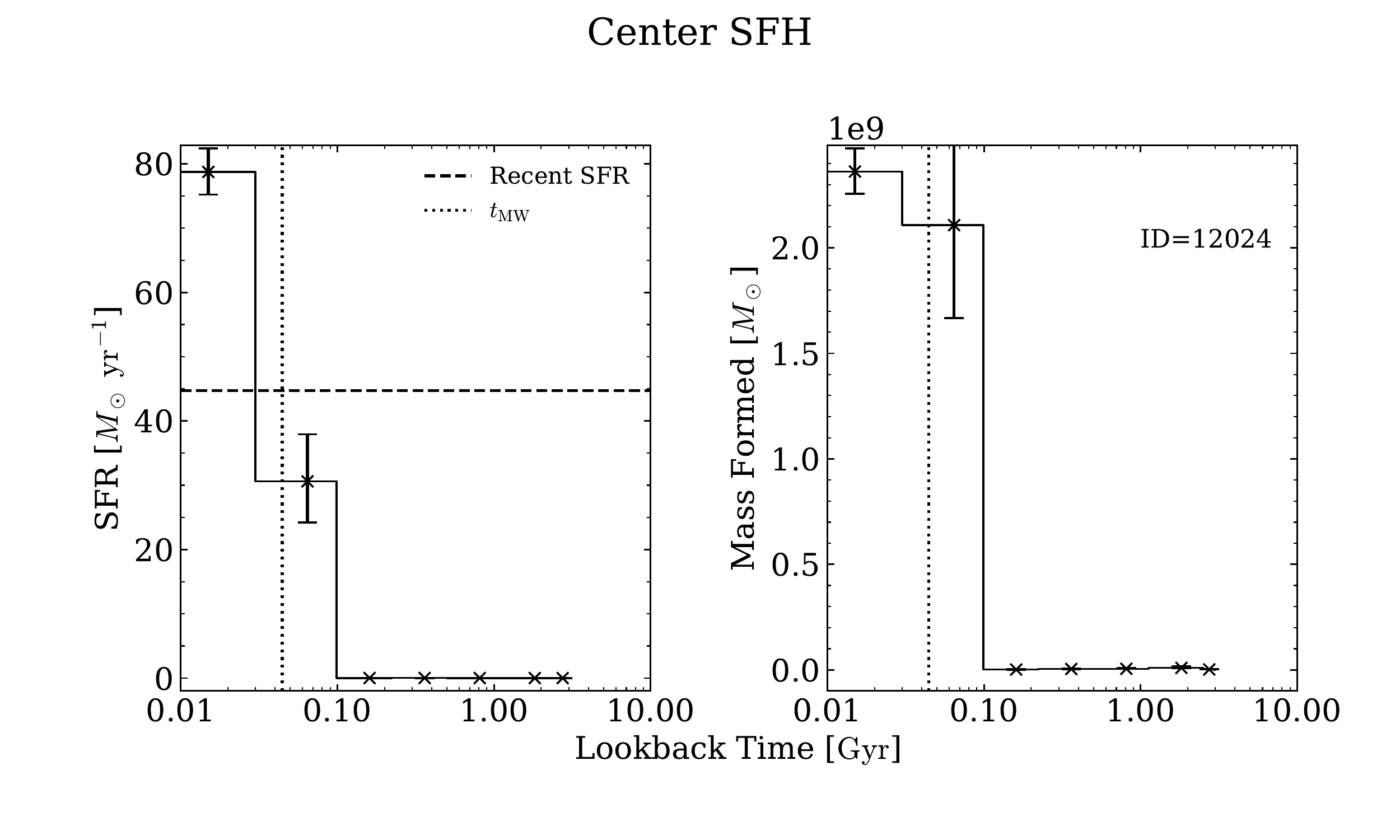}
    \includegraphics[width=\linewidth,trim={0 0 0 0cm},clip]{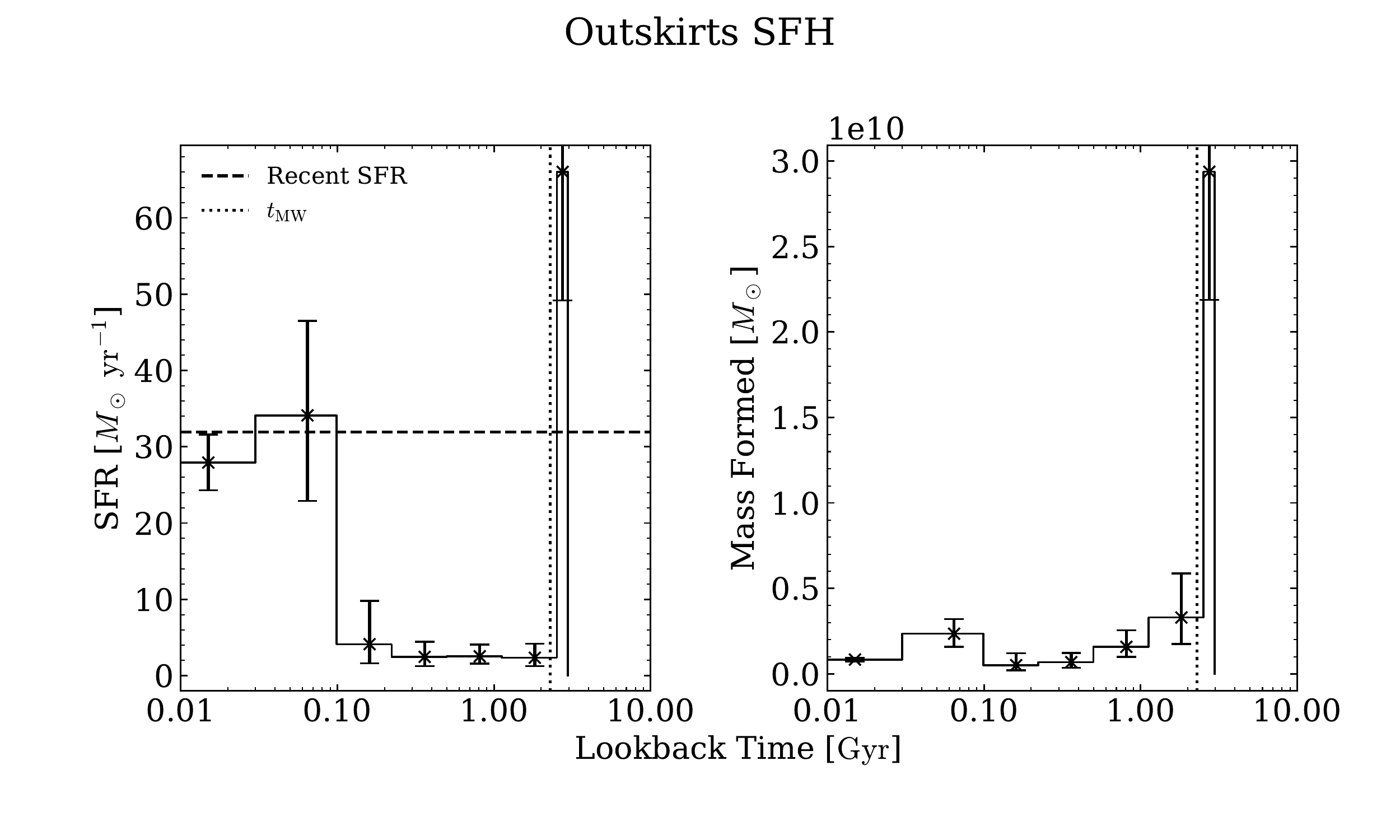}
    \includegraphics[width=\linewidth,trim={0 0 0 0cm},clip]{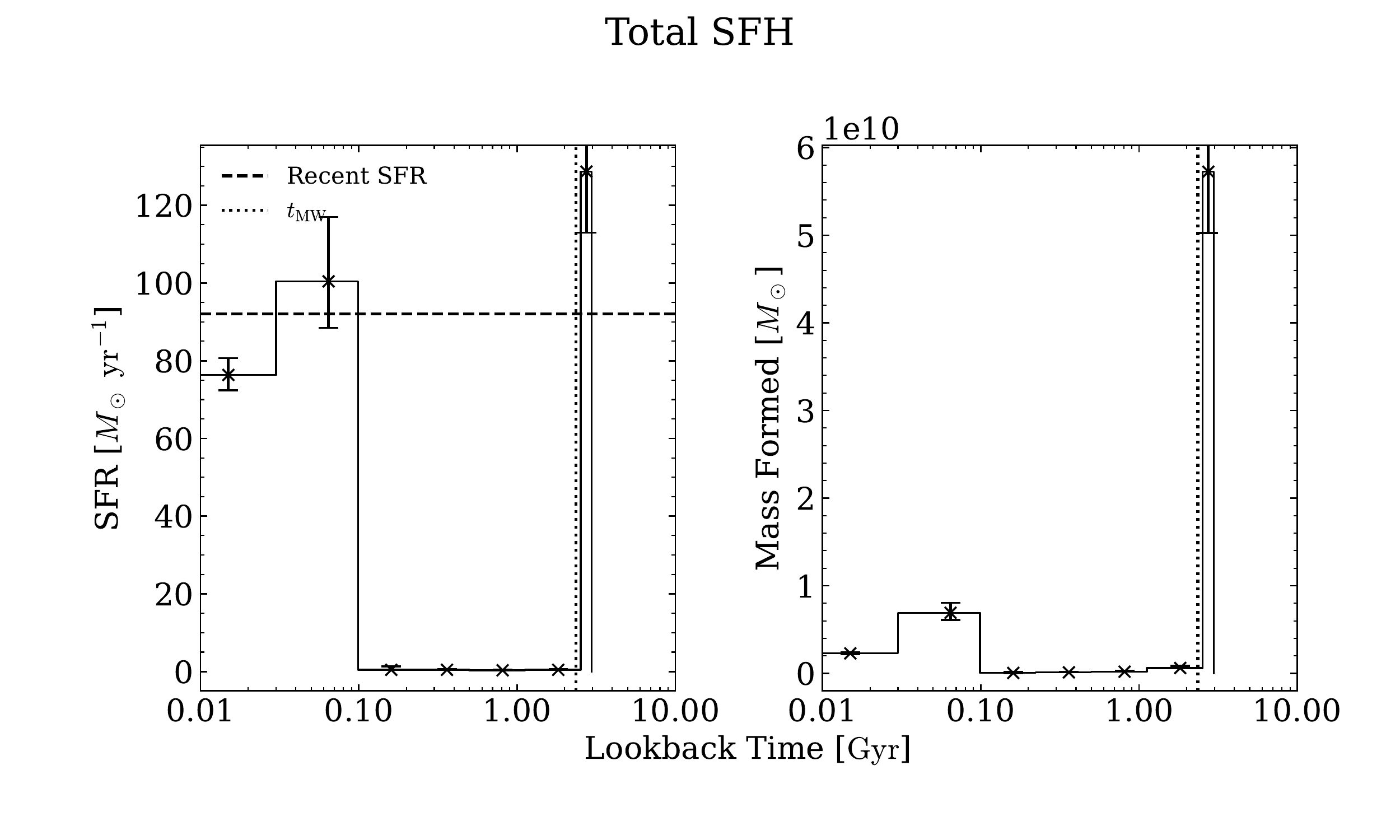}
    \caption{Example best-fit SFHs for a galaxy from our sample. The three figures show the SFHs for the center, outskirts, and total galaxy from top to bottom. For each figure, the left and right panels show the SFR and mass formed in each age bin as a function of lookback time from observation (i.e. a lookback time of 0 Gyr is the galaxy at the observed redshift). The errorbar in each age bin shows the 16$-$84 percentile range. The dotted and dashed lines show the mass-weighted ages and recent (past 100 Myr) SFRs for the 2 subcomponents and total galaxy.}
    \label{fig:SFH_ex}
\end{figure}

SFHs are modeled nonparametrically using the continuity prior described in \cite{Leja19}. This prior fits directly for the ratio of star formation rates (SFRs) between bins ($\Delta\log(\mathrm{SFR})$), which weights against discontinuities in the SFH (via sharp jumps or drops in the SFR). Each individual ratio ($\log(\mathrm{SFR}_i/\mathrm{SFR}_{i+1})$) is drawn from a Student's-t distribution, as in \cite{Leja19}, with an initial value of $\log(\mathrm{SFR}_i/\mathrm{SFR}_{i+1})=0$ for all SFR bins $i$. The SFH is computed over 7 fixed age bins (in lookback time from the observed redshift): 
\begin{align}
    0<&t<30~\mathrm{Myr}\nonumber\\
    30<&t<99~\mathrm{Myr}\nonumber\\
    99<&t<218~\mathrm{Myr}\nonumber\\
    218<&t<479~\mathrm{Myr}\\
    479~\mathrm{Myr}<&t<1.06~\mathrm{Gyr}\nonumber\\
    1.06<&t<2.32~\mathrm{Gyr}\nonumber\\
    2.32<&t<\sim2.7~\mathrm{Gyr},\nonumber
\end{align}
where the last bin is adjusted to account for the age of the universe at the observed redshift. Only 5 age bins are used when decomposing the IRAC and $K$-band photometry, as discussed in the next section. The smaller size of the oldest bin also allows for a maximally old population \citep[see][]{Leja19}. The total mass formed in the best-fit SFH ($M_F$) is left as a free parameter and an initial guess is taken from the CANDELS/SHARDS measurements from FAST \citep{Barro19}. For subcomponents, this initial guess is scaled by the fraction of light in F160W:
\begin{align}
    M_{x}=M_\mathrm{tot}\left(\frac{f_\mathrm{F160W,x}}{f_\mathrm{F160W,tot}}\right),
\end{align}
where $x$ refers to either the central or outer component. The mass-weighted age ($t_\mathrm{M}$ or $\langle t_{\star}\rangle_M$) is determined from the total mass formed and the SFH and the total mass formed is corrected to a stellar mass ($M_\star$) using the approximation
\begin{align}
    \log M_\star=\log M_F+1.06-0.24\log(t_\mathrm{M})+0.01\log(t_\mathrm{M})^2
\end{align}
with masses in solar units and ages in years \citep[Eq. 2]{Leja13}. This mass accounts for the total mass in stars and stellar remnants at observation (i.e. excluding mass lost during supernovae, winds, etc.), rather than all the mass formed throughout the galaxy's history.

\begin{figure}
    \centering
    \includegraphics[width=\linewidth]{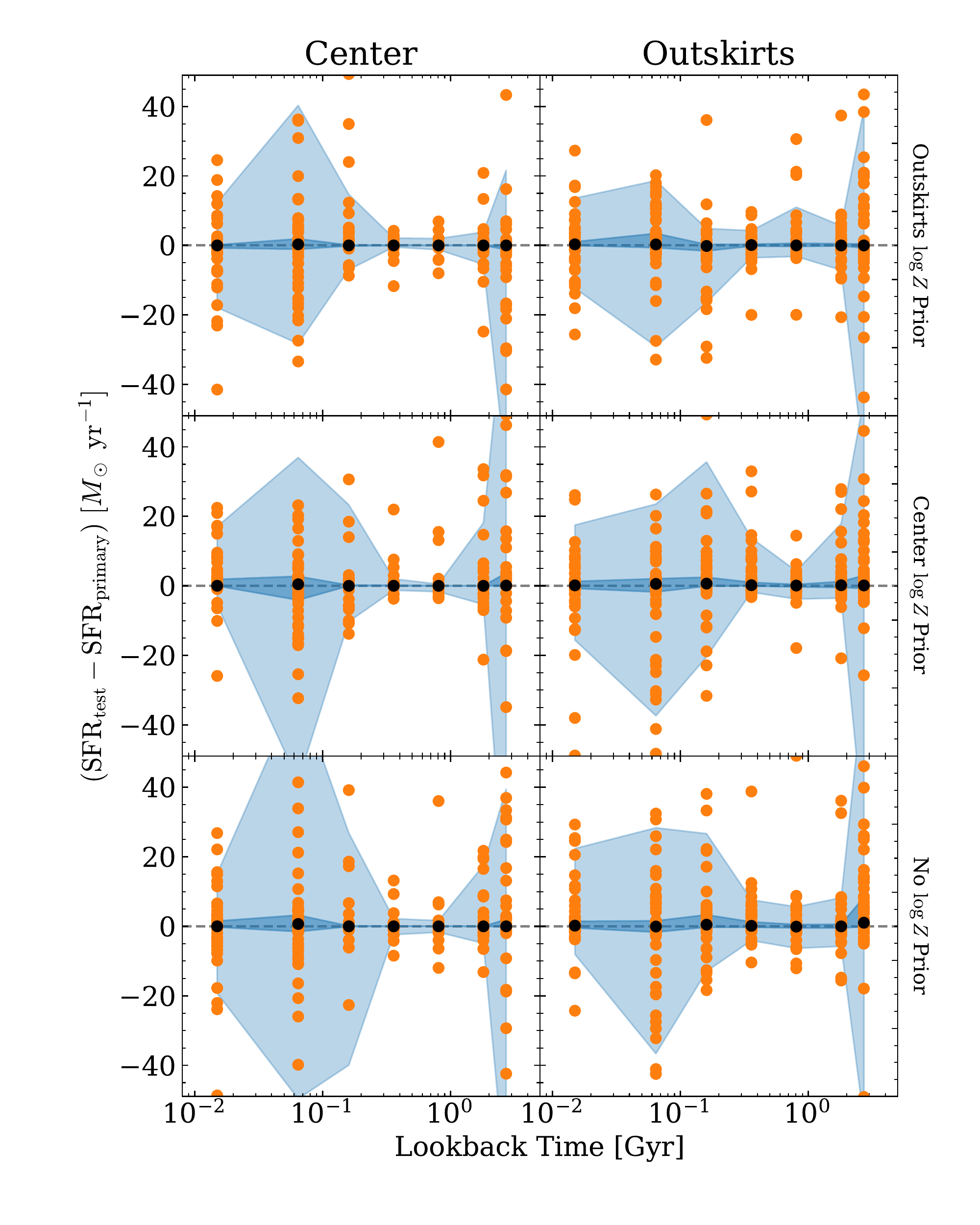}
    \caption{Comparison of star-formation histories for different metallicity prior test runs to the primary run. The resulting SFHs from SED fitting are robust to the choice of metallicity prior. For the primary run, Gaussian priors based on \cite{Sanders18} metallicity measurements are applied to both components. For the three test runs, the \cite{Sanders18} metallicity priors are applied to the outskirts only (top row), the center only (middle row), and neither subcomponent (bottom row). Orange points indicated the difference in the SFH between the test and primary runs at each time bin. Black points show the median SFH difference and light- and dark-blue shaded regions indicate the $16-84$ and $5-95$ percentile range, respectively.}
    \label{fig:Zpriorcomp}
\end{figure}

SEDs are fit to all filters described in Section \ref{sec:phot} where available for the total galaxy. For subcomponents, the $U$-band is not fit because it cannot be resolved (the unresolved $K$-band and IRAC data is dealt with in an iterative process explained in Sec. \ref{sec:iter}). Best-fit parameters are first determined by pure maximum-likelihood estimation via Levenberg-Marquardt (LM). Because much of the likelihood space in SED fitting is non-Gaussian and ill-conditioned, optimization methods like LM are not recommended for actually determining parameter values \citep{Johnson21}. Instead, the LM-determined parameters are used as an initial guess for Monte-Carlo sampling of the posterior  probability distribution function (PDF). Monte-Carlo sampling is done using Monte-Carlo Markov Chains (MCMC) in the \textsc{emcee} code \citep{ForemanMackey13} with 64 walkers and 256 iterations (32 and 128 when decomposing IRAC photometry, see Sec. \ref{sec:iter}). Best-fit parameters are then determined by finding the maximum a posteriori (MAP) sample with uncertainties from the 16th and 84th percentiles of the chain. Example \textsc{Prospector} results for the total galaxy as well as center and outskirts components are shown in Figures \ref{fig:SED_ex} (SEDs) and \ref{fig:SFH_ex} (SFHs), and in Appendix \ref{sec:excov} (covariances). For this example, the SFHs of the outskirts and total galaxy show two distinct peaks of star formation, compared to the single peak in the center. This, combined with the potential tidal features visible in the F160W and F850LP images of the galaxy (Fig. \ref{fig:decomp}), suggest this galaxy may have undergone a recent merger resulting in the increase in star formation in the outskirts. A merger could also have caused the burst of star formation in the center, since mergers may play a significant role in central star formation by driving gas to the center of galaxies \citep[e.g.,][]{Puglisi19,Puglisi21}.\\

\subsubsection{Impact of Adopted Assumptions of Metallicity Priors}\label{sec:met}
Global metallicity measurements from \cite{Sanders18} are used as priors in the SED fitting. Since these measurements are derived from the integrated spectrum, using them as priors for the resolved components requires specific assumptions. In the previous section, we describe how the metallicity prior is applied to both subcomponents equally. Since the metallicities are derived from spectra of the whole galaxy, it should be representative of the metallicity of both subcomponents as well. Moreover, the majority of metallicity gradients in star-forming galaxies are flat at $z\sim2$ \citep{Simons21}, so there shouldn't be significant differences between the center and outskirts metallicities. On the other hand, the centers of galaxies in this sample are brighter and larger than the outskirts on average, and slit losses should effect the outskirts more significantly than the centers. As such, the metallicity measurements would then tend to be better priors for the centers as opposed to the outskirts. To test the effect of various metallicity prior prescriptions, we rerun the entire sample with 3 test runs: applying the \cite{Sanders18} metallicity prior to the outskirts and not the center, applying the prior to the center and not the outskirts, and applying the prior to neither subcomponent. For the component(s) without the \cite{Sanders18} metallicity prior, we apply a top-hat prior ranging from $-3<\log(Z/Z_\odot)<0.2$. These test runs are compared to the ``primary'' run (\cite{Sanders18} metallicity prior applied to both components) and the resulting impact on the SFHs is shown in Figure \ref{fig:Zpriorcomp}. Applying different priors prescriptions has no significant effect on the SFHs on average for any of the test runs, and the $16-84$ percentile range is small ($<5~M_\odot~\mathrm{yr}^{-1}$). While the $5-95$ percentile range is larger, especially for certain time bins, it is still $<40~M_\odot~\mathrm{yr}^{-1}$, significantly less than the peak SFRs measured ($>100~M_\odot~\mathrm{yr}^{-1}$). Thus, we consider the results where the \cite{Sanders18} metallicity prior is applied to both subcomponents, as in the original procedure.\\

\subsection{Iterative Method for Decomposed IRAC Photometry}\label{sec:iter}
The resolution of HST imaging allows for direct measurements of the decomposed center and outskirts fluxes. However, data from other instruments, namely ground-based $K$-band from Subaru/MOIRCS and 4 MIR bands from \textit{Spitzer}/IRAC, is too low resolution for resolved measurements. While fitting to the HST photometry only is possible, the constraints provided by the observed NIR/MIR are crucial in measuring the stellar mass and age, since this wavelength regime traces the rest-frame optical and NIR at $z\sim2.3$. In order to incorporate these important filters into the decomposed SEDs, we use a simple iterative method to estimate the $K$-band and IRAC flux of both the center and outskirts. This also motivates our use of a simple 2-component decomposition in lieu of more complicated techniques \citep[e.g. Voronoi tesselation, as in][]{Fetherolf20}, as the system of equations becomes more difficult to solve with more components. Note the $U$-band photometry is excluded from the decomposed components and this iterative method.

The goal of this scaling is to use the shape of the HST-only SED to estimate initial $K$-band/IRAC SED, as well as the relative contribution of the total $K$-band/IRAC photometry to each component, and iteratively apply these estimated fluxes to future resolved SED fits. This utilizes the resolved information from the HST photometry as well as constraints from the total $K$-band/IRAC fluxes. First, SEDs are fit using fewer walkers and iterations (32 and 128, respectively) to the HST filters only (the number of SFH bins is reduced to 5 for this step so the number of data points is greater than the number of free parameters) in order to get MAP photometry estimates for all HST filters as well as the $K$-/IRAC bands. We then define the corrected $K$-band/IRAC photometry as the MAP predicted photometry multiplied by the ratio of the observed to MAP F160W flux (the longest wavelength HST filter available):
\begin{align}
    f_\mathrm{x,corr}(\lambda)=f_\mathrm{x,MAP}(\lambda)\left(\frac{f_\mathrm{x,obs}(\mathrm{F160W})}{f_\mathrm{x,MAP}(\mathrm{F160W})}\right),
\end{align}
where $\lambda$ refers to one of the $K$-band or IRAC filters and $x$ indicates the center or outskirts subcomponent. The ``MAP'' and ``obs'' labels indicate these are fluxes estimated by \textsc{Prospector} or observed photometries, respectively. This first scaling ensures the estimated $K$-band/IRAC fluxes are consistent with the observed flux of the longest resolve wavelength, as significant residuals between the observed and predicted F160W flux could result in an incorrect correction to the total $K$-band/IRAC flux and increase the number of iterations needed for convergence. 

The corrected $K$-band/IRAC fluxes are then scaled to agree with the observed $K$-band/IRAC fluxes of the total galaxy:
\begin{align}
    f'_\mathrm{x}(\lambda)=f_\mathrm{x,corr}(\lambda)\left(\frac{f_\mathrm{obs}(\lambda)}{\sum_\mathrm{x}f_\mathrm{x,corr}(\lambda)}\right),
\end{align}
where the summation is over the center and outskirts components and $f'_\mathrm{x}(\lambda)$ is the ``observed'' flux of the subcomponents to be fit in the next iteration. The uncertainty in these ``observed'' fluxes is given by
\begin{align}
    \delta f'_\mathrm{x}(\lambda)=\delta f_\mathrm{obs}(\lambda)\sqrt{\frac{f_\mathrm{obs}(\lambda)}{f_\mathrm{x,corr}(\lambda)}}.
\end{align}
This step matches the estimated $K$-band/IRAC flux of the total galaxy via the SED fits to the two components to the observed flux. By doing this, the shape of the IR SED, as inferred by the HST-only fit, is combined with the correct normalization to observed IR photometry. 

\begin{figure}[t!]
    \centering
    \includegraphics[width=1\linewidth]{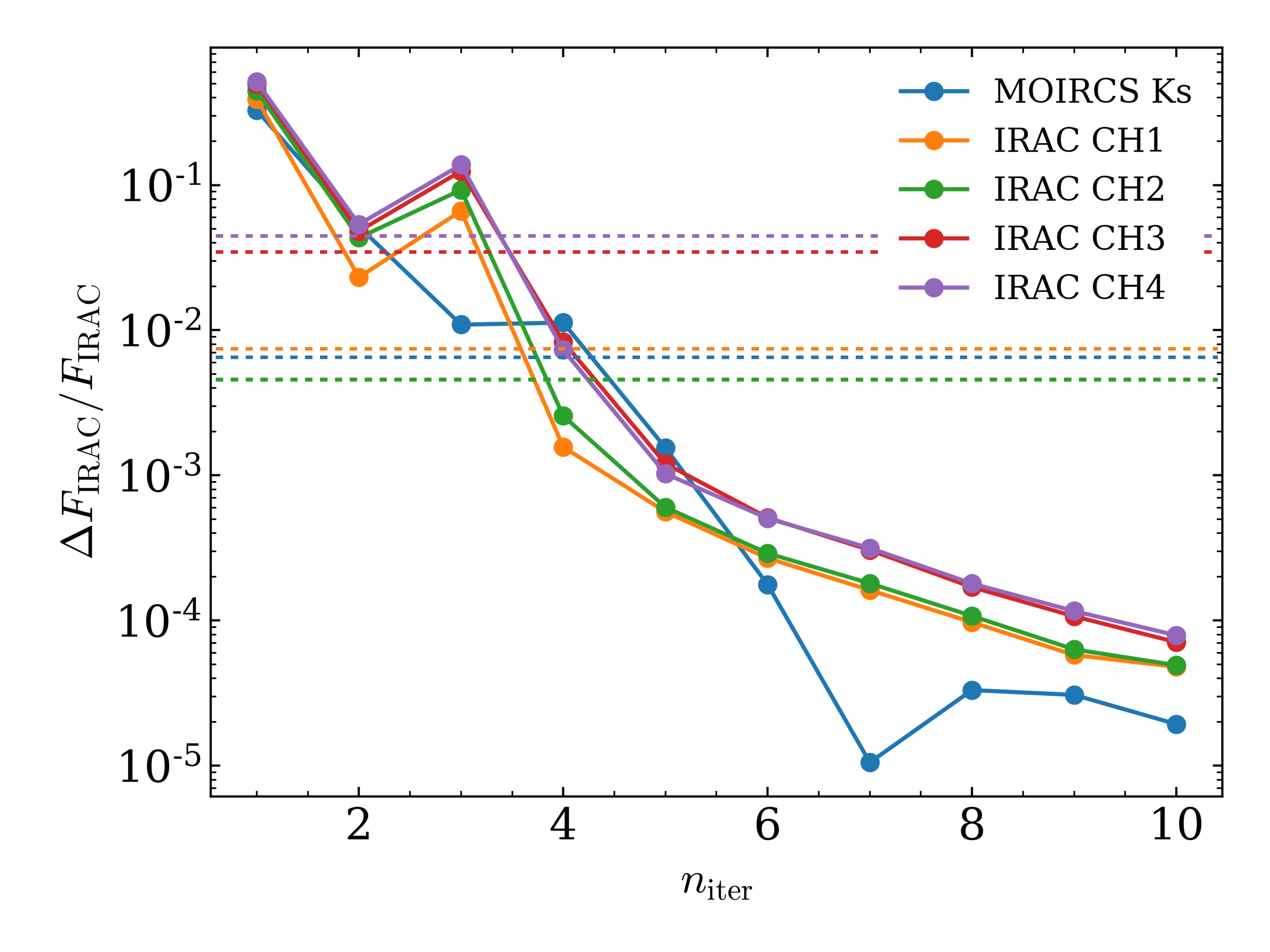}
    \caption{Convergence of the decomposed photometry in the Subaru/MOIRCS $K$-band and \textit{Spitzer}/IRAC channels for the example galaxy from Figures \ref{fig:SED_ex}, \ref{fig:SFH_ex}, and Appendix \ref{sec:excov}. The fractional change in the summed (i.e., $\sum_x f_{x,\mathrm{corr}}$) $K$-band/IRAC photometry is compared to the iteration number, and shows that after 4 iterations the behavior in all but one band is consistent. Dashed lines indicate the fractional error in each band for the example galaxy. After 4 iterations the fractional change in the flux is less than the error in the flux for all IRAC bands.}
    \label{fig:iter}
\end{figure}

The SED is then fit again, this time including the ``observed'' center and outskirts photometries and uncertainties. The process is repeated until the estimated total $K$-band/IRAC flux changes by $<5\%$ in all but one filter (this allows for flexibility when dealing with bands with very large photometric errors). This convergence is generally quick, though due to the longer computational times required to run \textsc{Prospector}, we limit this method to a maximum of 10 iterations. Only 2 of the 60 galaxies reach this limit. The behavior of the fractional change in the total $K$-band/IRAC flux over 10 iterations is show in Figure \ref{fig:iter} (solid lines and points) and compared to the fractional photometric error in each band (dashed lines). After 4 iterations, all bands are below the $5\%$ threshold for the change in flux. At the same times, all bands except for the $K$-band have changed by less than the corresponding fractional error, indicating that further refining the photometry in future iterations is unnecessary.

Finally, \textsc{Prospector} is run with the full number of walkers, iterations, and SFH bins (64, 256, and 7, respectively) on the components using these converged values for the $K$-band/IRAC flux. The summed center and outskirts photometry in the NIR and MIR is shown with black crosses in Figure \ref{fig:SED_ex}. In general, there is good agreement between the observed $K$-band/IRAC flux of the total galaxy and the predicted photometry from this iterative analysis. As mentioned before, including rest-frame optical/NIR is crucial in constraining the age, stellar mass, and SFH of the galaxy \citep[e.g.,][]{Papovich01,Papovich04,Pforr12,Pforr13,Conroy13,Mobasher15}. Moreover, fitting our sample with resolved photometry only results in stellar masses and metallicities that differ from fits including $K$-band/IRAC by $>0.2$ dex, while the mass-weighted ages and recent SFRs can differ by $>20~M_\odot~\mathrm{yr}^{-1}$ and $>1$ Gyr, respectively. As such, including the unresolved $K$-band/IRAC photometry is crucial in our analysis.

\begin{figure}[t!]
    \centering
    \includegraphics[width=1.\linewidth]{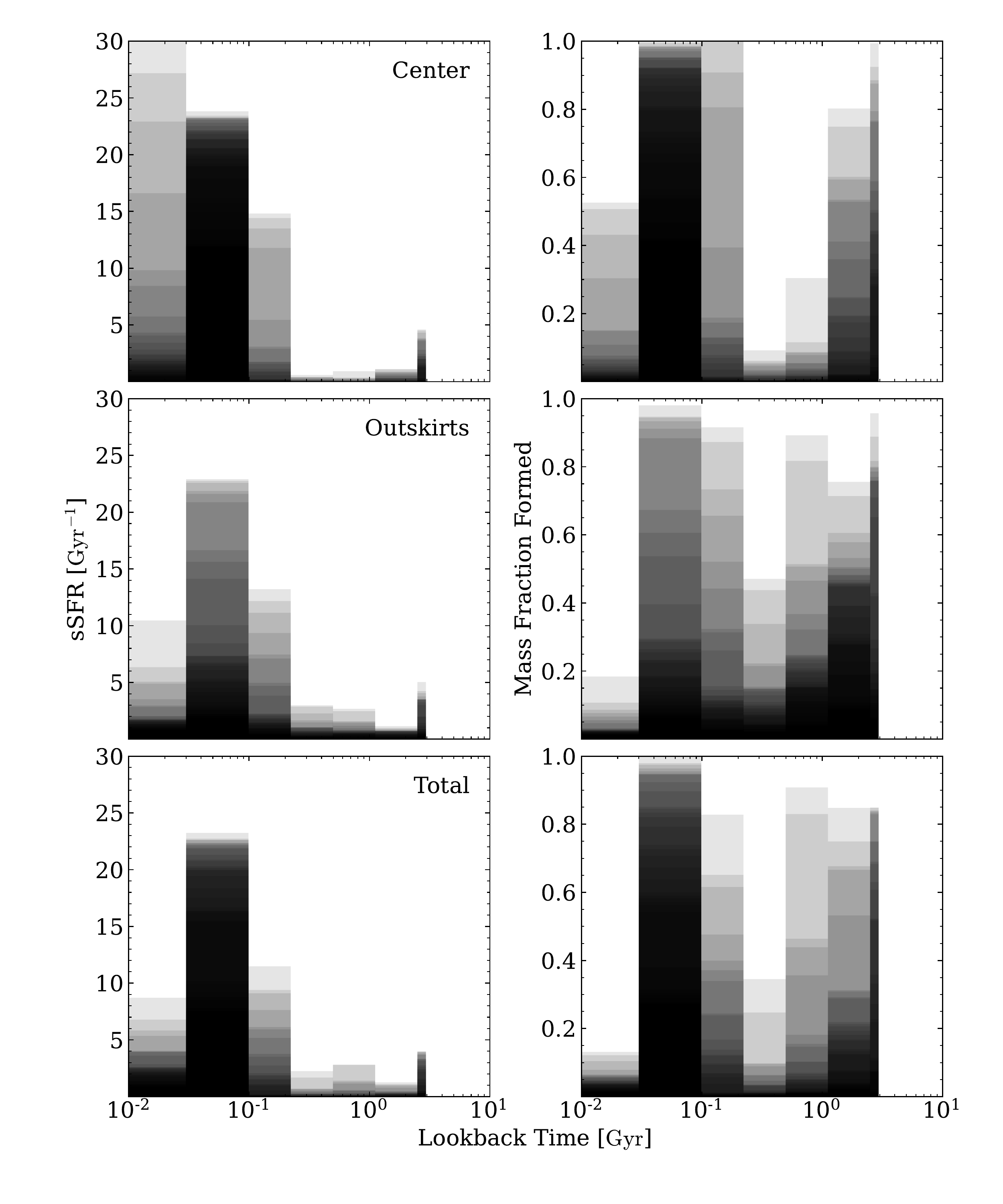}
    \caption{Center SFHs show a strong, recent burst of star formation. SFHs for the center (top) and outskirts (middle) components, as well as the total galaxy (bottom), are shown for all 60 galaxies in the sample. Specific SFRs (sSFR$=$SFR$/M_\star$) are shown in the left column and the fraction of mass formed (mass formed in an age bin divided by total mass formed) is shown on the right. Darker regions indicate a higher density of galaxies. The lookback time is relative to the redshift of observation (i.e. a lookback time of 1 Gyr refers to an epoch 1 Gyr earlier than $z\sim2.3$).}
    \label{fig:sfh_dens}
\end{figure}

\begin{figure*}
    \centering
    \includegraphics[width=\linewidth]{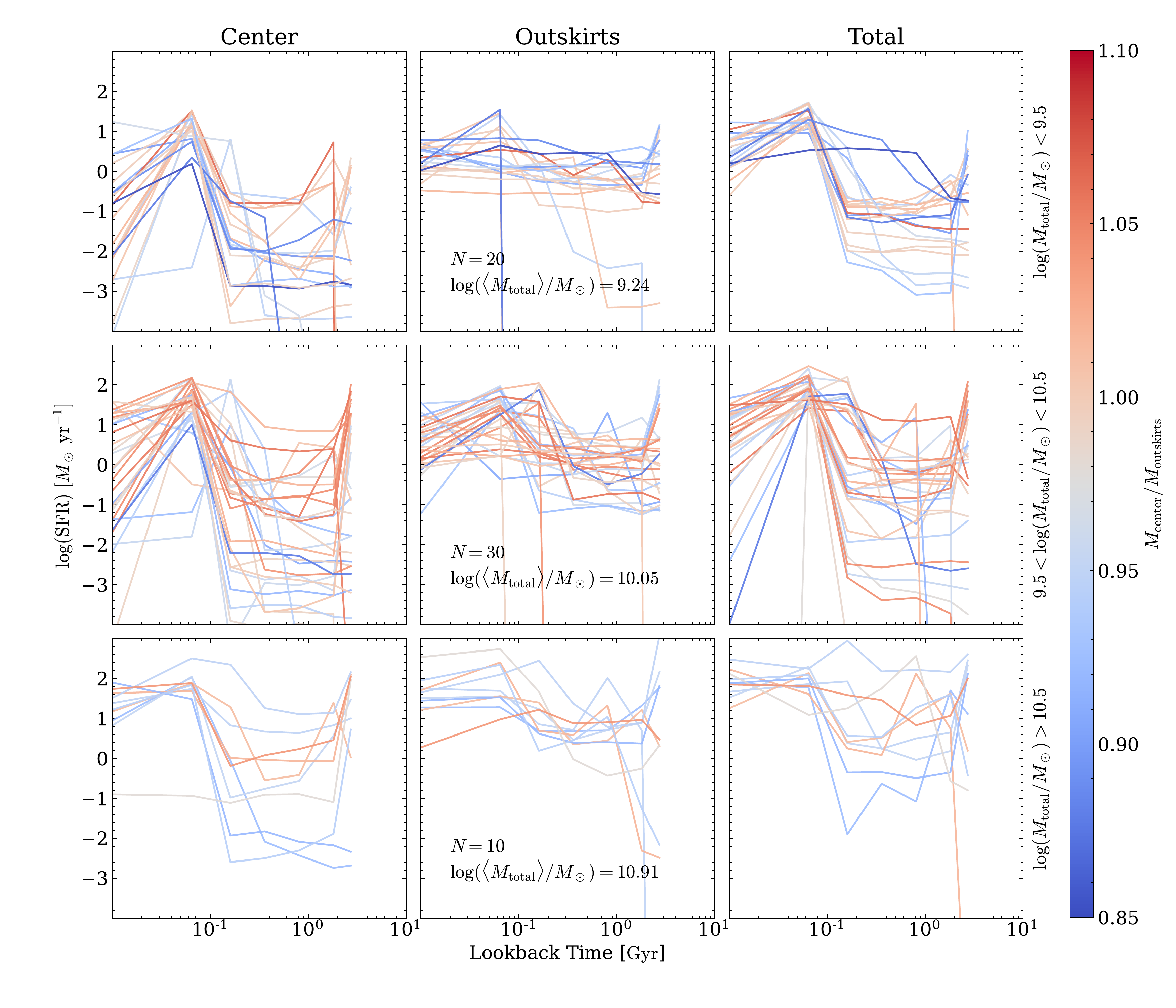}
    \caption{Individual SFHs for all galaxies in the sample binned by total stellar mass. There is a wide variety in SFHs for these galaxies and their subcomponents. The color indicates the center-to-outer mass ratio. From left to right, the columns show central, outer, and total galaxy SFRs as a function of lookback time from $z\sim2.3$. The rows show SFHs for galaxies in one of 3 mass bins: $\log(M_\star/M_\odot)<9.5$ (low mass, top), $9.5<\log(M_\star/M_\odot)<10.5$ (intermediate mass, middle), and squares for $\log(M_\star/M_\odot)>10.5$ (high mass, bottom). For each mass bin, the number and mean mass of galaxies in the bin is shown in the second column.}
    \label{fig:sfh_dens_class}
\end{figure*}

\section{Star Formation Histories}\label{sec:sfhs}
The measured central SFHs of our sample may provide insight into how dense stellar centers in Milky Way progenitors may have formed. If centers in these lower mass systems formed first, followed by inside-out growth of the surrounding regions \citep[e.g.,][]{vanDokkum10,Carrasco10,Nelson16}, then we would expect to see significant levels of star formation early on. In the cold mode accretion/clump merger scenario \citep{Dekel09} or other scenarios where the center grows in a coeval way with the rest of the galaxy \citep[e.g.,][]{Kormendy04}, star formation rates should be more constant and the SFH should have a similar shape to that of the outskirts, though an increase in the clump accretion rate \citep[e.g.,][]{vanDokkum13} could result in a sharp peak in the bulge SFH. Similarly, a burst of star formation could imply a rapid growth of the center via wet disk contraction \citep{Dekel14,Zolotov15,Tacchella18} and the galaxies with the strongest bursts of star formation should appear the most compact.

Figure \ref{fig:sfh_dens} shows the SFHs (sSFR and fraction of mass formed) of the center (top), outskirts (middle), and total galaxy (bottom) for all 60 galaxies in the sample. The shading indicates how many galaxies have a similar SFR in a given age bin. A prominent feature is a strong burst of star formation between a lookback time 30 and 100 Myr before $z\sim2.3$. This indicates that centers in these galaxies are younger and have not built slowly over time. For most of the galaxies, both the outskirts and total galaxy build up more of their mass earlier on than the center and all 3 show a quenching event 0-10 Myr before observation. This quenching event agrees with results that suggest bulge formation can morphologically quench galaxies by stabilizing the disk against future star formation \citep{Martig09} or that rapid gas consumption/AGN fueling can temporarily suppress star formation after compaction \citep{Tacchella16}.

In Figure \ref{fig:sfh_dens_class}, we show the same SFHs, now as individual SFRs, separated in rows by the total stellar mass of the galaxy. This highlights the diversity in SFHs in this sample. A strong burst of SFR in the center is more prominent in lower mass galaxies, which also show relatively constant SFR in the outskirts and a rising SFH for the integrated galaxy. At higher masses, a single burst of star formation in the center is less common, with higher star formation rates usually distributed over a wider range of lookback times. The total SFR in these galaxies is also higher and more constant in these galaxies, compared to the increasing total SFHs in lower mass galaxies. This suggests that these galaxies may be undergoing inside out growth \citep{vanDokkum10,Carrasco10,vanDokkum14,Nelson16}, forming a larger fraction of the center mass at earlier times.

Figure \ref{fig:sfh_dens_class} also illustrates the difference between the SFHs of the centers and outskirts of these galaxies. In most of sample, the center exhibits a large burst in star formation at late times while the outskirts form stars more steadily and have higher SFRs in general. This highlights the differential formation histories of the inner and outer regions of star forming galaxies and establishes the existence of a formation pathway distinct from the canonical inside-out growth mechanism used for massive galaxy formation.

\subsection{On the Robustness of the Reconstruction of the SFH}
The robustness of the reconstruction of the SFH in non-parametric form made by \textsc{Prospector} has been extensively tested in a number of previous work \citep{Johnson21,Leja21,Tacchella21,Ji22}. In particular, these works have used synthetic galaxies from the IllustrisTNG cosmological hydrodynamical simulations \citep{Springel18,NelsonD18,Pillepich18,Naiman18,Marinacci18,NelsonD19,Pillepich19} to directly compared the \textsc{Prospector} SFH output to the input galaxies' SFH. They have also tested the stability of the results against assumed priors, particular priors on the time dependence of the discretized SFH itself (e.g. continuity vs. Dirichlet priors). The general conclusions from these works is that the non-parametric SFHs derived by \textsc{Prospector} are robust and stable when good quality photometry covering a broad range of the rest-frame SED, from UV to near-IR is available, such as ours, and when some parameters, such as spectroscopic redshift and metallicity, are independently known and not left as free parameters during the fitting procedure. Here, although we do not repeat their tests and follow them in adopting the Continuity prior when deriving the SFH, we do test the stability of our results against the input photometry, the time binning adopted for the SFH reconstruction and the adoption of the metallicity prior. In Appendix \ref{sec:sfhcomp} we have compared the SFHs of the integrated galaxies shown in Figure \ref{fig:sfh_dens_class} with SFHs measures obtained from a much expanded photometric data set that comprises 42 photometric bands, with different time binning (9 time bins vs. our adopted 7 bins), and with and without assuming a strong metallicity prior. As discussed in Appendix \ref{sec:sfhcomp}, we conduct our test for the integrated SFHs only, and not for the centers and outskirts as well, because the extended photometric data is derived from ground-based images in natural seeing. The complexity of such data set, therefore, prevents us from running our decomposition analysis in a robust way in this case. The conclusions from our test is that the integrated SFH is robust and stable, however, which adds strong support that the decomposed SFHs of centers and outskirts, obtained with the same data sets, \textsc{Prospector} settings and priors, are robust as well.

\begin{figure}
    \centering
    \includegraphics[width=\linewidth]{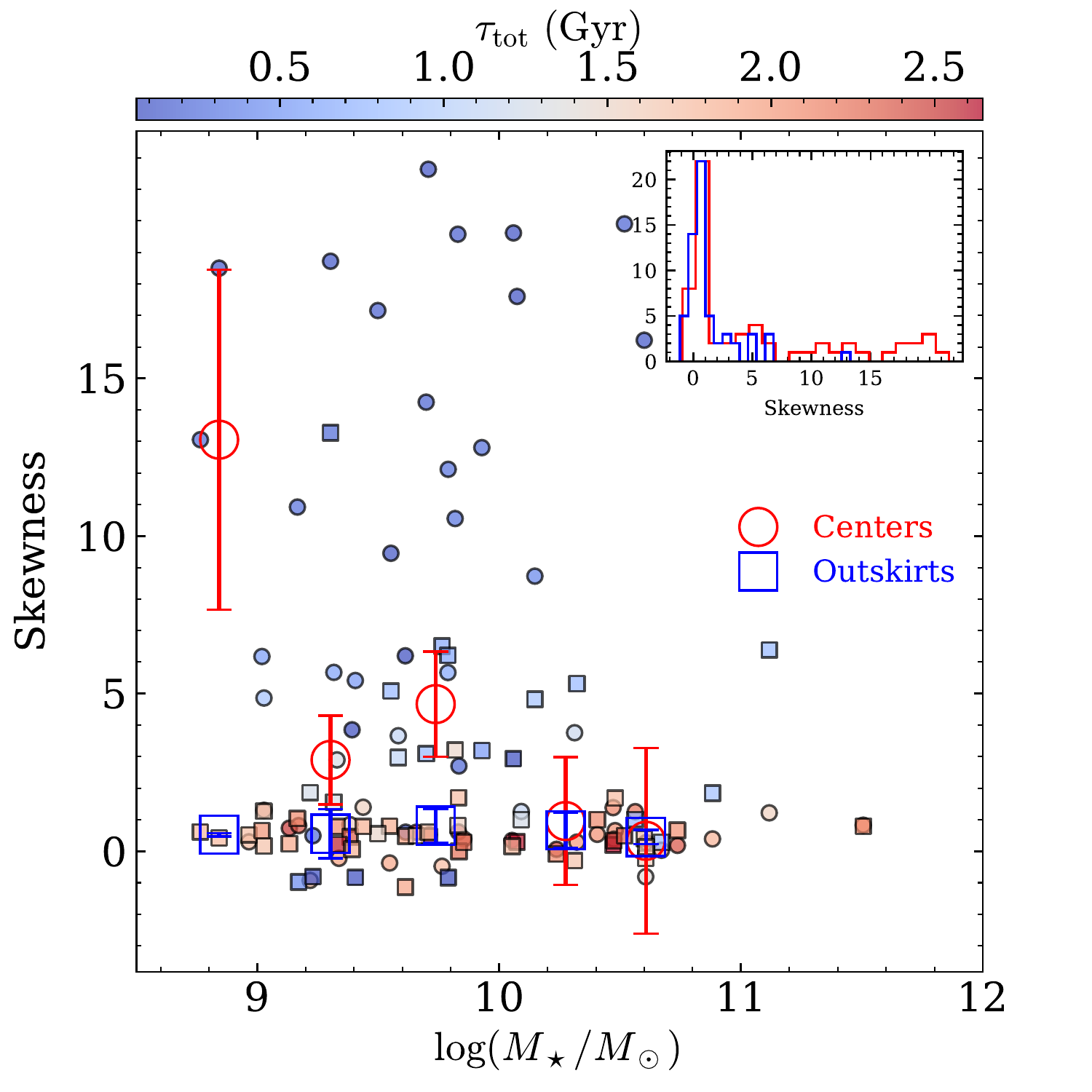}
    \caption{The centers of star-forming galaxies form most of their stars late in a short burst when compared with their outer regions. Circular points indicate the skewness (computed using Eq. \ref{eq:skew}) of the centers of galaxies and square points represent the outskirts. The masses shown are the total stellar mass of the galaxy. Points are colored based on the total timescale ($\tau_\mathrm{tot}$, Eq. \ref{eq:tau}). Red and blue open points indicate the running median for the centers and outskirts, respectively, with error bars showing the standard error in the mean. The inset shows histograms of the skewness for the centers (red) and outskirts (blue) of all galaxies in the sample.}
    \label{fig:timescales}
\end{figure}

\subsection{Star Formation Timescales}\label{sec:timescales}
To further illustrate the different formation histories of galactic centers and outskirts, we compare various star-formation timescales for the two components. In particular, we compute two values: the total galaxy timescale (equivalent to twice the standard deviation of the SFH) via
\begin{align}\label{eq:tau}
    \tau_\mathrm{tot}=2\int_{t_\mathrm{obs}}^{t_\mathrm{univ}}(t-t_\mathrm{age})^2\frac{\mathrm{SFR}(t)}{M_{\star,\mathrm{tot}}}dt,
\end{align}
and the skewness of the SFH
\begin{align}\label{eq:skew}
    \mathrm{Skew}=\frac{8}{\tau_\mathrm{tot}^3}\int_{t_\mathrm{obs}}^{t_\mathrm{univ}}(t-t_\mathrm{age})^3\frac{\mathrm{SFR}(t)}{M_{\star,\mathrm{tot}}}dt.
\end{align}
The total timescale describes how concentrated the star formation activity in the galaxy is, where a short timescale indicates the star formation is concentrated to a single burst while a long timescale would represent gradual, continuous build-up of stellar mass. The skewness describes how late or early the star-formation is occurring, with a large positive skewness indicating the star formation occurs late in the galaxy's history while a large negative skewness indicates significant early star formation. Conversely, a skewness near zero implies the star formation rate is evenly distributed throughout the galaxy's history.

In Figure \ref{fig:timescales} we compare the skewness of galaxy centers (circles) and outskirts (squares) over a range of total stellar masses, also including a third parameter in the total timescale ($\tau_\mathrm{tot}$). Generally, galaxy centers have a large positive skewness indicating most of the star formation in the core occurs at later times. Conversely, the outskirts have low skewness for all masses. The difference between the inner (red) and outer (blue) regions is also apparent in the inset histogram. The short timescales on which this star formation occurs in the cores is also indicative of a late burst of star formation, which is reflected in the SFHs in Figures \ref{fig:sfh_dens} and \ref{fig:sfh_dens_class}. Conversely, the  outskirts appear to form in much more gradual fashion, assembling their stellar mass over longer times, crudely $\approx5\times$ longer than the center, with slowly increasing SFR. Notably, the skewness of the centers decreases on average with increasing mass (red, open circles) while that of the outskirts stays constant (blue, open squares), further suggesting that  higher-mass star-forming galaxies in the sample have a different central formation history than their low-mass counter parts. 

\begin{figure}
    \centering
    \includegraphics[width=0.9\linewidth]{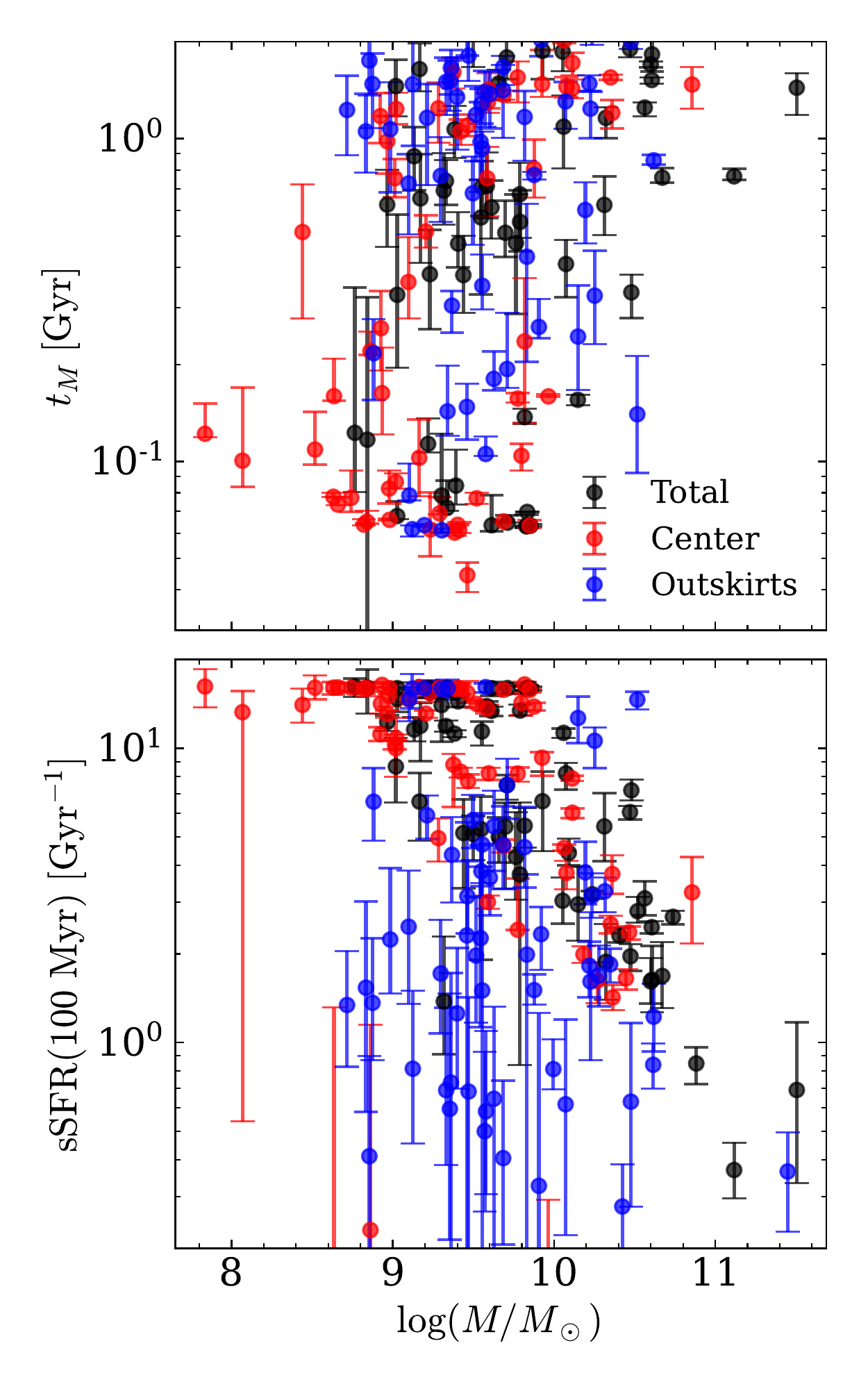}
    \caption{The age and sSFR of galaxies in the sample decreases with increasing stellar mass. The mass-weighted age (top) and sSFR in most recent 100 Myr (bottom) are compared to the mass for the center (red), outskirts (blue), and total galaxy (black).}
    \label{fig:cpt_ages}
\end{figure}

\begin{figure*}
    \centering
    \includegraphics[width=0.9\linewidth]{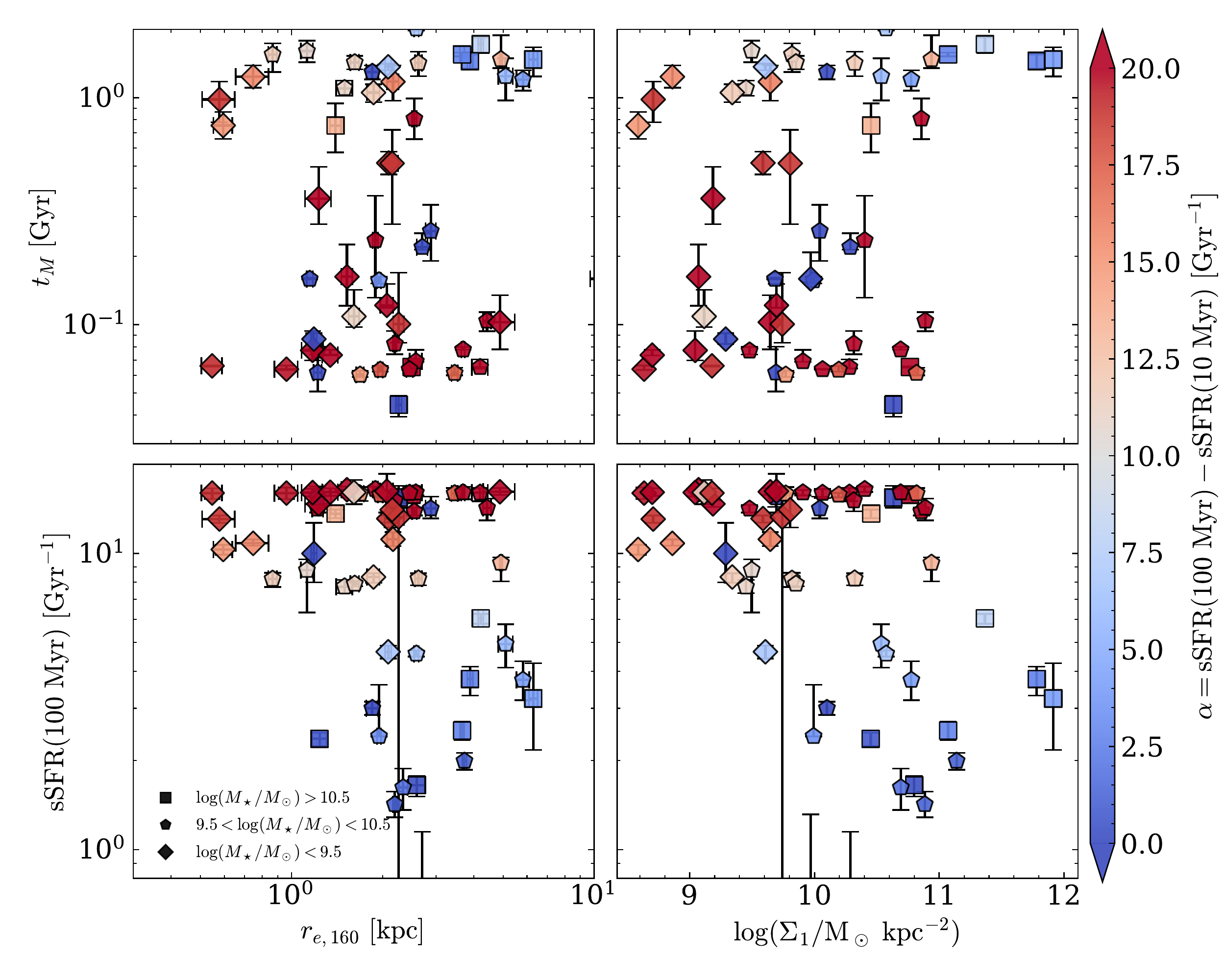}
    \caption{Lack of trends with size suggest that the build up of the galaxy center does not lead to a significant compaction of the galaxy. The mass-weighted age (top row) and sSFR in most recent 100 Myr (bottom row) in the galaxy center is compared to the half-light radius of the total galaxy (left column) and $\Sigma_1$ (right column). In the middle and left columns, symbols indicate galaxies fall in one of 3 mass bins: diamonds for low mass ($\log(M_\star/M_\odot)<9.5$), pentagons for intermediate mass ($9.5<\log(M_\star/M_\odot)<10.5$), and squares for high mass ($\log(M_\star/M_\odot)>10.5$). Points in these columns are colored by the difference in their sSFR at 100 Myr and 10 Myr prior to observation, denoted by $\alpha$. A higher value indicates the sSFR has decreased more during this period, while a lower value indicates the sSFR has remained relatively constant (or increased, if negative).}
    \label{fig:alpha_chg}
\end{figure*}

\subsection{Recent Star Formation and Compaction}\label{sec:recent_sf}
The SFH of the centers often shows large enhancements of SFR in the two most recent time bins at 10 and 100 Myr, indicative of a burst, during which time they form a substantial fraction of their stellar mass (72\% on average). During the same time period, for most galaxies the SFR of the outskirts remains approximately constant and their stellar mass increases substantially less, by only 16\%. In other words, is most galaxies the centers become proportionally more massive than the outskirts, with an increased mass growth rate of 450\% over the center, and thus become more compact. This is consistent with the general features of the``compaction'' phenomenon predicted, for example, in presence of dissipative gas accretion in unstable disks \citep[e.g.,][]{Dekel14,Zolotov15,Tacchella16,Tacchella18}. 

\begin{figure*}
    \centering
    \includegraphics[width=0.8\linewidth]{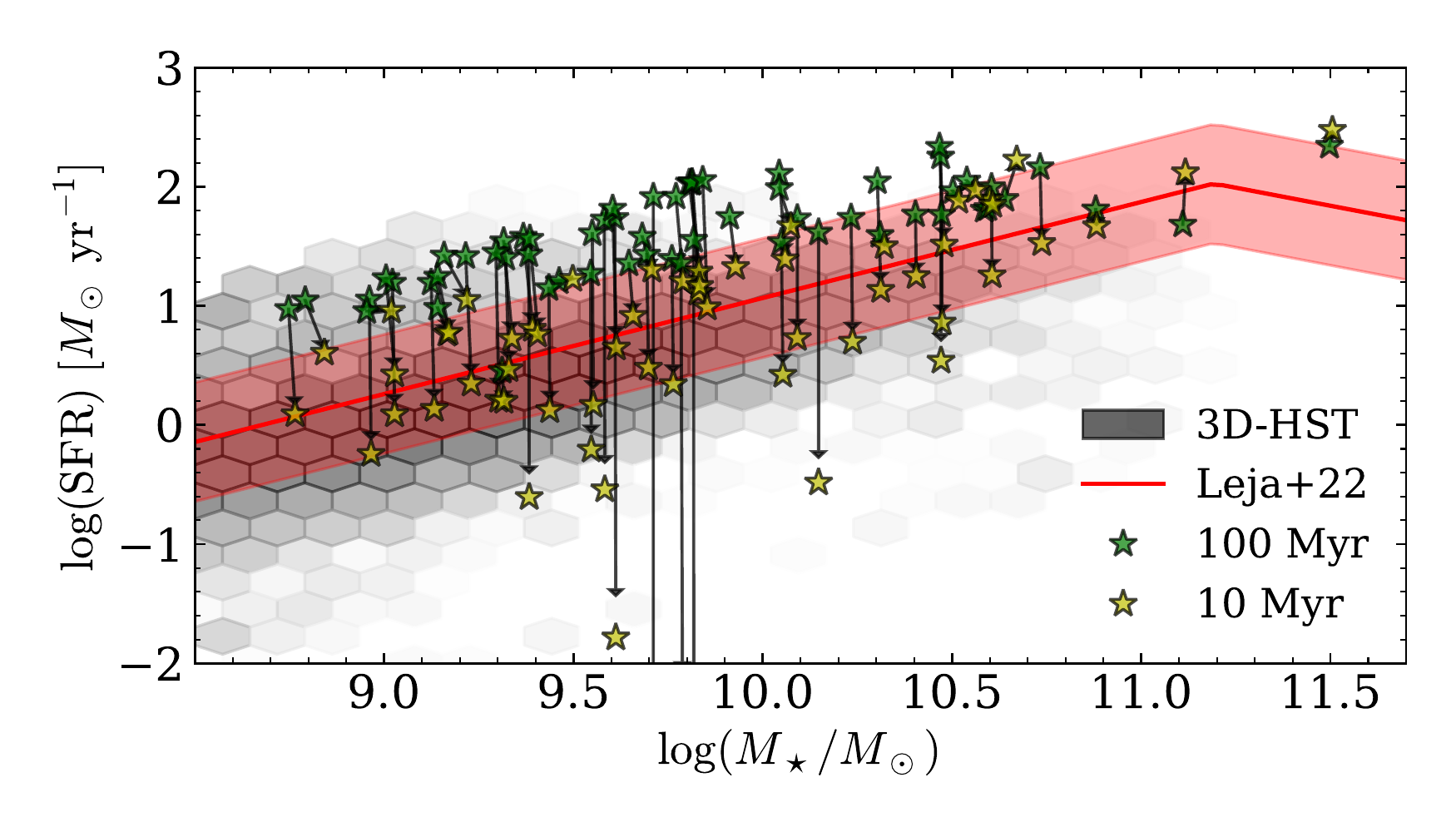}
    \label{fig:msevol}
    \caption{The majority of galaxies in our sample remain on the MS even after the central burst in SFR. Galaxies are plotted on the SFMS with their mass and SFR during two time bins, 100 and 10 Myr prior to observation, with yellow and green stars, respectively. Arrows show the SFR-stellar mass evolution in each galaxy between these two bins. The grayscale hexbins show an underlying distribution of star-forming galaxies from 3D-HST \citep{Brammer12}. The red line shows the empirical SFMS relation from \cite{Leja21}, with the shaded region indicating a 0.5 dex scatter.}
\end{figure*}

To further examine the possibility that we are observing the centers of our galaxies during a compaction event, we compare the mass-weighted age and sSFR in the most recent 100 Myr with stellar mass, size $r_e$, and projected stellar mass density within 1 kpc  \citep[$\Sigma_1$, see Eqn. 3 in][]{Barro17}, used as a measure of compactness, in Figures \ref{fig:cpt_ages} and \ref{fig:alpha_chg}. In Figure \ref{fig:cpt_ages}, we compare the mass-weighted age (top) and recent sSFR (bottom) with the mass of the center (red), outskirts (blue), and total galaxy (black). For the majority of galaxies, the centers are indeed substantially younger than the outskirts, with the age of the integrated galaxies being intermediate. The sSFR mirrors this behavior, with the centers having the largest values and the outskirt the lowest. More massive galaxies also tend to have lower sSFR in the past 100 Myr and their stellar populations are older, in both centers and outskirts, as well as for the integrated system. 

In Figure \ref{fig:alpha_chg}, the left and right columns compare the age and sSFR to size and central density $\Sigma_1$ \citep[see][]{Barro17,Lee18} of the integrated systems, respectively, with both quantities acting as proxies of the compactness. Galaxies are separated into three different bins of total stellar mass, each indicated by different symbols: diamonds for low mass ($\log(M_\star/M_\odot)<9.5$), pentagons for intermediate mass ($9.5<\log(M_\star/M_\odot)<10.5$), and squares for high mass ($\log(M_\star/M_\odot)>10.5$). The color of a galaxy indicates its $\alpha$, where $\alpha\equiv\mathrm{SFR(100~Myr)}-\mathrm{SFR(10~Myr)}$. A larger value of $\alpha$ reflects a greater decrease in the sSFR, i.e., onset of a decrease in SFR that potentially leads to quenching, while a smaller $\alpha$ shows little change in the sSFR (or an increase if $\alpha<0$). Although there is considerable scatter in these plots, some trends seem discernible.

The top left panel of Figure \ref{fig:alpha_chg} does not show any overall correlation between stellar age and size. Since such correlation has been observed for quiescent galaxies \citep{Fagioli16,Williams17,Ji22}, this would suggest that these galaxies are not close to initiating the quenching phase. The color-coding suggests however, that the galaxies for which $\alpha$ is larger, i.e., the SFR in the most recent time bin is small than that in the previous bin, have younger stellar populations, which would be consistent with a bursty behavior. The bottom left panel shows that galaxies with large $\alpha$ are at the top of the sSFR distribution and span the full range of observed size, which would be expected for bursty systems that have not yet started the structural transformations that appear to accompany the quenching process. We note that galaxies with low $\alpha$ seem to preferentially populate the low sSFR and large-size region of the plot. Both the bottom left and bottom right panels show a decrease of sSFR with increasing size and density, respectively. This is likely a mass effect, as more massive galaxies intrinsically have more dense centers \citep[e.g.,][]{Tacchella17}, are larger (due to the size mass relation), and have lower sSFRs for a given SFR (since sSFR$\propto M_\odot^{-1}$). The top right panel does not appear contain any trend. 

Thus, the questions of whether the bursts of star formation that occurred in the centers during the $\sim100$ Myr prior to observation does or does not result in more compact galaxies, as expected during a gas compaction event \citep{Dekel14,Zolotov15,Tacchella16,Tacchella18} or merging of star forming clumps \citep{Dekel09,vanDokkum13}, cannot be answered by this analysis. Compaction may still result in the increased SFRs and build up of a dense and compact central structure; our sample and our analysis simply do not provide any evidence in favor or against it, even if we do appear to systematically detect bursting centers surrounded by more steadily star-forming outskirts. Finally, also note that more massive galaxies also tend to be older and experience less of a drop off in sSFR after the increased star formation levels. As evidenced by Figure \ref{fig:sfh_dens_class}, these galaxies may be members of a more massive population (compared to the majority of our galaxies that occupy the middle of the SFMS). These galaxies may exhibit inside-out growth and quenching \citep{vanDokkum10,Carrasco10,vanDokkum14}, though there is also evidence for outside-in growth \citep{Tadaki17,Tadaki20}.

\subsection{Main Sequence Evolution}
From the SFH we can also predict the evolutionary paths of our galaxies in the SFR vs. $M_*$ plane, and their position relative to the Main Sequence. This is done in Figure \ref{fig:msevol}, which shows the evolution of our sample galaxies over the last two SFH time bins. Most galaxies appear slightly above the SFMS (red shaded region) at 100 Myr prior to observation (green points), consistent with being caught during a substantial burst of star formation. Subsequently, in the next time bin, as the burst subsides they evolve onto the main sequence by (yellow points). The majority of galaxies in the sample are also found on the SFMS close to observation (10 Myr bin). This behavior is similar to the confinement of galaxies in the SFMS shown in \cite{Tacchella16}. During the so-called ``blue-nugget'' (compact, star-forming) phase, galaxies move up to the top of the main-sequence as SFRs increase, which is seen in the 100 Myr SFRs. \cite{Tacchella16} find that this is followed by movement down to the lower end of the SFMS as gas consumption and feedback stall future star formation. They suggest this behavior occurs on timescales of roughly $\sim100$ Myr, which is supported by the location of our galaxies on the SFMS in the 10 Myr bin, roughly 100 Myr after the increase in SFRs.

The predominance of main-sequence galaxies, combined with the notable burst in star formation just before observation suggest that the appearance of the burst in most of these galaxies is a selection effect of studying star-forming galaxies. The fact the galaxies in our sample are solidly within the MS, even after most of them experienced a burst, which brought them above the MS, perhaps explains why we find no evidence of structural transformations, i.e. the shrinking of size and increase of central stellar density, that appear to accompany galaxies as they descend below the MS \citep{Cheung12,Barro17,Whitaker17,Lee18,Ji22}: our galaxies are not yet quenching. 

\begin{figure*}
    \centering
    \includegraphics[width=\linewidth]{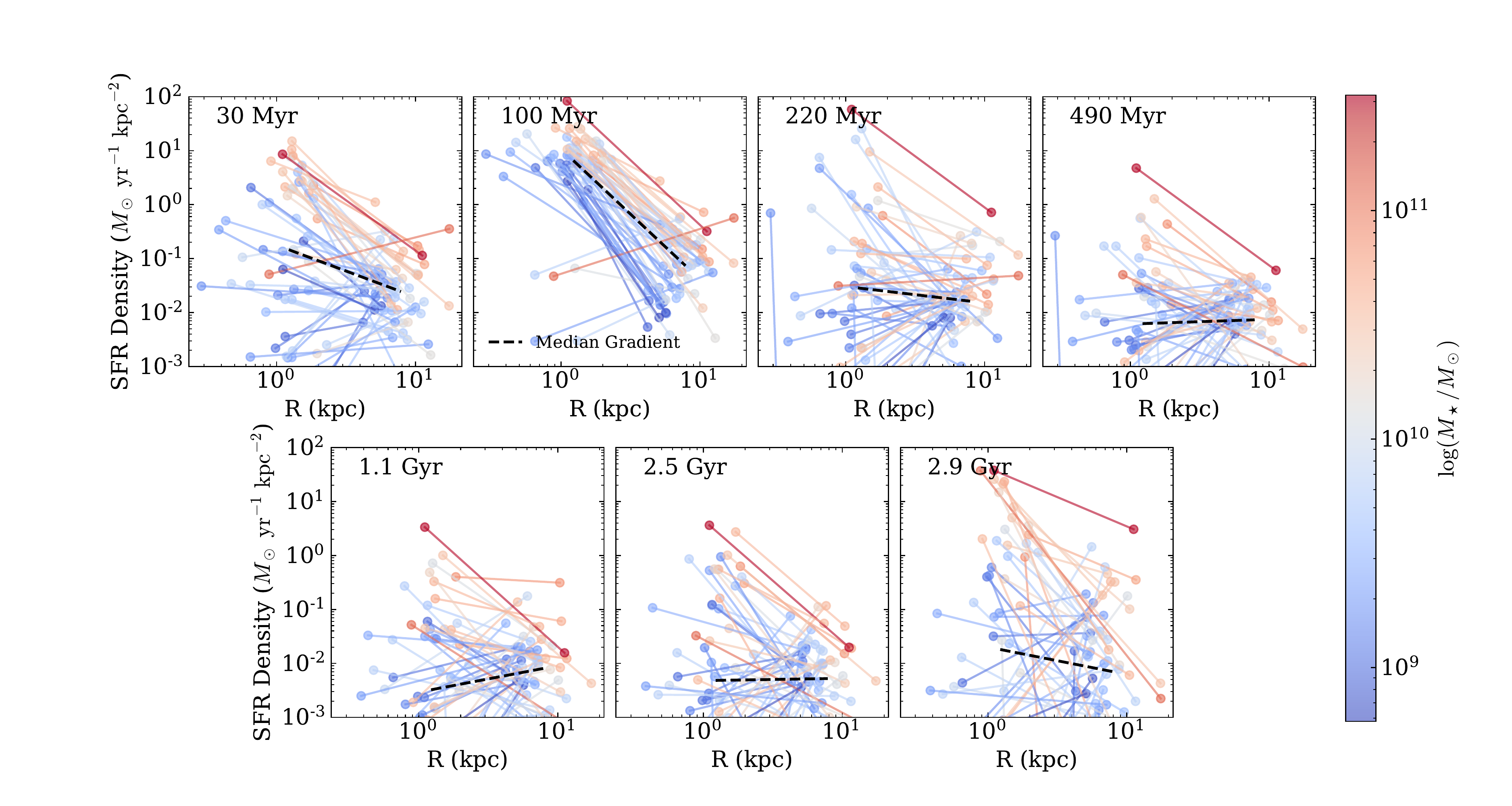}
    \caption{On average, the star formation rate gradient is flat for much of the history of galaxies in this sample. The seven panels show the individual SFR density (SFR/Area) gradients over the formation history of each galaxy in the sample with colored lines. The inner radius is defined as half the center radius $R$ and the outer radius is the average between $R$ and the Kron radius measured in the CANDELS/SHARDS photometric catalogs. The 7 panels indicate the upper edge of the corresponding age bin from the measured star-formation histories. The color of the gradients indicates the total stellar mass of the galaxy and the dashed black line shows the median SFR gradient.}
    \label{fig:gradients}
\end{figure*}

\subsection{Star Formation Rate Gradients}\label{sec:gradients}
Using the two-bin, spatially-resolved SFHs of our galaxies, we can also attempt a measure the time evolution of the SFR gradients. SFR density gradients are computed by dividing the SFR of each component by the area of that component. For this purpose, we define the inner and outer components as concentric circular regions with central radius $R$ (i.e., the radius used for the photometric decomposition) and Kron radius \citep{Kron80}, respectively. Figure \ref{fig:gradients} shows the SFR density gradients (colored lines) for all galaxies in the sample across all seven time bins. The sample median gradients are also shown as black dashed lines. As Figure \ref{fig:gradients} suggests, for much of a star-forming galaxy's history, the gradients are relatively flat; namely, star formation builds up the inner and outer parts at approximately equal rates. This agrees with previous studies that find significant mass evolution at all radii in Milky Way progenitors \citep[e.g.,][]{vanDokkum13}. However, significant negative gradients in star formation rate density are present across the main-sequence at $z\sim1$ \citep{Nelson16}, a significant difference from the roughly equal growth at all radii that we seem to be observing at higher redshifts. 

\begin{figure*}
    \centering
    \includegraphics[width=0.9\linewidth]{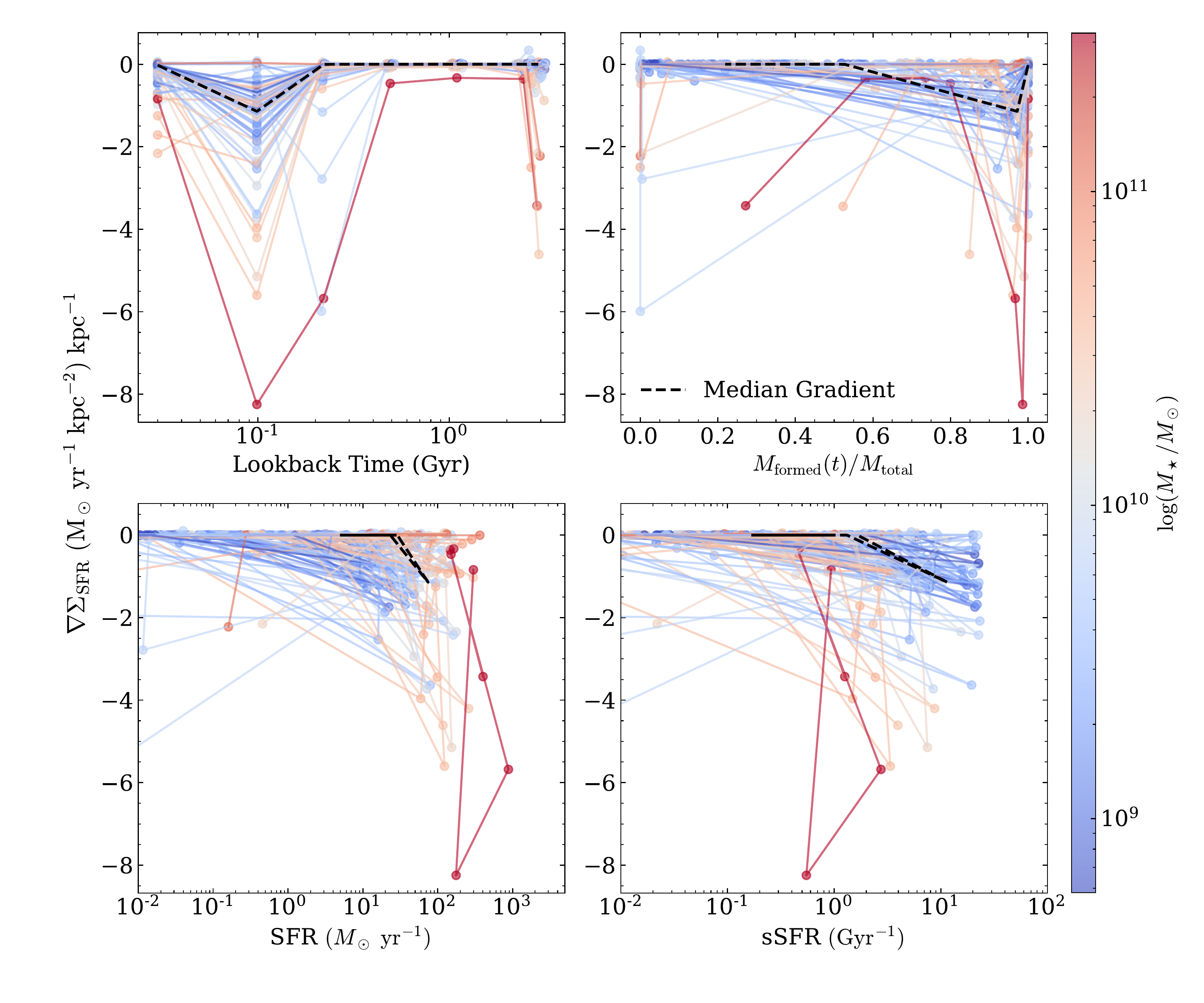}
    \caption{The evolution of the SFR density gradient ($\nabla\Sigma_\mathrm{SFR}\equiv\Delta\Sigma_\mathrm{SFR}/\Delta R$) with lookback time (top left), cumulative mass formed (top right), SFR (bottom left), and sSFR (bottom right). The symbols and coloring is the same as Figure \ref{fig:gradients}.}
    \label{fig:gradtrends}
\end{figure*}

In our sample of main-sequence galaxies, we find a dramatic shift to negative SFR gradients $\sim100$ Myr prior to observation (2nd panel, Fig. \ref{fig:gradients}a), which we associate with a rapid build-up of the central regions in the galaxies. This is also reflected in Figure \ref{fig:gradients}b, where on average the gradients become significantly negative at a lookback time of $\sim 100$ Myr. These negative gradients, with approximately the same slope as the gradients reported in \cite{Nelson16}, persist even after the largest increase in SFR at 100-300 Myr lookback time observed in several of our sample galaxies. Figure \ref{fig:gradtrends} examines the evolution of the gradients versus that of lookback time, integrated stellar mass, SFR and sSFR. The negative gradients are the steepest when the galaxies experience the largest SFR which, as we have seen from the SFH, happens at relatively recent lookback time from the observations, at 100-300 Myr. By this time, the galaxies have also assembled nearly all of the stellar mass found at the time of observation. As the bottom panel of Figure \ref{fig:gradtrends} shows, however, this is not the time when the galaxies have the largest sSFR, but rather an intermediate value (in log space).

The change in SFR gradient also supports observed gas-phase metallicity gradients in star-forming galaxies. \cite{Simons21} find that the vast majority of star-forming galaxies at $0.6<z<2.6$ have flat or slightly positive metallicity gradients. In order to achieve these flat gradients, most of the mass in the galaxy must be built up evenly across the galaxy, since metals are produced in stars and through stellar evolution. We measure flat SFR gradients on average across $\sim96\%$ of the galaxies' lifetimes, which supports the formation of a flat metallicity gradient. Moreover, \cite{Simons21} find galaxies at $z\sim0$ have negative metallicity gradients for most masses, which suggests these galaxies must also have negative metallicity gradients for much of their lifetimes. In this case, these may galaxies may have shifted to a negative gradient and started to build up their centers, which produced increased gas-phase metallicities in the center.

\section{Summary}\label{sec:summary}

In this paper, we examine the formation of dense central regions in star-forming galaxies at $z\sim2.3$, an epoch that coincides with the formation of the bulge and chemical thick disk in the MW \citep{Queiroz21,Miglio21}. A sample of 60 galaxies is selected from MOSDEF \citep{Kriek15,Sanders18} and the GOODS-N CANDELS/SHARDS photometric catalogs \citep{Barro19} with accurate photometry, spectroscopic redshifts, and metallicities. Galaxies are decomposed into central and outer components via a color-selected circular aperture, from which resolved HST photometry is measured. Using the \textsc{Prospector} code \citep{Johnson21}, we fit SEDs to each galaxy in the sample using an iterative method to account for unresolved light in the ground-based $K$-band and \textit{Spitzer}/IRAC bands.

The formation histories of these components provide an interesting insight into the differential formation of ``normal'' galaxies near the peak of cosmic star formation. These galaxies show strongly peaked SFHs for all but the most massive galaxies. While this means we may be observing the inside out growth of galaxies above $\log(M_\star/M_\odot)>10.5$, it also suggests that a rapid increase in the SFR is responsible for the formation of the centers of lower mass galaxies. Analysis of the timescales and skewness (Eq. \ref{eq:tau} and \ref{eq:skew}, respectively) of galaxies indicates that outskirts tend to have more uniform SFHs and much longer total timescales. Conversely, galactic centers have much more uneven SFHs with short timescales, indicative of a formation history dominated by a late burst of star formation. 

This increase in SFR may be due to a a gas compaction event \citep{Dekel14,Zolotov15,Tacchella16,Tacchella18} or an increase in the rate at which star-forming clumps are accreted into the galactic center \citep{Dekel09,vanDokkum13}, both of which should be reflected in the smaller sizes and larger central densities of galaxies with high sSFRs 100 Myr prior to observation. However, we find no trends in the age or sSFR of the central parts of galaxies with size and central density. Compaction of gas or increased clump accretion may still be possible mechanisms, but the subsequent change in morphology would not result in a more compact system.

Analysis of the SFR gradients also reveals flat gradients on average for the majority of the galactic lifetimes, in support of previous studies of the mass evolution of Milky Way progenitors \citep{vanDokkum13}. These galaxies then transitions to a steep negative gradient at $\sim100$ Myr before observation, mirroring the inside-out growth found in studies of resolved H$\alpha$ emission \citep{Nelson16}. This evolution in the SFR gradient may also provide an explanation for the mostly flat observed metallicity gradients in main-sequence galaxies \citep{Simons21}, which may be due to the long period of time over which these galaxies have flat SFR gradients.

\section*{Acknowledgements}
We are grateful to Avishai Dekel and Cristina Chiappini for reading the manuscript and for very useful comments. We acknowledge use of observations with the NASA/ESA \textit{Hubble Space Telescope} obtained from the MAST Data Archive at the Space Telescope Science Institute, which is operated by the Association of Universities for Research in Astronomy, Incorporated under NASA contract NAS5-26555. Support for Program number HST-AR-15798 was provided through a grant from the STScI under NASA contract NAS5-26555. This research made use of Montage. It is funded by the National Science Foundation under grant No. ACI-1440620 and was previously funded by the National Aeronautics and Space Administrations Earth Science Technology Office, Computation Technologies Project, under Cooperative Agreement No. NCC5-626 between NASA and the California Institute of Technology.

\facilities{HST (ACS and WFC3), KPNO (Mosaic), Subaru (MOIRCS), \textit{Spitzer} (IRAC), GTC (OSIRIS)}

\software{
\textsc{Astropy} \citep{astropy:2013,astropy:2018,astropy:2022},
\textsc{Prospector} \citep{Johnson21},
\textsc{FSPS} \citep{Conroy09,Conroy10a,Conroy10b},
\textsc{Python-FSPS} \citep{ForemanMackey14},
\textsc{emcee} \citep{ForemanMackey13},
\textsc{Montage} (\url{montage.ipac.caltech.edu}),
\textsc{GALFIT} \citep{Peng02,Peng10}
\textsc{Photutils} \citep{photutils},
\textsc{matplotlib} \citep{matplotlib2007},
\textsc{numpy} \citep{numpy2011}
}

\appendix
\restartappendixnumbering
\section{Sample Galaxy Images and Center Apertures}\label{sec:montage}
\begin{figure*}
    \centering
    \includegraphics[width=\linewidth]{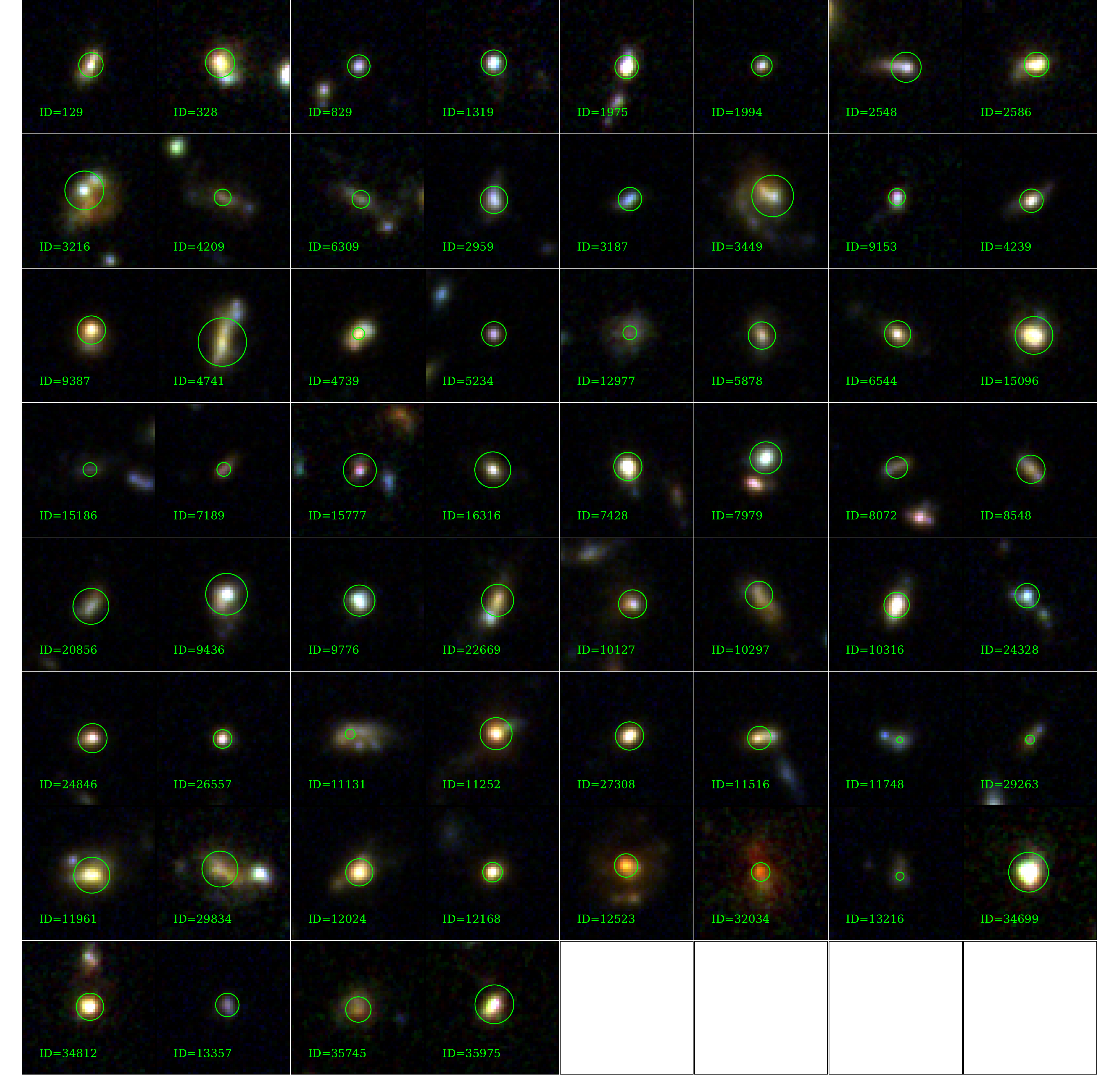}
    \caption{Three-color images (red=F160W, green=F125W, blue=F850LP) of all 60 main-sequence galaxies in our sample. The center apertures determined by the procedure in Section \ref{sec:decomp} are shown in green. The IDs are from the MOSDEF survey \citep{Kriek15}.}
    \label{fig:cutouts}
\end{figure*}

Figure \ref{fig:cutouts} shows three-color images of the 60 galaxies in our sample and the center apertures determined in Section \ref{sec:decomp}. The color-selected centers agree with the apparent structural centers of these galaxies.

\restartappendixnumbering
\section{Example Posterior Distributions and Covariances}\label{sec:excov}
\begin{figure*}[t!]
    \centering
    \includegraphics[width=1\linewidth]{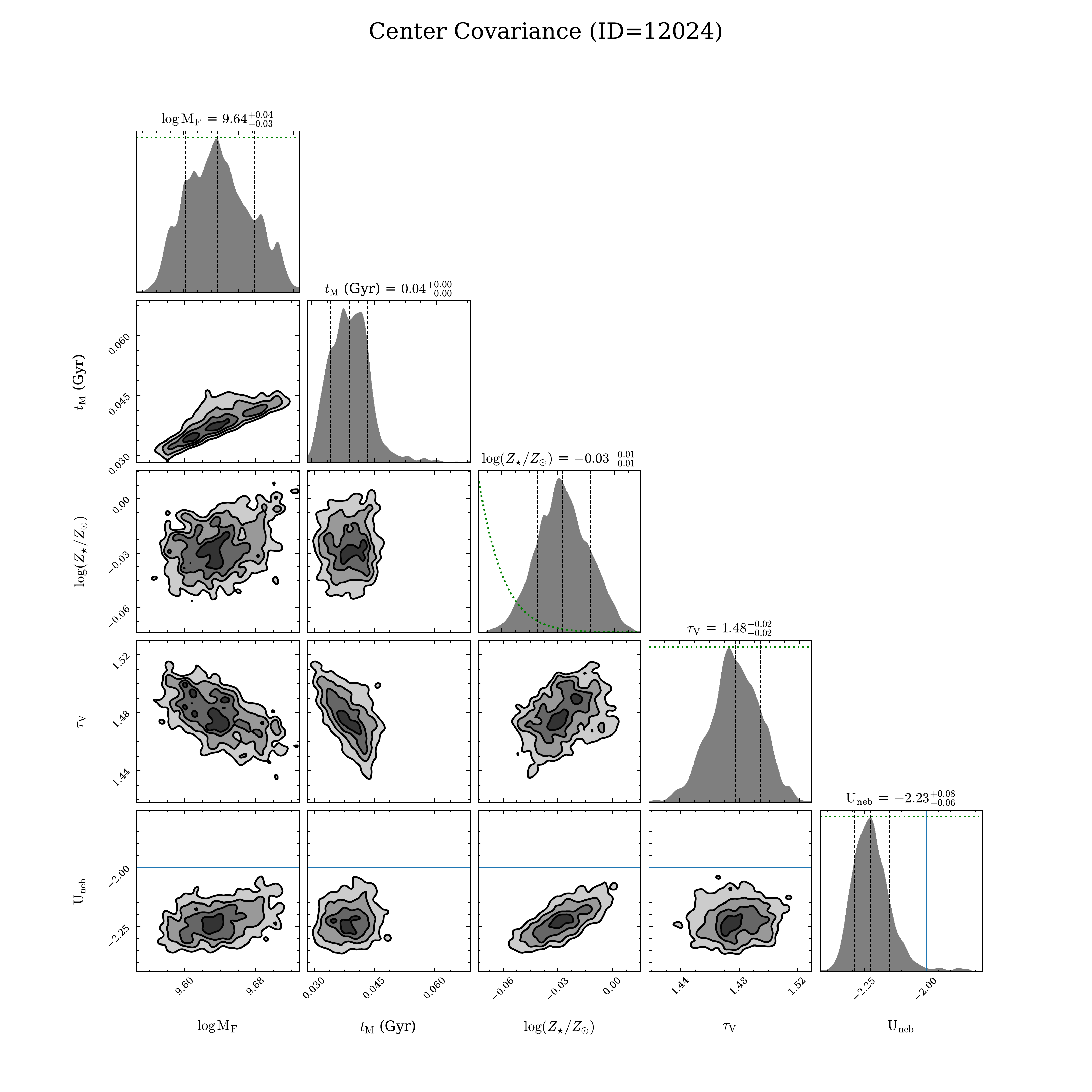}
    \caption{Corner plots showing posterior distributions of \textsc{Prospector} stellar population parameters for the central part of the galaxy. The included parameters are (by column, from left to right) the stellar mass ($M_\star$, not the mass formed), the mass-weighted age ($t_\mathrm{M}$, transformed from the SFH), the metallicity ($\log Z$), the $V$-band optical depth ($\tau_V$), and the ionization parameter ($U_\mathrm{neb}$). Histograms and contours show the projected 1- and 2-D posterior PDFs for the listed parameters. The median value of each parameter and its uncertainties (from the 16th and 84th percentiles) are given at the top of each column. Blue lines and points indicate the initial guess given to \textsc{Prospector} (e.g. MOSDEF metallicities, GOODS-N masses, etc.). Priors are shown with dotted green lines. Note the stellar mass and mass weighted age do not have initial guesses or priors because these are transformed parameters.}
    \label{fig:covar_ex_bulge}
\end{figure*}

\begin{figure*}[b!]
    \centering
    \includegraphics[width=1\linewidth]{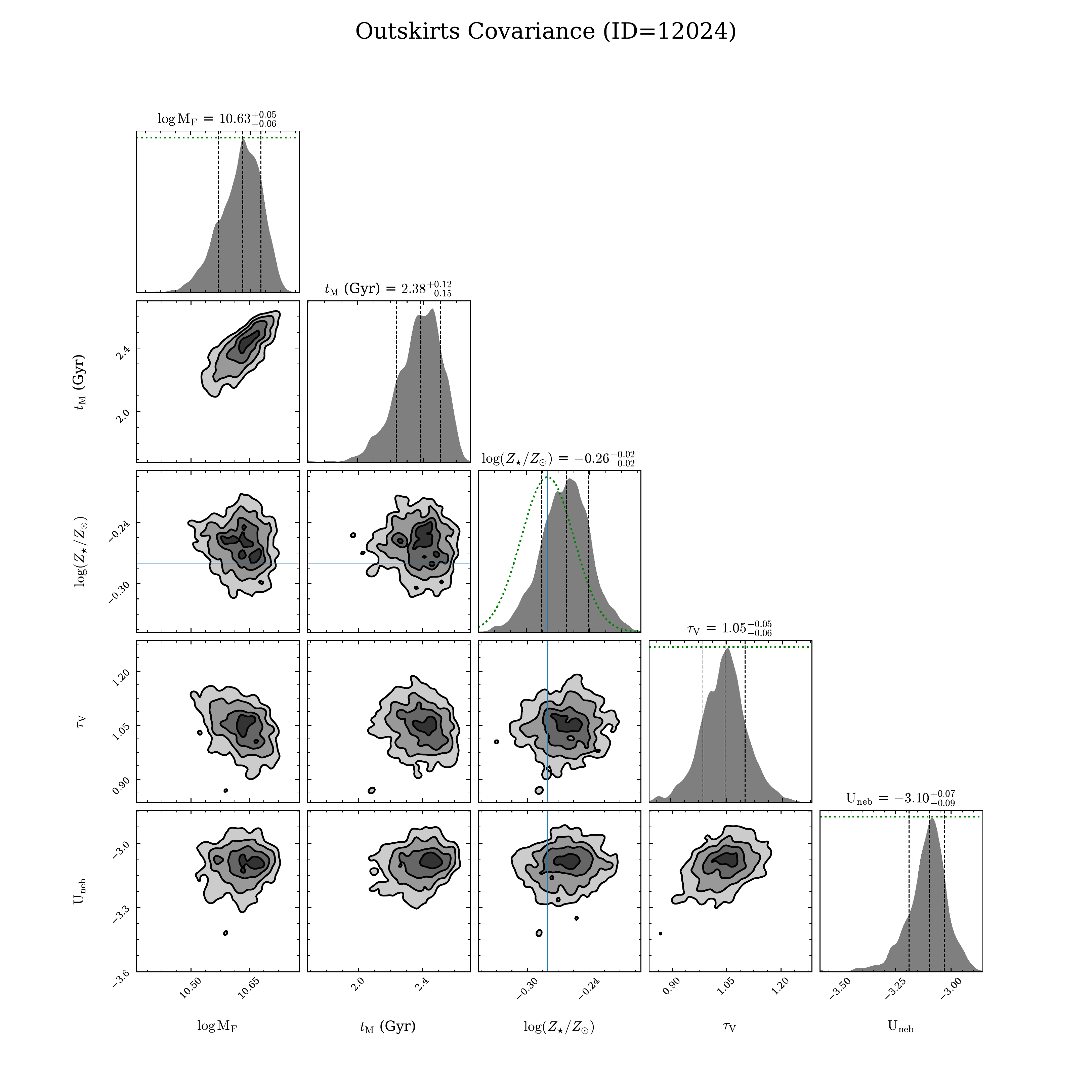}
    \caption{Same as Fig. \ref{fig:covar_ex_bulge} but showing the outer component.}
    \label{fig:covar_ex_disk}
\end{figure*}

\begin{figure*}[b!]
    \centering
    \includegraphics[width=1\linewidth]{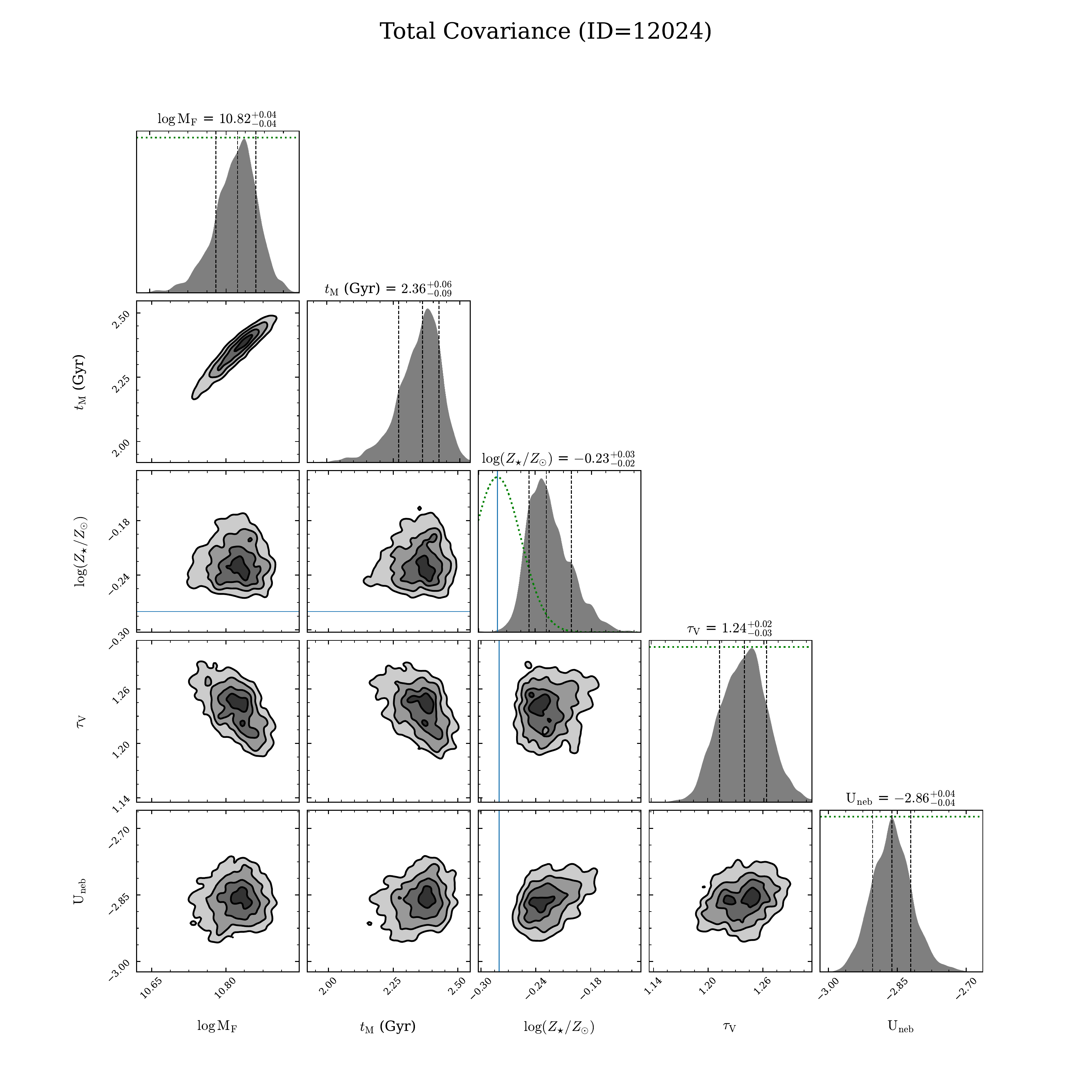}
    \caption{Same as Fig. \ref{fig:covar_ex_bulge} but showing the population parameters for the integrated galaxy.}
    \label{fig:covar_ex_tot}
\end{figure*}

Figures \ref{fig:covar_ex_bulge}, \ref{fig:covar_ex_disk}, and \ref{fig:covar_ex_tot} show posterior distributions for the inner, outer, and integrated galaxy components for the example galaxy SED in Figure \ref{fig:SED_ex}. In general, all stellar population parameters are well constrained for both components and the total galaxy. 


\restartappendixnumbering
\section{Verifying the Robustness of \textsc{Prospector} Star Formation Histories}\label{sec:sfhcomp}

We have tested the stability and robustness of our reconstruction of the integrated SFHs, i.e., the SFH of each galaxy as a whole and not of centers and outskirts, for our samples against the input photometry, the adoption of the metallicity prior and the sampling of the lookback time: the 6 bins adopted in the primary runs vs. the 9 bins adopted here. 
Specifically, for each galaxy in the sample we have re-run \textsc{Prospector} (test runs) utilizing the expanded CANDELS photometric catalogs in the GOODS-N field which, in addition to the HST data in the GOODS \citep{Giavalisco04} and CANDELS \citep{Grogin11,Koekemoer11} surveys, the ground- and space-based ancillary data from UV to FIR, includes the 25 medium-band photometry at optical wavelengths acquired during the SHARDS survey \citep{PerezGonzalez13} with the OSIRIS instrument at the 10.4-m telescope Gran Telescopio Canarias (GTC). The photometric apertures of the SHARDS data have been matched to those of the existing CANDELS data \citep{Barro19}. The total number of photometric bands used in the \textsc{Prospector} SED modeling for the test runs is 42. We have then compared the SFHs obtained during these tests run with the corresponding integrated SFHs obtained with the same settings used for the SFH of the centers and outskirts (primary runs). 

We have done two test runs, namely with and without ad prior on gas-phase metallicity, using the MOSDEF measures that we used to the primary runs. Also, for all test runs we have sampled the SFH in 9 bins of lookback time to test the stability of the SFH with respect to the choice of the time bins. To compare the 9-bin SFH of the test runs with the 6-bin one for the primary runs, we perform a linear interpolation of the test-run SFHs at the central value of each primary-run time bin. 

\begin{figure*}[b!]
    \centering
    \includegraphics[width=0.85\linewidth]{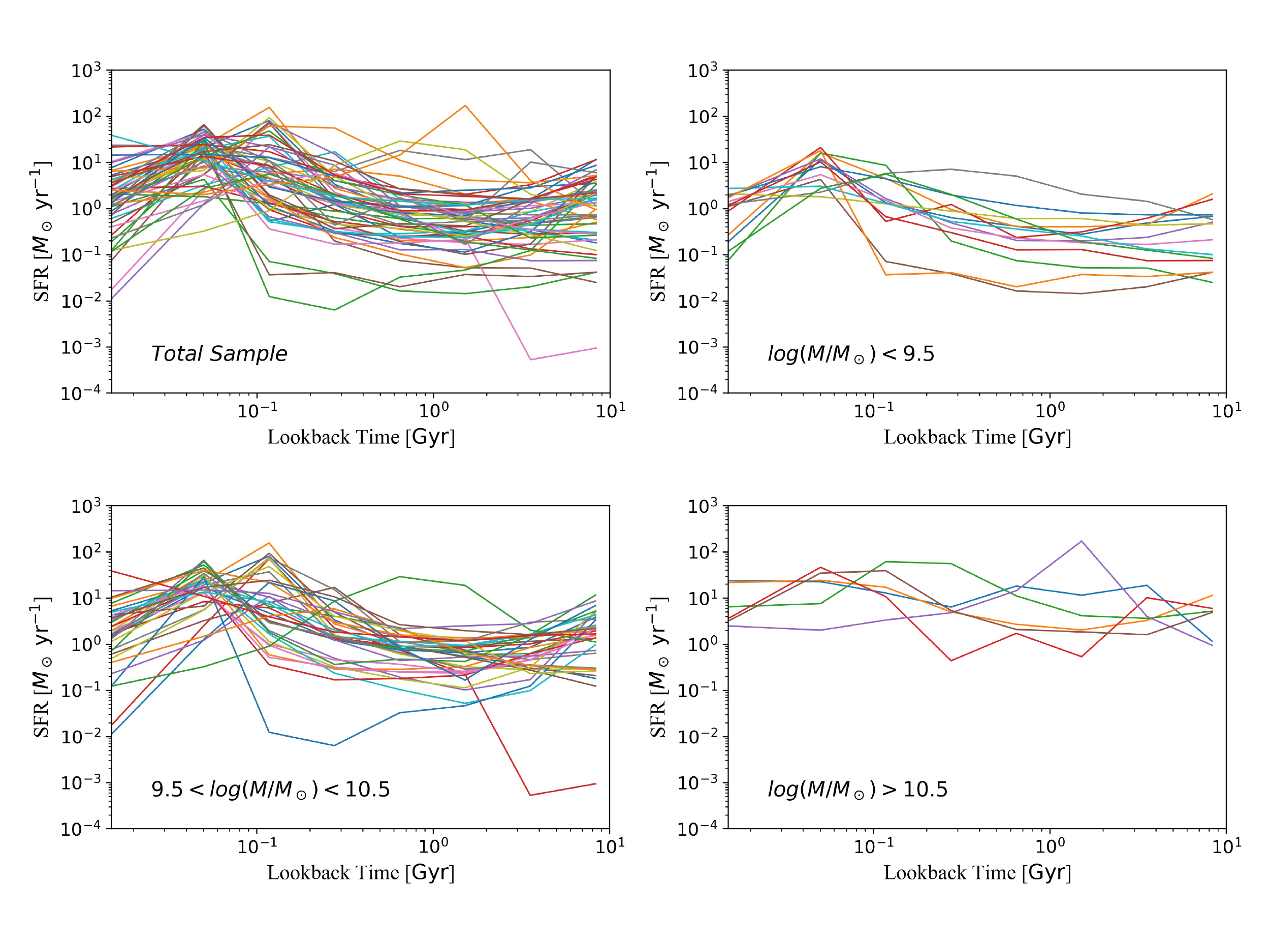}
    \caption{The individual SFHs of sample galaxies obtained during the test run for the total sample of star-forming galaxies (top left), as well as various mass bins. These test runs are fit with an increased number of photometric data points, more time bins, and no metallicity prior to help validate the robustness of our semi-resolved SFHs. Due the very large number of photometric bands at low angular resolution, we did not attempt to decompose the photometry into ``centers'' and ``outskirts'' and only the integrated SFH has been derived. Overall, there is excellent qualitative agreement of the shape of the SFH from the two sets of runs.}
    \label{fig:sfh_xphoto}
\end{figure*}

\begin{figure*}[b!]
    \centering
    \includegraphics[width=0.85\linewidth]{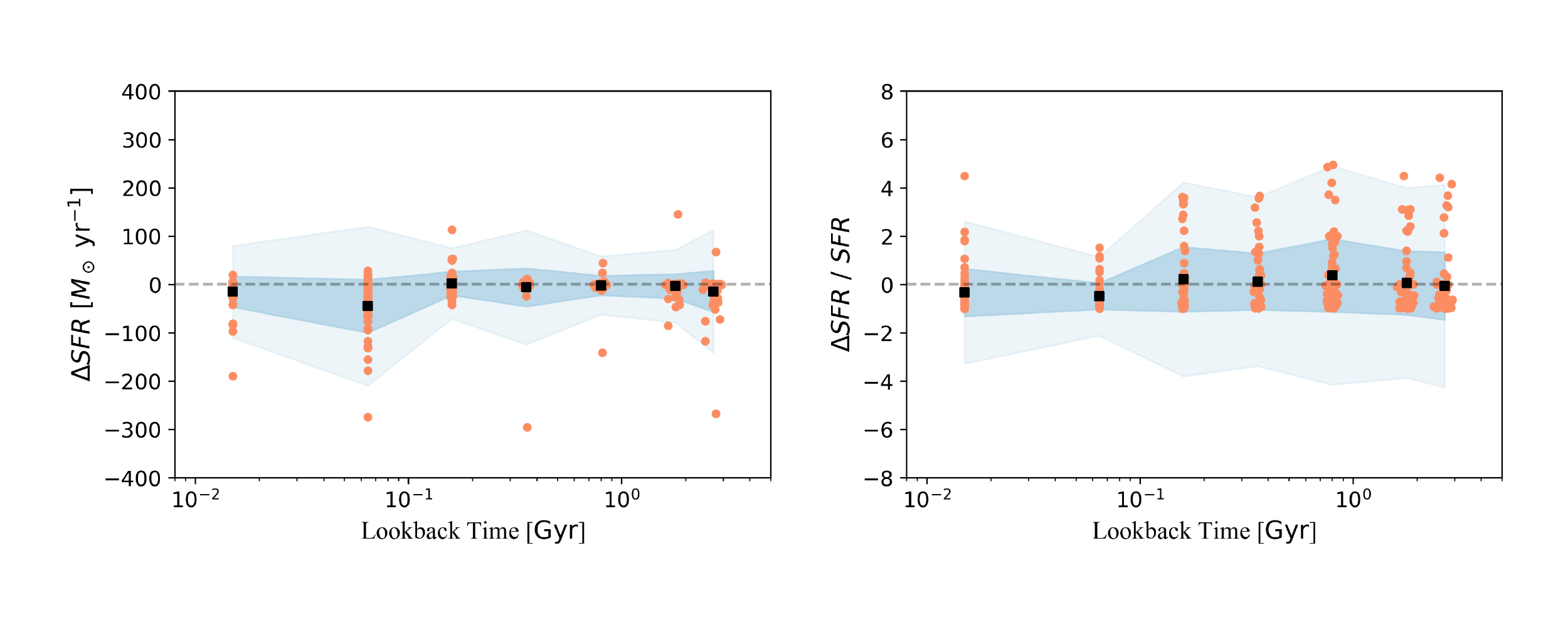}
    \caption{Overall, no systematic deviations are observed between the SFHs in our analysis (primary runs) and the SFHs from test runs with more photometry and higher time resolution (with no metallicity prior), suggesting that on average the shape of the reconstructed SFH is robust. The left panel shows the absolute difference between the SFH derived during the primary runs and the test runs, while the right panel shows the fractional difference. The value of the SFR of the test runs at the time bin of the primary runs has been interpolated from adjacent bins. The orange points represent the individual galaxies, the black points the median, and shaded areas show the 68 and 95 percentile.}
    \label{fig:sfh_diff}
\end{figure*}

Figure \ref{fig:sfh_xphoto} shows the SFHs derived during the test runs for the whole sample and in the three mass bins adopted for the primary runs. A visual comparison with the integrated SFHs derived from the primary runs shows that the shape of the SFHs from the two sets of runs are in good qualitative agreement, suggesting that the primary-run SFHs are robust. Figure \ref{fig:sfh_diff} quantifies the difference between the output of the sets of run: the left panel shows the absolute difference of the SFH of each galaxy in the two sets of runs, while the left panel the fractional difference. It can be seen that the median difference between the two sets of runs is essentially zero for all values of the look-back time, while the scatter remains small for the three most recent time bins and it only increases at large values of the look-back time, highlighting the difficulty of reconstructing the earliest phases of the SFH. Fortunately, the key results of this work are based on the difference of SFH of ``centers'' and ``outskirts'' during later stages of their evolution, at look-back times close to the time of observation. Overall, the agreement between the primary and test runs appears to be very good, demonstrating that the overall shape of the SFH is insensitive to the details of the input SED. 

\begin{figure*}[t!]
    \centering
    \includegraphics[width=0.9\linewidth]{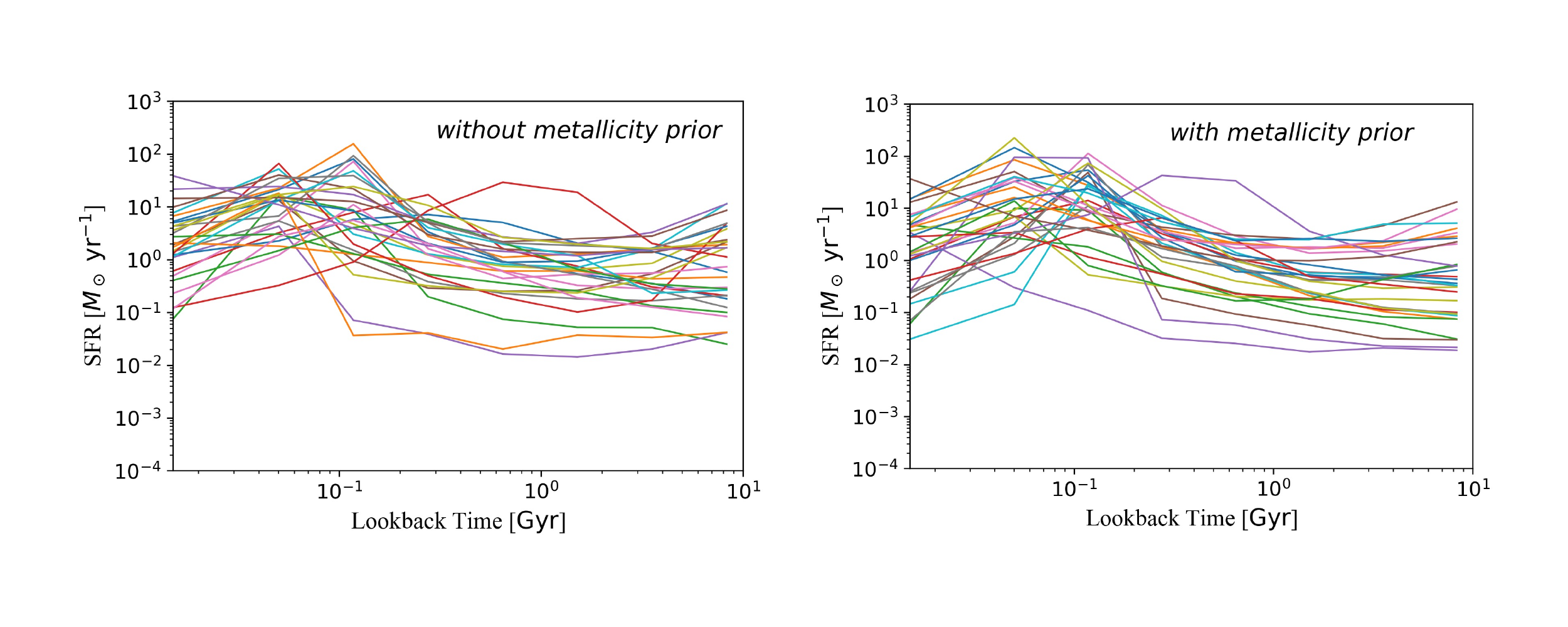}
    \includegraphics[width=0.9\linewidth]{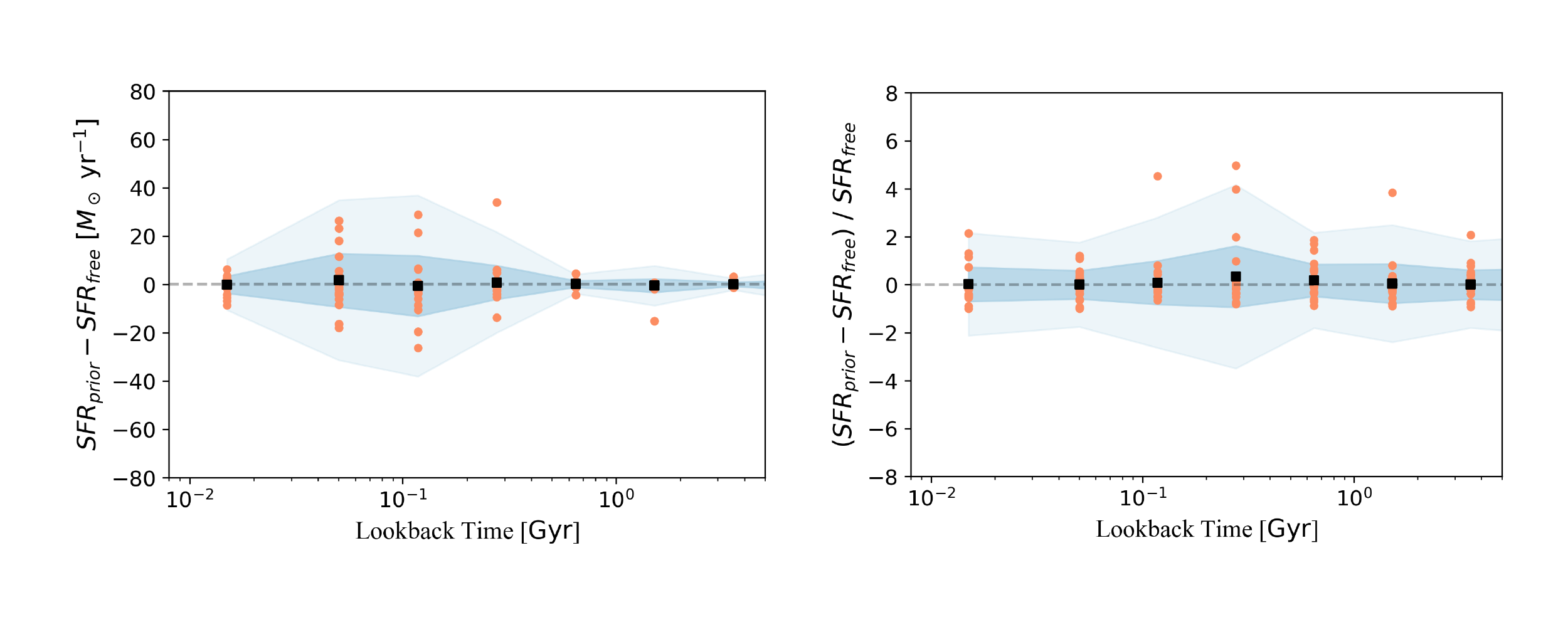}
    \caption{No systematic difference is observed between runs with and without a prior on the metallicity. The top panels show individual SFHs obtained during the test runs, with (right) and without (left) adopting a metallicity prior. Overall, the shape of the SFH is qualitatively fairly insensitive to the prior, and only small differences are observed. The bottom panels show the absolute (left) and fractional (right) differences in the SFHs, respectively, with and without the prior. The shaded areas indicate the 68th (blue) and 95th (light blue) percentile ranges.}
    \label{fig:sfh_zprior}
\end{figure*}

Finally, Figure \ref{fig:sfh_zprior} illustrates the sensitivity of the output SFH to the gas-phase metallicity prior. This test is relevant for this work because during the primary runs the same metallicity prior is used for both the centers and the outskirts. A strong dependence of the SFH on the metallicity prior would have diminished the significance of the difference that we have observed for the two regions of the galaxies. As the figures illustrate, the test reveals only small differences in the output SFH, suggesting a small sensitivity to the metallicity prior. These differences are substantially smaller than the differences of the SFHs of centers and outskirts, suggesting that they are very unlikely the result of of the over-simplification of assigning the same metallicity to both regions of the galaxies. The bottom panels of Figure \ref{fig:sfh_zprior} show the difference between the SFHs derived adopting a strong prior on the gas phase metallicity and without such a prior. No systematic difference is observed, on average, between the two cases, with the scatter of the fractional difference (bottom left panel) remaining approximately constant with look-back time.

Overall, the test runs show that the overall reconstruction of the SFHs of the sample galaxies is robust against the input photometry and the assumption of the metallicity prior, supporting the validity of our conclusions. 

\bibliographystyle{aasjournal}
\bibliography{references}

\begin{thebibliography}{}
\expandafter\ifx\csname natexlab\endcsname\relax\def\natexlab#1{#1}\fi
\providecommand{\url}[1]{\href{#1}{#1}}
\providecommand{\dodoi}[1]{doi:~\href{http://doi.org/#1}{\nolinkurl{#1}}}
\providecommand{\doeprint}[1]{\href{http://ascl.net/#1}{\nolinkurl{http://ascl.net/#1}}}
\providecommand{\doarXiv}[1]{\href{https://arxiv.org/abs/#1}{\nolinkurl{https://arxiv.org/abs/#1}}}

\bibitem[{{Abramson} {et~al.}(2014){Abramson}, {Kelson}, {Dressler},
  {Poggianti}, {Gladders}, {Oemler}, \& {Vulcani}}]{Abramson14}
{Abramson}, L.~E., {Kelson}, D.~D., {Dressler}, A., {et~al.} 2014, \apjl, 785,
  L36, \dodoi{10.1088/2041-8205/785/2/L36}

\bibitem[{{Ashby} {et~al.}(2013){Ashby}, {Willner}, {Fazio}, {Huang}, {Arendt},
  {Barmby}, {Barro}, {Bell}, {Bouwens}, {Cattaneo}, {Croton}, {Dav{\'e}},
  {Dunlop}, {Egami}, {Faber}, {Finlator}, {Grogin}, {Guhathakurta},
  {Hernquist}, {Hora}, {Illingworth}, {Kashlinsky}, {Koekemoer}, {Koo},
  {Labb{\'e}}, {Li}, {Lin}, {Moseley}, {Nandra}, {Newman}, {Noeske}, {Ouchi},
  {Peth}, {Rigopoulou}, {Robertson}, {Sarajedini}, {Simard}, {Smith}, {Wang},
  {Wechsler}, {Weiner}, {Wilson}, {Wuyts}, {Yamada}, \& {Yan}}]{Ashby13}
{Ashby}, M.~L.~N., {Willner}, S.~P., {Fazio}, G.~G., {et~al.} 2013, \apj, 769,
  80, \dodoi{10.1088/0004-637X/769/1/80}

\bibitem[{{Asplund} {et~al.}(2009){Asplund}, {Grevesse}, {Sauval}, \&
  {Scott}}]{Asplund09}
{Asplund}, M., {Grevesse}, N., {Sauval}, A.~J., \& {Scott}, P. 2009, \araa, 47,
  481, \dodoi{10.1146/annurev.astro.46.060407.145222}

\bibitem[{{Astropy Collaboration} {et~al.}(2013){Astropy Collaboration},
  {Robitaille}, {Tollerud}, {Greenfield}, {Droettboom}, {Bray}, {Aldcroft},
  {Davis}, {Ginsburg}, {Price-Whelan}, {Kerzendorf}, {Conley}, {Crighton},
  {Barbary}, {Muna}, {Ferguson}, {Grollier}, {Parikh}, {Nair}, {Unther},
  {Deil}, {Woillez}, {Conseil}, {Kramer}, {Turner}, {Singer}, {Fox}, {Weaver},
  {Zabalza}, {Edwards}, {Azalee Bostroem}, {Burke}, {Casey}, {Crawford},
  {Dencheva}, {Ely}, {Jenness}, {Labrie}, {Lim}, {Pierfederici}, {Pontzen},
  {Ptak}, {Refsdal}, {Servillat}, \& {Streicher}}]{astropy:2013}
{Astropy Collaboration}, {Robitaille}, T.~P., {Tollerud}, E.~J., {et~al.} 2013,
  \aap, 558, A33, \dodoi{10.1051/0004-6361/201322068}

\bibitem[{{Astropy Collaboration} {et~al.}(2018){Astropy Collaboration},
  {Price-Whelan}, {Sip{\H{o}}cz}, {G{\"u}nther}, {Lim}, {Crawford}, {Conseil},
  {Shupe}, {Craig}, {Dencheva}, {Ginsburg}, {Vand erPlas}, {Bradley},
  {P{\'e}rez-Su{\'a}rez}, {de Val-Borro}, {Aldcroft}, {Cruz}, {Robitaille},
  {Tollerud}, {Ardelean}, {Babej}, {Bach}, {Bachetti}, {Bakanov}, {Bamford},
  {Barentsen}, {Barmby}, {Baumbach}, {Berry}, {Biscani}, {Boquien}, {Bostroem},
  {Bouma}, {Brammer}, {Bray}, {Breytenbach}, {Buddelmeijer}, {Burke},
  {Calderone}, {Cano Rodr{\'\i}guez}, {Cara}, {Cardoso}, {Cheedella}, {Copin},
  {Corrales}, {Crichton}, {D'Avella}, {Deil}, {Depagne}, {Dietrich}, {Donath},
  {Droettboom}, {Earl}, {Erben}, {Fabbro}, {Ferreira}, {Finethy}, {Fox},
  {Garrison}, {Gibbons}, {Goldstein}, {Gommers}, {Greco}, {Greenfield},
  {Groener}, {Grollier}, {Hagen}, {Hirst}, {Homeier}, {Horton}, {Hosseinzadeh},
  {Hu}, {Hunkeler}, {Ivezi{\'c}}, {Jain}, {Jenness}, {Kanarek}, {Kendrew},
  {Kern}, {Kerzendorf}, {Khvalko}, {King}, {Kirkby}, {Kulkarni}, {Kumar},
  {Lee}, {Lenz}, {Littlefair}, {Ma}, {Macleod}, {Mastropietro}, {McCully},
  {Montagnac}, {Morris}, {Mueller}, {Mumford}, {Muna}, {Murphy}, {Nelson},
  {Nguyen}, {Ninan}, {N{\"o}the}, {Ogaz}, {Oh}, {Parejko}, {Parley}, {Pascual},
  {Patil}, {Patil}, {Plunkett}, {Prochaska}, {Rastogi}, {Reddy Janga},
  {Sabater}, {Sakurikar}, {Seifert}, {Sherbert}, {Sherwood-Taylor}, {Shih},
  {Sick}, {Silbiger}, {Singanamalla}, {Singer}, {Sladen}, {Sooley},
  {Sornarajah}, {Streicher}, {Teuben}, {Thomas}, {Tremblay}, {Turner},
  {Terr{\'o}n}, {van Kerkwijk}, {de la Vega}, {Watkins}, {Weaver}, {Whitmore},
  {Woillez}, {Zabalza}, \& {Astropy Contributors}}]{astropy:2018}
{Astropy Collaboration}, {Price-Whelan}, A.~M., {Sip{\H{o}}cz}, B.~M., {et~al.}
  2018, \aj, 156, 123, \dodoi{10.3847/1538-3881/aabc4f}

\bibitem[{{Astropy Collaboration} {et~al.}(2022){Astropy Collaboration},
  {Price-Whelan}, {Lim}, {Earl}, {Starkman}, {Bradley}, {Shupe}, {Patil},
  {Corrales}, {Brasseur}, {N{"o}the}, {Donath}, {Tollerud}, {Morris},
  {Ginsburg}, {Vaher}, {Weaver}, {Tocknell}, {Jamieson}, {van Kerkwijk},
  {Robitaille}, {Merry}, {Bachetti}, {G{"u}nther}, {Aldcroft},
  {Alvarado-Montes}, {Archibald}, {B{'o}di}, {Bapat}, {Barentsen}, {Baz{'a}n},
  {Biswas}, {Boquien}, {Burke}, {Cara}, {Cara}, {Conroy}, {Conseil}, {Craig},
  {Cross}, {Cruz}, {D'Eugenio}, {Dencheva}, {Devillepoix}, {Dietrich},
  {Eigenbrot}, {Erben}, {Ferreira}, {Foreman-Mackey}, {Fox}, {Freij}, {Garg},
  {Geda}, {Glattly}, {Gondhalekar}, {Gordon}, {Grant}, {Greenfield}, {Groener},
  {Guest}, {Gurovich}, {Handberg}, {Hart}, {Hatfield-Dodds}, {Homeier},
  {Hosseinzadeh}, {Jenness}, {Jones}, {Joseph}, {Kalmbach}, {Karamehmetoglu},
  {Ka{l}uszy{'n}ski}, {Kelley}, {Kern}, {Kerzendorf}, {Koch}, {Kulumani},
  {Lee}, {Ly}, {Ma}, {MacBride}, {Maljaars}, {Muna}, {Murphy}, {Norman},
  {O'Steen}, {Oman}, {Pacifici}, {Pascual}, {Pascual-Granado}, {Patil},
  {Perren}, {Pickering}, {Rastogi}, {Roulston}, {Ryan}, {Rykoff}, {Sabater},
  {Sakurikar}, {Salgado}, {Sanghi}, {Saunders}, {Savchenko}, {Schwardt},
  {Seifert-Eckert}, {Shih}, {Jain}, {Shukla}, {Sick}, {Simpson},
  {Singanamalla}, {Singer}, {Singhal}, {Sinha}, {Sip{H{o}}cz}, {Spitler},
  {Stansby}, {Streicher}, {{{S}}umak}, {Swinbank}, {Taranu}, {Tewary},
  {Tremblay}, {Val-Borro}, {Van Kooten}, {Vasovi{'c}}, {Verma}, {de Miranda
  Cardoso}, {Williams}, {Wilson}, {Winkel}, {Wood-Vasey}, {Xue}, {Yoachim},
  {Zhang}, {Zonca}, \& {Astropy Project Contributors}}]{astropy:2022}
{Astropy Collaboration}, {Price-Whelan}, A.~M., {Lim}, P.~L., {et~al.} 2022,
  apj, 935, 167, \dodoi{10.3847/1538-4357/ac7c74}

\bibitem[{{Barbuy} {et~al.}(2018){Barbuy}, {Chiappini}, \&
  {Gerhard}}]{Barbuy18}
{Barbuy}, B., {Chiappini}, C., \& {Gerhard}, O. 2018, \araa, 56, 223,
  \dodoi{10.1146/annurev-astro-081817-051826}

\bibitem[{{Barro} {et~al.}(2017){Barro}, {Faber}, {Koo}, {Dekel}, {Fang},
  {Trump}, {P{\'e}rez-Gonz{\'a}lez}, {Pacifici}, {Primack}, {Somerville},
  {Yan}, {Guo}, {Liu}, {Ceverino}, {Kocevski}, \& {McGrath}}]{Barro17}
{Barro}, G., {Faber}, S.~M., {Koo}, D.~C., {et~al.} 2017, \apj, 840, 47,
  \dodoi{10.3847/1538-4357/aa6b05}

\bibitem[{{Barro} {et~al.}(2019){Barro}, {P{\'e}rez-Gonz{\'a}lez}, {Cava},
  {Brammer}, {Pandya}, {Eliche Moral}, {Esquej}, {Dom{\'\i}nguez-S{\'a}nchez},
  {Alcalde Pampliega}, {Guo}, {Koekemoer}, {Trump}, {Ashby}, {Cardiel},
  {Castellano}, {Conselice}, {Dickinson}, {Dolch}, {Donley}, {Espino Briones},
  {Faber}, {Fazio}, {Ferguson}, {Finkelstein}, {Fontana}, {Galametz},
  {Gardner}, {Gawiser}, {Giavalisco}, {Grazian}, {Grogin}, {Hathi}, {Hemmati},
  {Hern{\'a}n-Caballero}, {Kocevski}, {Koo}, {Kodra}, {Lee}, {Lin}, {Lucas},
  {Mobasher}, {McGrath}, {Nandra}, {Nayyeri}, {Newman}, {Pforr}, {Peth},
  {Rafelski}, {Rodr{\'\i}guez-Munoz}, {Salvato}, {Stefanon}, {van der Wel},
  {Willner}, {Wiklind}, \& {Wuyts}}]{Barro19}
{Barro}, G., {P{\'e}rez-Gonz{\'a}lez}, P.~G., {Cava}, A., {et~al.} 2019, \apjs,
  243, 22, \dodoi{10.3847/1538-4365/ab23f2}

\bibitem[{{Bensby} {et~al.}(2017){Bensby}, {Feltzing}, {Gould}, {Yee},
  {Johnson}, {Asplund}, {Mel{\'e}ndez}, {Lucatello}, {Howes}, {McWilliam},
  {Udalski}, {Szyma{\'n}ski}, {Soszy{\'n}ski}, {Poleski}, {Wyrzykowski},
  {Ulaczyk}, {Koz{\l}owski}, {Pietrukowicz}, {Skowron}, {Mr{\'o}z}, {Pawlak},
  {Abe}, {Asakura}, {Bhattacharya}, {Bond}, {Bennett}, {Hirao}, {Nagakane},
  {Koshimoto}, {Sumi}, {Suzuki}, \& {Tristram}}]{Bensby17}
{Bensby}, T., {Feltzing}, S., {Gould}, A., {et~al.} 2017, \aap, 605, A89,
  \dodoi{10.1051/0004-6361/201730560}

\bibitem[{{Bradley} {et~al.}(2022){Bradley}, {Sip{\H{o}}cz}, {Robitaille},
  {Tollerud}, {Vin{\'\i}cius}, {Deil}, {Barbary}, {Wilson}, {Busko}, {Donath},
  {G{\"u}nther}, {Cara}, {Lim}, {Me{\ss}linger}, {Conseil}, {Bostroem},
  {Droettboom}, {Bray}, {Andersen Bratholm}, {Barentsen}, {Craig}, {Rathi},
  {Pascual}, {Perren}, {Georgiev}, {De Val-Borro}, {Kerzendorf}, {Bach},
  {Quint}, \& {Souchereau}}]{photutils}
{Bradley}, L., {Sip{\H{o}}cz}, B., {Robitaille}, T., {et~al.} 2022,
  {astropy/photutils: 1.5.0}, 1.5.0, Zenodo,  Zenodo,
  \dodoi{10.5281/zenodo.6825092}

\bibitem[{{Brammer} {et~al.}(2012){Brammer}, {van Dokkum}, {Franx},
  {Fumagalli}, {Patel}, {Rix}, {Skelton}, {Kriek}, {Nelson}, {Schmidt},
  {Bezanson}, {da Cunha}, {Erb}, {Fan}, {F{\"o}rster Schreiber}, {Illingworth},
  {Labb{\'e}}, {Leja}, {Lundgren}, {Magee}, {Marchesini}, {McCarthy},
  {Momcheva}, {Muzzin}, {Quadri}, {Steidel}, {Tal}, {Wake}, {Whitaker}, \&
  {Williams}}]{Brammer12}
{Brammer}, G.~B., {van Dokkum}, P.~G., {Franx}, M., {et~al.} 2012, \apjs, 200,
  13, \dodoi{10.1088/0067-0049/200/2/13}

\bibitem[{{Capak} {et~al.}(2004){Capak}, {Cowie}, {Hu}, {Barger}, {Dickinson},
  {Fernandez}, {Giavalisco}, {Komiyama}, {Kretchmer}, {McNally}, {Miyazaki},
  {Okamura}, \& {Stern}}]{Capak04}
{Capak}, P., {Cowie}, L.~L., {Hu}, E.~M., {et~al.} 2004, \aj, 127, 180,
  \dodoi{10.1086/380611}

\bibitem[{{Carrasco} {et~al.}(2010){Carrasco}, {Conselice}, \&
  {Trujillo}}]{Carrasco10}
{Carrasco}, E.~R., {Conselice}, C.~J., \& {Trujillo}, I. 2010, \mnras, 405,
  2253, \dodoi{10.1111/j.1365-2966.2010.16645.x}

\bibitem[{{Ceverino} {et~al.}(2015){Ceverino}, {Dekel}, {Tweed}, \&
  {Primack}}]{Ceverino15}
{Ceverino}, D., {Dekel}, A., {Tweed}, D., \& {Primack}, J. 2015, \mnras, 447,
  3291, \dodoi{10.1093/mnras/stu2694}

\bibitem[{{Chen} {et~al.}(2020){Chen}, {Faber}, {Koo}, {Somerville}, {Primack},
  {Dekel}, {Rodr{\'\i}guez-Puebla}, {Guo}, {Barro}, {Kocevski}, {van der Wel},
  {Woo}, {Bell}, {Fang}, {Ferguson}, {Giavalisco}, {Huertas-Company}, {Jiang},
  {Kassin}, {Lin}, {Liu}, {Luo}, {Luo}, {Pacifici}, {Pandya}, {Salim}, {Shu},
  {Tacchella}, {Terrazas}, \& {Yesuf}}]{Chen20}
{Chen}, Z., {Faber}, S.~M., {Koo}, D.~C., {et~al.} 2020, \apj, 897, 102,
  \dodoi{10.3847/1538-4357/ab9633}

\bibitem[{{Cheung} {et~al.}(2012){Cheung}, {Faber}, {Koo}, {Dutton}, {Simard},
  {McGrath}, {Huang}, {Bell}, {Dekel}, {Fang}, {Salim}, {Barro}, {Bundy},
  {Coil}, {Cooper}, {Conselice}, {Davis}, {Dom{\'\i}nguez}, {Kassin},
  {Kocevski}, {Koekemoer}, {Lin}, {Lotz}, {Newman}, {Phillips}, {Rosario},
  {Weiner}, \& {Willmer}}]{Cheung12}
{Cheung}, E., {Faber}, S.~M., {Koo}, D.~C., {et~al.} 2012, \apj, 760, 131,
  \dodoi{10.1088/0004-637X/760/2/131}

\bibitem[{{Coil} {et~al.}(2015){Coil}, {Aird}, {Reddy}, {Shapley}, {Kriek},
  {Siana}, {Mobasher}, {Freeman}, {Price}, \& {Shivaei}}]{Coil15}
{Coil}, A.~L., {Aird}, J., {Reddy}, N., {et~al.} 2015, \apj, 801, 35,
  \dodoi{10.1088/0004-637X/801/1/35}

\bibitem[{{Conroy}(2013)}]{Conroy13}
{Conroy}, C. 2013, \araa, 51, 393, \dodoi{10.1146/annurev-astro-082812-141017}

\bibitem[{{Conroy} \& {Gunn}(2010)}]{Conroy10b}
{Conroy}, C., \& {Gunn}, J.~E. 2010, \apj, 712, 833,
  \dodoi{10.1088/0004-637X/712/2/833}

\bibitem[{{Conroy} {et~al.}(2009){Conroy}, {Gunn}, \& {White}}]{Conroy09}
{Conroy}, C., {Gunn}, J.~E., \& {White}, M. 2009, \apj, 699, 486,
  \dodoi{10.1088/0004-637X/699/1/486}

\bibitem[{{Conroy} {et~al.}(2010){Conroy}, {White}, \& {Gunn}}]{Conroy10a}
{Conroy}, C., {White}, M., \& {Gunn}, J.~E. 2010, \apj, 708, 58,
  \dodoi{10.1088/0004-637X/708/1/58}

\bibitem[{{Cutler} {et~al.}(2022){Cutler}, {Whitaker}, {Mowla}, {Brammer}, {van
  der Wel}, {Marchesini}, {van Dokkum}, {Momcheva}, {Song}, {Akhshik},
  {Nelson}, {Bezanson}, {Franx}, {Kriek}, {Lange-Vagle}, {Leja}, {MacKenty},
  {Muzzin}, \& {Shipley}}]{Cutler22}
{Cutler}, S.~E., {Whitaker}, K.~E., {Mowla}, L.~A., {et~al.} 2022, \apj, 925,
  34, \dodoi{10.3847/1538-4357/ac341c}

\bibitem[{{Dekel} \& {Burkert}(2014)}]{Dekel14}
{Dekel}, A., \& {Burkert}, A. 2014, \mnras, 438, 1870,
  \dodoi{10.1093/mnras/stt2331}

\bibitem[{{Dekel} {et~al.}(2019){Dekel}, {Lapiner}, \& {Dubois}}]{Dekel19}
{Dekel}, A., {Lapiner}, S., \& {Dubois}, Y. 2019, arXiv e-prints,
  arXiv:1904.08431.
\newblock \doarXiv{1904.08431}

\bibitem[{{Dekel} {et~al.}(2009{\natexlab{a}}){Dekel}, {Sari}, \&
  {Ceverino}}]{Dekel09a}
{Dekel}, A., {Sari}, R., \& {Ceverino}, D. 2009{\natexlab{a}}, \apj, 703, 785,
  \dodoi{10.1088/0004-637X/703/1/785}

\bibitem[{{Dekel} {et~al.}(2009{\natexlab{b}}){Dekel}, {Birnboim}, {Engel},
  {Freundlich}, {Goerdt}, {Mumcuoglu}, {Neistein}, {Pichon}, {Teyssier}, \&
  {Zinger}}]{Dekel09}
{Dekel}, A., {Birnboim}, Y., {Engel}, G., {et~al.} 2009{\natexlab{b}}, \nat,
  457, 451, \dodoi{10.1038/nature07648}

\bibitem[{{Dickinson} {et~al.}(2003){Dickinson}, {Bergeron}, {Casertano},
  {Cesarsky}, {Chary}, {Cristiani}, {Eisenhardt}, {Elbaz}, {Elston}, {Fall},
  {Ferguson}, {Fosbury}, {Giacconi}, {Giavalisco}, {Grogin}, {Hauser},
  {Hanisch}, {Hook}, {J{\~a}{\c{R}}gensen}, {Koekemoer}, {Ledlow}, {Livio},
  {Mobasher}, {Padovani}, {Papovich}, {Primack}, {Rauscher}, {Reach},
  {Renzini}, {Rieke}, {Rosati}, {Roth}, {Roy}, {Schreier}, {Stern},
  {Stiavelli}, {Takamiya}, {Tollestrup}, {Urry}, {Williams}, {Winge}, \&
  {Wright}}]{Dickinson03}
{Dickinson}, M., {Bergeron}, J., {Casertano}, S., {et~al.} 2003, {Great
  Observatories Origins Deep Survey (GOODS) Validation Observations}, Spitzer
  Proposal

\bibitem[{{Dimauro} {et~al.}(2022){Dimauro}, {Daddi}, {Shankar}, {Cattaneo},
  {Huertas-Company}, {Bernardi}, {Caro}, {Dupke}, {H{\"a}u{\ss}ler},
  {Johnston}, {Cortesi}, {Mei}, \& {Peletier}}]{Dimauro22}
{Dimauro}, P., {Daddi}, E., {Shankar}, F., {et~al.} 2022, \mnras, 513, 256,
  \dodoi{10.1093/mnras/stac884}

\bibitem[{{Elbaz} {et~al.}(2018){Elbaz}, {Leiton}, {Nagar}, {Okumura},
  {Franco}, {Schreiber}, {Pannella}, {Wang}, {Dickinson}, {D{\'\i}az-Santos},
  {Ciesla}, {Daddi}, {Bournaud}, {Magdis}, {Zhou}, \& {Rujopakarn}}]{Elbaz18}
{Elbaz}, D., {Leiton}, R., {Nagar}, N., {et~al.} 2018, \aap, 616, A110,
  \dodoi{10.1051/0004-6361/201732370}

\bibitem[{{Elmegreen} {et~al.}(2008){Elmegreen}, {Bournaud}, \&
  {Elmegreen}}]{Elmegreen08}
{Elmegreen}, B.~G., {Bournaud}, F., \& {Elmegreen}, D.~M. 2008, \apj, 688, 67,
  \dodoi{10.1086/592190}

\bibitem[{{Fagioli} {et~al.}(2016){Fagioli}, {Carollo}, {Renzini}, {Lilly},
  {Onodera}, \& {Tacchella}}]{Fagioli16}
{Fagioli}, M., {Carollo}, C.~M., {Renzini}, A., {et~al.} 2016, \apj, 831, 173,
  \dodoi{10.3847/0004-637X/831/2/173}

\bibitem[{{Fetherolf} {et~al.}(2020){Fetherolf}, {Reddy}, {Shapley}, {Kriek},
  {Siana}, {Coil}, {Mobasher}, {Freeman}, {Sanders}, {Price}, {Shivaei},
  {Azadi}, {de Groot}, {Leung}, \& {Zick}}]{Fetherolf20}
{Fetherolf}, T., {Reddy}, N.~A., {Shapley}, A.~E., {et~al.} 2020, \mnras, 498,
  5009, \dodoi{10.1093/mnras/staa2775}

\bibitem[{{Foreman-Mackey} {et~al.}(2013){Foreman-Mackey}, {Hogg}, {Lang}, \&
  {Goodman}}]{ForemanMackey13}
{Foreman-Mackey}, D., {Hogg}, D.~W., {Lang}, D., \& {Goodman}, J. 2013, \pasp,
  125, 306, \dodoi{10.1086/670067}

\bibitem[{{Foreman-Mackey} {et~al.}(2014){Foreman-Mackey}, {Sick}, \&
  {Johnson}}]{ForemanMackey14}
{Foreman-Mackey}, D., {Sick}, J., \& {Johnson}, B. 2014, {python-fsps: Python
  bindings to FSPS (v0.1.1)}, v0.1.1,  Zenodo, \dodoi{10.5281/zenodo.12157}

\bibitem[{{Franco} {et~al.}(2020){Franco}, {Elbaz}, {Zhou}, {Magnelli},
  {Schreiber}, {Ciesla}, {Dickinson}, {Nagar}, {Magdis}, {Alexander},
  {B{\'e}thermin}, {Demarco}, {Daddi}, {Wang}, {Mullaney}, {Sargent}, {Inami},
  {Shu}, {Bournaud}, {Chary}, {Coogan}, {Ferguson}, {Finkelstein},
  {Giavalisco}, {G{\'o}mez-Guijarro}, {Iono}, {Juneau}, {Lagache}, {Lin},
  {Motohara}, {Okumura}, {Pannella}, {Papovich}, {Pope}, {Rujopakarn},
  {Silverman}, \& {Xiao}}]{Franco20}
{Franco}, M., {Elbaz}, D., {Zhou}, L., {et~al.} 2020, \aap, 643, A30,
  \dodoi{10.1051/0004-6361/202038312}

\bibitem[{{Genzel} {et~al.}(2017){Genzel}, {F{\"o}rster Schreiber},
  {{\"U}bler}, {Lang}, {Naab}, {Bender}, {Tacconi}, {Wisnioski}, {Wuyts},
  {Alexander}, {Beifiori}, {Belli}, {Brammer}, {Burkert}, {Carollo}, {Chan},
  {Davies}, {Fossati}, {Galametz}, {Genel}, {Gerhard}, {Lutz}, {Mendel},
  {Momcheva}, {Nelson}, {Renzini}, {Saglia}, {Sternberg}, {Tacchella},
  {Tadaki}, \& {Wilman}}]{Genzel17}
{Genzel}, R., {F{\"o}rster Schreiber}, N.~M., {{\"U}bler}, H., {et~al.} 2017,
  \nat, 543, 397, \dodoi{10.1038/nature21685}

\bibitem[{{Genzel} {et~al.}(2020){Genzel}, {Price}, {{\"U}bler}, {F{\"o}rster
  Schreiber}, {Shimizu}, {Tacconi}, {Bender}, {Burkert}, {Contursi}, {Coogan},
  {Davies}, {Davies}, {Dekel}, {Herrera-Camus}, {Lee}, {Lutz}, {Naab}, {Neri},
  {Nestor}, {Renzini}, {Saglia}, {Schuster}, {Sternberg}, {Wisnioski}, \&
  {Wuyts}}]{Genzel20}
{Genzel}, R., {Price}, S.~H., {{\"U}bler}, H., {et~al.} 2020, \apj, 902, 98,
  \dodoi{10.3847/1538-4357/abb0ea}

\bibitem[{{Giavalisco}(2002)}]{Giavalisco02}
{Giavalisco}, M. 2002, \araa, 40, 579,
  \dodoi{10.1146/annurev.astro.40.121301.111837}

\bibitem[{{Giavalisco} {et~al.}(1996){Giavalisco}, {Steidel}, \&
  {Macchetto}}]{Giavalisco96}
{Giavalisco}, M., {Steidel}, C.~C., \& {Macchetto}, F.~D. 1996, \apj, 470, 189,
  \dodoi{10.1086/177859}

\bibitem[{{Giavalisco} {et~al.}(2004){Giavalisco}, {Ferguson}, {Koekemoer},
  {Dickinson}, {Alexander}, {Bauer}, {Bergeron}, {Biagetti}, {Brandt},
  {Casertano}, {Cesarsky}, {Chatzichristou}, {Conselice}, {Cristiani}, {Da
  Costa}, {Dahlen}, {de Mello}, {Eisenhardt}, {Erben}, {Fall}, {Fassnacht},
  {Fosbury}, {Fruchter}, {Gardner}, {Grogin}, {Hook}, {Hornschemeier}, {Idzi},
  {Jogee}, {Kretchmer}, {Laidler}, {Lee}, {Livio}, {Lucas}, {Madau},
  {Mobasher}, {Moustakas}, {Nonino}, {Padovani}, {Papovich}, {Park},
  {Ravindranath}, {Renzini}, {Richardson}, {Riess}, {Rosati}, {Schirmer},
  {Schreier}, {Somerville}, {Spinrad}, {Stern}, {Stiavelli}, {Strolger},
  {Urry}, {Vandame}, {Williams}, \& {Wolf}}]{Giavalisco04}
{Giavalisco}, M., {Ferguson}, H.~C., {Koekemoer}, A.~M., {et~al.} 2004, \apjl,
  600, L93, \dodoi{10.1086/379232}

\bibitem[{{G{\'o}mez-Guijarro} {et~al.}(2022){G{\'o}mez-Guijarro}, {Elbaz},
  {Xiao}, {Kokorev}, {Magdis}, {Magnelli}, {Daddi}, {Valentino}, {Sargent},
  {Dickinson}, {B{\'e}thermin}, {Franco}, {Pope}, {Kalita}, {Ciesla},
  {Demarco}, {Inami}, {Rujopakarn}, {Shu}, {Wang}, {Zhou}, {Alexander},
  {Bournaud}, {Chary}, {Ferguson}, {Finkelstein}, {Giavalisco}, {Iono},
  {Juneau}, {Kartaltepe}, {Lagache}, {Le Floc'h}, {Leiton}, {Leroy}, {Lin},
  {Motohara}, {Mullaney}, {Okumura}, {Pannella}, {Papovich}, \&
  {Treister}}]{Gomez-Guijarro22}
{G{\'o}mez-Guijarro}, C., {Elbaz}, D., {Xiao}, M., {et~al.} 2022, \aap, 659,
  A196, \dodoi{10.1051/0004-6361/202142352}

\bibitem[{{Grogin} {et~al.}(2011){Grogin}, {Kocevski}, {Faber}, {Ferguson},
  {Koekemoer}, {Riess}, {Acquaviva}, {Alexander}, {Almaini}, {Ashby}, {Barden},
  {Bell}, {Bournaud}, {Brown}, {Caputi}, {Casertano}, {Cassata}, {Castellano},
  {Challis}, {Chary}, {Cheung}, {Cirasuolo}, {Conselice}, {Roshan Cooray},
  {Croton}, {Daddi}, {Dahlen}, {Dav{\'e}}, {de Mello}, {Dekel}, {Dickinson},
  {Dolch}, {Donley}, {Dunlop}, {Dutton}, {Elbaz}, {Fazio}, {Filippenko},
  {Finkelstein}, {Fontana}, {Gardner}, {Garnavich}, {Gawiser}, {Giavalisco},
  {Grazian}, {Guo}, {Hathi}, {H{\"a}ussler}, {Hopkins}, {Huang}, {Huang},
  {Jha}, {Kartaltepe}, {Kirshner}, {Koo}, {Lai}, {Lee}, {Li}, {Lotz}, {Lucas},
  {Madau}, {McCarthy}, {McGrath}, {McIntosh}, {McLure}, {Mobasher},
  {Moustakas}, {Mozena}, {Nandra}, {Newman}, {Niemi}, {Noeske}, {Papovich},
  {Pentericci}, {Pope}, {Primack}, {Rajan}, {Ravindranath}, {Reddy}, {Renzini},
  {Rix}, {Robaina}, {Rodney}, {Rosario}, {Rosati}, {Salimbeni}, {Scarlata},
  {Siana}, {Simard}, {Smidt}, {Somerville}, {Spinrad}, {Straughn}, {Strolger},
  {Telford}, {Teplitz}, {Trump}, {van der Wel}, {Villforth}, {Wechsler},
  {Weiner}, {Wiklind}, {Wild}, {Wilson}, {Wuyts}, {Yan}, \& {Yun}}]{Grogin11}
{Grogin}, N.~A., {Kocevski}, D.~D., {Faber}, S.~M., {et~al.} 2011, \apjs, 197,
  35, \dodoi{10.1088/0067-0049/197/2/35}

\bibitem[{{Guo} {et~al.}(2015){Guo}, {Zheng}, {Wang}, \& {Fu}}]{Guo15}
{Guo}, K., {Zheng}, X.~Z., {Wang}, T., \& {Fu}, H. 2015, \apjl, 808, L49,
  \dodoi{10.1088/2041-8205/808/2/L49}

\bibitem[{{Guo} {et~al.}(2012){Guo}, {Giavalisco}, {Ferguson}, {Cassata}, \&
  {Koekemoer}}]{Guo12}
{Guo}, Y., {Giavalisco}, M., {Ferguson}, H.~C., {Cassata}, P., \& {Koekemoer},
  A.~M. 2012, \apj, 757, 120, \dodoi{10.1088/0004-637X/757/2/120}

\bibitem[{{H{\"a}ring} \& {Rix}(2004)}]{Haring04}
{H{\"a}ring}, N., \& {Rix}, H.-W. 2004, \apjl, 604, L89, \dodoi{10.1086/383567}

\bibitem[{{Huertas-Company} {et~al.}(2020){Huertas-Company}, {Guo}, {Ginzburg},
  {Lee}, {Mandelker}, {Metter}, {Primack}, {Dekel}, {Ceverino}, {Faber}, {Koo},
  {Koekemoer}, {Snyder}, {Giavalisco}, \& {Zhang}}]{Huertas-Company20}
{Huertas-Company}, M., {Guo}, Y., {Ginzburg}, O., {et~al.} 2020, \mnras, 499,
  814, \dodoi{10.1093/mnras/staa2777}

\bibitem[{{Hunter}(2007)}]{matplotlib2007}
{Hunter}, J.~D. 2007, Computing in Science Engineering, 9, 90,
  \dodoi{10.1109/MCSE.2007.55}

\bibitem[{{Inoue} \& {Inutsuka}(2016)}]{Inoue16}
{Inoue}, T., \& {Inutsuka}, S.-i. 2016, \apj, 833, 10,
  \dodoi{10.3847/0004-637X/833/1/10}

\bibitem[{{Ji} \& {Giavalisco}(2022)}]{Ji22}
{Ji}, Z., \& {Giavalisco}, M. 2022, arXiv e-prints, arXiv:2204.02414.
\newblock \doarXiv{2204.02414}

\bibitem[{{Johnson} {et~al.}(2021){Johnson}, {Leja}, {Conroy}, \&
  {Speagle}}]{Johnson21}
{Johnson}, B.~D., {Leja}, J., {Conroy}, C., \& {Speagle}, J.~S. 2021, \apjs,
  254, 22, \dodoi{10.3847/1538-4365/abef67}

\bibitem[{{Kajisawa} {et~al.}(2011){Kajisawa}, {Ichikawa}, {Tanaka}, {Yamada},
  {Akiyama}, {Suzuki}, {Tokoku}, {Katsuno Uchimoto}, {Konishi}, {Yoshikawa},
  {Nishimura}, {Omata}, {Ouchi}, {Iwata}, {Hamana}, \& {Onodera}}]{Kajisawa11}
{Kajisawa}, M., {Ichikawa}, T., {Tanaka}, I., {et~al.} 2011, \pasj, 63, 379,
  \dodoi{10.1093/pasj/63.sp2.S379}

\bibitem[{{Kauffmann} {et~al.}(2003){Kauffmann}, {Heckman}, {White}, {Charlot},
  {Tremonti}, {Brinchmann}, {Bruzual}, {Peng}, {Seibert}, {Bernardi},
  {Blanton}, {Brinkmann}, {Castander}, {Cs{\'a}bai}, {Fukugita}, {Ivezic},
  {Munn}, {Nichol}, {Padmanabhan}, {Thakar}, {Weinberg}, \&
  {York}}]{Kauffmann03}
{Kauffmann}, G., {Heckman}, T.~M., {White}, S. D.~M., {et~al.} 2003, \mnras,
  341, 33, \dodoi{10.1046/j.1365-8711.2003.06291.x}

\bibitem[{{Kere{\v{s}}} {et~al.}(2005){Kere{\v{s}}}, {Katz}, {Weinberg}, \&
  {Dav{\'e}}}]{Keres05}
{Kere{\v{s}}}, D., {Katz}, N., {Weinberg}, D.~H., \& {Dav{\'e}}, R. 2005,
  \mnras, 363, 2, \dodoi{10.1111/j.1365-2966.2005.09451.x}

\bibitem[{{Koekemoer} {et~al.}(2011){Koekemoer}, {Faber}, {Ferguson}, {Grogin},
  {Kocevski}, {Koo}, {Lai}, {Lotz}, {Lucas}, {McGrath}, {Ogaz}, {Rajan},
  {Riess}, {Rodney}, {Strolger}, {Casertano}, {Castellano}, {Dahlen},
  {Dickinson}, {Dolch}, {Fontana}, {Giavalisco}, {Grazian}, {Guo}, {Hathi},
  {Huang}, {van der Wel}, {Yan}, {Acquaviva}, {Alexander}, {Almaini}, {Ashby},
  {Barden}, {Bell}, {Bournaud}, {Brown}, {Caputi}, {Cassata}, {Challis},
  {Chary}, {Cheung}, {Cirasuolo}, {Conselice}, {Roshan Cooray}, {Croton},
  {Daddi}, {Dav{\'e}}, {de Mello}, {de Ravel}, {Dekel}, {Donley}, {Dunlop},
  {Dutton}, {Elbaz}, {Fazio}, {Filippenko}, {Finkelstein}, {Frazer}, {Gardner},
  {Garnavich}, {Gawiser}, {Gruetzbauch}, {Hartley}, {H{\"a}ussler},
  {Herrington}, {Hopkins}, {Huang}, {Jha}, {Johnson}, {Kartaltepe},
  {Khostovan}, {Kirshner}, {Lani}, {Lee}, {Li}, {Madau}, {McCarthy},
  {McIntosh}, {McLure}, {McPartland}, {Mobasher}, {Moreira}, {Mortlock},
  {Moustakas}, {Mozena}, {Nandra}, {Newman}, {Nielsen}, {Niemi}, {Noeske},
  {Papovich}, {Pentericci}, {Pope}, {Primack}, {Ravindranath}, {Reddy},
  {Renzini}, {Rix}, {Robaina}, {Rosario}, {Rosati}, {Salimbeni}, {Scarlata},
  {Siana}, {Simard}, {Smidt}, {Snyder}, {Somerville}, {Spinrad}, {Straughn},
  {Telford}, {Teplitz}, {Trump}, {Vargas}, {Villforth}, {Wagner}, {Wandro},
  {Wechsler}, {Weiner}, {Wiklind}, {Wild}, {Wilson}, {Wuyts}, \&
  {Yun}}]{Koekemoer11}
{Koekemoer}, A.~M., {Faber}, S.~M., {Ferguson}, H.~C., {et~al.} 2011, \apjs,
  197, 36, \dodoi{10.1088/0067-0049/197/2/36}

\bibitem[{{Kormendy} \& {Ho}(2013)}]{Kormendy13}
{Kormendy}, J., \& {Ho}, L.~C. 2013, \araa, 51, 511,
  \dodoi{10.1146/annurev-astro-082708-101811}

\bibitem[{{Kormendy} \& {Kennicutt}(2004)}]{Kormendy04}
{Kormendy}, J., \& {Kennicutt}, Robert~C., J. 2004, \araa, 42, 603,
  \dodoi{10.1146/annurev.astro.42.053102.134024}

\bibitem[{{Kriek} {et~al.}(2009){Kriek}, {van Dokkum}, {Labb{\'e}}, {Franx},
  {Illingworth}, {Marchesini}, \& {Quadri}}]{Kriek09}
{Kriek}, M., {van Dokkum}, P.~G., {Labb{\'e}}, I., {et~al.} 2009, \apj, 700,
  221, \dodoi{10.1088/0004-637X/700/1/221}

\bibitem[{{Kriek} {et~al.}(2015){Kriek}, {Shapley}, {Reddy}, {Siana}, {Coil},
  {Mobasher}, {Freeman}, {de Groot}, {Price}, {Sanders}, {Shivaei}, {Brammer},
  {Momcheva}, {Skelton}, {van Dokkum}, {Whitaker}, {Aird}, {Azadi}, {Kassis},
  {Bullock}, {Conroy}, {Dav{\'e}}, {Kere{\v{s}}}, \& {Krumholz}}]{Kriek15}
{Kriek}, M., {Shapley}, A.~E., {Reddy}, N.~A., {et~al.} 2015, \apjs, 218, 15,
  \dodoi{10.1088/0067-0049/218/2/15}

\bibitem[{{Kron}(1980)}]{Kron80}
{Kron}, R.~G. 1980, \apjs, 43, 305, \dodoi{10.1086/190669}

\bibitem[{{Kroupa}(2001)}]{Kroupa01}
{Kroupa}, P. 2001, \mnras, 322, 231, \dodoi{10.1046/j.1365-8711.2001.04022.x}

\bibitem[{{Lee} {et~al.}(2018){Lee}, {Giavalisco}, {Whitaker}, {Williams},
  {Ferguson}, {Acquaviva}, {Koekemoer}, {Straughn}, {Guo}, {Kartaltepe},
  {Lotz}, {Pacifici}, {Croton}, {Somerville}, \& {Lu}}]{Lee18}
{Lee}, B., {Giavalisco}, M., {Whitaker}, K., {et~al.} 2018, \apj, 853, 131,
  \dodoi{10.3847/1538-4357/aaa40f}

\bibitem[{{Leja} {et~al.}(2019){Leja}, {Carnall}, {Johnson}, {Conroy}, \&
  {Speagle}}]{Leja19}
{Leja}, J., {Carnall}, A.~C., {Johnson}, B.~D., {Conroy}, C., \& {Speagle},
  J.~S. 2019, \apj, 876, 3, \dodoi{10.3847/1538-4357/ab133c}

\bibitem[{{Leja} {et~al.}(2013){Leja}, {van Dokkum}, {Momcheva}, {Brammer},
  {Skelton}, {Whitaker}, {Andrews}, {Franx}, {Kriek}, {van der Wel},
  {Bezanson}, {Conroy}, {F{\"o}rster Schreiber}, {Nelson}, \& {Patel}}]{Leja13}
{Leja}, J., {van Dokkum}, P.~G., {Momcheva}, I., {et~al.} 2013, \apjl, 778,
  L24, \dodoi{10.1088/2041-8205/778/2/L24}

\bibitem[{{Leja} {et~al.}(2022){Leja}, {Speagle}, {Ting}, {Johnson}, {Conroy},
  {Whitaker}, {Nelson}, {Dokkum}, \& {Franx}}]{Leja21}
{Leja}, J., {Speagle}, J.~S., {Ting}, Y.-S., {et~al.} 2022, \apj, 936, 165,
  \dodoi{10.3847/1538-4357/ac887d}

\bibitem[{{Liu} {et~al.}(2018){Liu}, {Daddi}, {Dickinson}, {Owen}, {Pannella},
  {Sargent}, {B{\'e}thermin}, {Magdis}, {Gao}, {Shu}, {Wang}, {Jin}, \&
  {Inami}}]{Liu18}
{Liu}, D., {Daddi}, E., {Dickinson}, M., {et~al.} 2018, \apj, 853, 172,
  \dodoi{10.3847/1538-4357/aaa600}

\bibitem[{{Mandelker} {et~al.}(2017){Mandelker}, {Dekel}, {Ceverino}, {DeGraf},
  {Guo}, \& {Primack}}]{Mandelker17}
{Mandelker}, N., {Dekel}, A., {Ceverino}, D., {et~al.} 2017, \mnras, 464, 635,
  \dodoi{10.1093/mnras/stw2358}

\bibitem[{{Marinacci} {et~al.}(2018){Marinacci}, {Vogelsberger}, {Pakmor},
  {Torrey}, {Springel}, {Hernquist}, {Nelson}, {Weinberger}, {Pillepich},
  {Naiman}, \& {Genel}}]{Marinacci18}
{Marinacci}, F., {Vogelsberger}, M., {Pakmor}, R., {et~al.} 2018, \mnras, 480,
  5113, \dodoi{10.1093/mnras/sty2206}

\bibitem[{{Martig} {et~al.}(2009){Martig}, {Bournaud}, {Teyssier}, \&
  {Dekel}}]{Martig09}
{Martig}, M., {Bournaud}, F., {Teyssier}, R., \& {Dekel}, A. 2009, \apj, 707,
  250, \dodoi{10.1088/0004-637X/707/1/250}

\bibitem[{{Miglio} {et~al.}(2021){Miglio}, {Chiappini}, {Mackereth}, {Davies},
  {Brogaard}, {Casagrande}, {Chaplin}, {Girardi}, {Kawata}, {Khan}, {Izzard},
  {Montalb{\'a}n}, {Mosser}, {Vincenzo}, {Bossini}, {Noels}, {Rodrigues},
  {Valentini}, \& {Mandel}}]{Miglio21}
{Miglio}, A., {Chiappini}, C., {Mackereth}, J.~T., {et~al.} 2021, \aap, 645,
  A85, \dodoi{10.1051/0004-6361/202038307}

\bibitem[{{Mobasher} {et~al.}(2015){Mobasher}, {Dahlen}, {Ferguson},
  {Acquaviva}, {Barro}, {Finkelstein}, {Fontana}, {Gruetzbauch}, {Johnson},
  {Lu}, {Papovich}, {Pforr}, {Salvato}, {Somerville}, {Wiklind}, {Wuyts},
  {Ashby}, {Bell}, {Conselice}, {Dickinson}, {Faber}, {Fazio}, {Finlator},
  {Galametz}, {Gawiser}, {Giavalisco}, {Grazian}, {Grogin}, {Guo}, {Hathi},
  {Kocevski}, {Koekemoer}, {Koo}, {Newman}, {Reddy}, {Santini}, \&
  {Wechsler}}]{Mobasher15}
{Mobasher}, B., {Dahlen}, T., {Ferguson}, H.~C., {et~al.} 2015, \apj, 808, 101,
  \dodoi{10.1088/0004-637X/808/1/101}

\bibitem[{{Mosleh} {et~al.}(2020){Mosleh}, {Hosseinnejad},
  {Hosseini-ShahiSavandi}, \& {Tacchella}}]{Mosleh20}
{Mosleh}, M., {Hosseinnejad}, S., {Hosseini-ShahiSavandi}, S.~Z., \&
  {Tacchella}, S. 2020, \apj, 905, 170, \dodoi{10.3847/1538-4357/abc7cc}

\bibitem[{{Naiman} {et~al.}(2018){Naiman}, {Pillepich}, {Springel},
  {Ramirez-Ruiz}, {Torrey}, {Vogelsberger}, {Pakmor}, {Nelson}, {Marinacci},
  {Hernquist}, {Weinberger}, \& {Genel}}]{Naiman18}
{Naiman}, J.~P., {Pillepich}, A., {Springel}, V., {et~al.} 2018, \mnras, 477,
  1206, \dodoi{10.1093/mnras/sty618}

\bibitem[{{Nelson} {et~al.}(2018){Nelson}, {Pillepich}, {Springel},
  {Weinberger}, {Hernquist}, {Pakmor}, {Genel}, {Torrey}, {Vogelsberger},
  {Kauffmann}, {Marinacci}, \& {Naiman}}]{NelsonD18}
{Nelson}, D., {Pillepich}, A., {Springel}, V., {et~al.} 2018, \mnras, 475, 624,
  \dodoi{10.1093/mnras/stx3040}

\bibitem[{{Nelson} {et~al.}(2019{\natexlab{a}}){Nelson}, {Pillepich},
  {Springel}, {Pakmor}, {Weinberger}, {Genel}, {Torrey}, {Vogelsberger},
  {Marinacci}, \& {Hernquist}}]{NelsonD19}
---. 2019{\natexlab{a}}, \mnras, 490, 3234, \dodoi{10.1093/mnras/stz2306}

\bibitem[{{Nelson} {et~al.}(2016){Nelson}, {van Dokkum}, {F{\"o}rster
  Schreiber}, {Franx}, {Brammer}, {Momcheva}, {Wuyts}, {Whitaker}, {Skelton},
  {Fumagalli}, {Hayward}, {Kriek}, {Labb{\'e}}, {Leja}, {Rix}, {Tacconi}, {van
  der Wel}, {van den Bosch}, {Oesch}, {Dickey}, \& {Ulf Lange}}]{Nelson16}
{Nelson}, E.~J., {van Dokkum}, P.~G., {F{\"o}rster Schreiber}, N.~M., {et~al.}
  2016, \apj, 828, 27, \dodoi{10.3847/0004-637X/828/1/27}

\bibitem[{{Nelson} {et~al.}(2019{\natexlab{b}}){Nelson}, {Tadaki}, {Tacconi},
  {Lutz}, {F{\"o}rster Schreiber}, {Cibinel}, {Wuyts}, {Lang}, {Leja},
  {Montes}, {Oesch}, {Belli}, {Davies}, {Davies}, {Genzel}, {Lippa}, {Price},
  {{\"U}bler}, \& {Wisnioski}}]{Nelson19}
{Nelson}, E.~J., {Tadaki}, K.-i., {Tacconi}, L.~J., {et~al.}
  2019{\natexlab{b}}, \apj, 870, 130, \dodoi{10.3847/1538-4357/aaf38a}

\bibitem[{{Newman} {et~al.}(2018){Newman}, {Belli}, {Ellis}, \&
  {Patel}}]{Newman18}
{Newman}, A.~B., {Belli}, S., {Ellis}, R.~S., \& {Patel}, S.~G. 2018, \apj,
  862, 126, \dodoi{10.3847/1538-4357/aacd4f}

\bibitem[{{Ocvirk} {et~al.}(2006){Ocvirk}, {Pichon}, {Lan{\c{c}}on}, \&
  {Thi{\'e}baut}}]{Ocvirk06}
{Ocvirk}, P., {Pichon}, C., {Lan{\c{c}}on}, A., \& {Thi{\'e}baut}, E. 2006,
  \mnras, 365, 46, \dodoi{10.1111/j.1365-2966.2005.09182.x}

\bibitem[{{Papovich} {et~al.}(2001){Papovich}, {Dickinson}, \&
  {Ferguson}}]{Papovich01}
{Papovich}, C., {Dickinson}, M., \& {Ferguson}, H.~C. 2001, \apj, 559, 620,
  \dodoi{10.1086/322412}

\bibitem[{{Papovich} {et~al.}(2004){Papovich}, {Dickinson}, {Ferguson},
  {Giavalisco}, {Lotz}, {Madau}, {Idzi}, {Kretchmer}, {Moustakas}, {de Mello},
  {Gardner}, {Rieke}, {Somerville}, \& {Stern}}]{Papovich04}
{Papovich}, C., {Dickinson}, M., {Ferguson}, H.~C., {et~al.} 2004, \apjl, 600,
  L111, \dodoi{10.1086/381075}

\bibitem[{{Peng} {et~al.}(2002){Peng}, {Ho}, {Impey}, \& {Rix}}]{Peng02}
{Peng}, C.~Y., {Ho}, L.~C., {Impey}, C.~D., \& {Rix}, H.-W. 2002, \aj, 124,
  266, \dodoi{10.1086/340952}

\bibitem[{{Peng} {et~al.}(2010){Peng}, {Ho}, {Impey}, \& {Rix}}]{Peng10}
---. 2010, \aj, 139, 2097, \dodoi{10.1088/0004-6256/139/6/2097}

\bibitem[{{P{\'e}rez-Gonz{\'a}lez} {et~al.}(2013){P{\'e}rez-Gonz{\'a}lez},
  {Cava}, {Barro}, {Villar}, {Cardiel}, {Ferreras}, {Rodr{\'\i}guez-Espinosa},
  {Alonso-Herrero}, {Balcells}, {Cenarro}, {Cepa}, {Charlot}, {Cimatti},
  {Conselice}, {Daddi}, {Donley}, {Elbaz}, {Espino}, {Gallego}, {Gobat},
  {Gonz{\'a}lez-Mart{\'\i}n}, {Guzm{\'a}n}, {Hern{\'a}n-Caballero},
  {Mu{\~n}oz-Tu{\~n}{\'o}n}, {Renzini}, {Rodr{\'\i}guez-Zaur{\'\i}n}, {Tresse},
  {Trujillo}, \& {Zamorano}}]{PerezGonzalez13}
{P{\'e}rez-Gonz{\'a}lez}, P.~G., {Cava}, A., {Barro}, G., {et~al.} 2013, \apj,
  762, 46, \dodoi{10.1088/0004-637X/762/1/46}

\bibitem[{{Pforr} {et~al.}(2012){Pforr}, {Maraston}, \& {Tonini}}]{Pforr12}
{Pforr}, J., {Maraston}, C., \& {Tonini}, C. 2012, \mnras, 422, 3285,
  \dodoi{10.1111/j.1365-2966.2012.20848.x}

\bibitem[{{Pforr} {et~al.}(2013){Pforr}, {Maraston}, \& {Tonini}}]{Pforr13}
---. 2013, \mnras, 435, 1389, \dodoi{10.1093/mnras/stt1382}

\bibitem[{{Pillepich} {et~al.}(2018){Pillepich}, {Nelson}, {Hernquist},
  {Springel}, {Pakmor}, {Torrey}, {Weinberger}, {Genel}, {Naiman}, {Marinacci},
  \& {Vogelsberger}}]{Pillepich18}
{Pillepich}, A., {Nelson}, D., {Hernquist}, L., {et~al.} 2018, \mnras, 475,
  648, \dodoi{10.1093/mnras/stx3112}

\bibitem[{{Pillepich} {et~al.}(2019){Pillepich}, {Nelson}, {Springel},
  {Pakmor}, {Torrey}, {Weinberger}, {Vogelsberger}, {Marinacci}, {Genel}, {van
  der Wel}, \& {Hernquist}}]{Pillepich19}
{Pillepich}, A., {Nelson}, D., {Springel}, V., {et~al.} 2019, \mnras, 490,
  3196, \dodoi{10.1093/mnras/stz2338}

\bibitem[{{Puglisi} {et~al.}(2019){Puglisi}, {Daddi}, {Liu}, {Bournaud},
  {Silverman}, {Circosta}, {Calabr{\`o}}, {Aravena}, {Cibinel}, {Dannerbauer},
  {Delvecchio}, {Elbaz}, {Gao}, {Gobat}, {Jin}, {Le Floc'h}, {Magdis},
  {Mancini}, {Riechers}, {Rodighiero}, {Sargent}, {Valentino}, \&
  {Zanisi}}]{Puglisi19}
{Puglisi}, A., {Daddi}, E., {Liu}, D., {et~al.} 2019, \apjl, 877, L23,
  \dodoi{10.3847/2041-8213/ab1f92}

\bibitem[{{Puglisi} {et~al.}(2021){Puglisi}, {Daddi}, {Valentino}, {Magdis},
  {Liu}, {Kokorev}, {Circosta}, {Elbaz}, {Bournaud}, {Gomez-Guijarro}, {Jin},
  {Madden}, {Sargent}, \& {Swinbank}}]{Puglisi21}
{Puglisi}, A., {Daddi}, E., {Valentino}, F., {et~al.} 2021, \mnras, 508, 5217,
  \dodoi{10.1093/mnras/stab2914}

\bibitem[{{Queiroz} {et~al.}(2020){Queiroz}, {Anders}, {Chiappini},
  {Khalatyan}, {Santiago}, {Steinmetz}, {Valentini}, {Miglio}, {Bossini},
  {Barbuy}, {Minchev}, {Minniti}, {Garc{\'\i}a Hern{\'a}ndez}, {Schultheis},
  {Beaton}, {Beers}, {Bizyaev}, {Brownstein}, {Cunha},
  {Fern{\'a}ndez-Trincado}, {Frinchaboy}, {Lane}, {Majewski}, {Nataf},
  {Nitschelm}, {Pan}, {Roman-Lopes}, {Sobeck}, {Stringfellow}, \&
  {Zamora}}]{Queiroz20}
{Queiroz}, A.~B.~A., {Anders}, F., {Chiappini}, C., {et~al.} 2020, \aap, 638,
  A76, \dodoi{10.1051/0004-6361/201937364}

\bibitem[{{Queiroz} {et~al.}(2021){Queiroz}, {Chiappini}, {Perez-Villegas},
  {Khalatyan}, {Anders}, {Barbuy}, {Santiago}, {Steinmetz}, {Cunha},
  {Schultheis}, {Majewski}, {Minchev}, {Minniti}, {Beaton}, {Cohen}, {da
  Costa}, {Fern{\'a}ndez-Trincado}, {Garcia-Hern{\'a}ndez}, {Geisler},
  {Hasselquist}, {Lane}, {Nitschelm}, {Rojas-Arriagada}, {Roman-Lopes},
  {Smith}, \& {Zasowski}}]{Queiroz21}
{Queiroz}, A.~B.~A., {Chiappini}, C., {Perez-Villegas}, A., {et~al.} 2021,
  \aap, 656, A156, \dodoi{10.1051/0004-6361/202039030}

\bibitem[{{Renzini}(2020)}]{Renzini20}
{Renzini}, A. 2020, \mnras, 495, L42, \dodoi{10.1093/mnrasl/slaa054}

\bibitem[{{Sanders} {et~al.}(2018){Sanders}, {Shapley}, {Kriek}, {Freeman},
  {Reddy}, {Siana}, {Coil}, {Mobasher}, {Dav{\'e}}, {Shivaei}, {Azadi},
  {Price}, {Leung}, {Fetherolf}, {de Groot}, {Zick}, {Fornasini}, \&
  {Barro}}]{Sanders18}
{Sanders}, R.~L., {Shapley}, A.~E., {Kriek}, M., {et~al.} 2018, \apj, 858, 99,
  \dodoi{10.3847/1538-4357/aabcbd}

\bibitem[{{Schreiber} {et~al.}(2016){Schreiber}, {Elbaz}, {Pannella}, {Ciesla},
  {Wang}, {Koekemoer}, {Rafelski}, \& {Daddi}}]{Schreiber16}
{Schreiber}, C., {Elbaz}, D., {Pannella}, M., {et~al.} 2016, \aap, 589, A35,
  \dodoi{10.1051/0004-6361/201527200}

\bibitem[{{Shapley} {et~al.}(2001){Shapley}, {Steidel}, {Adelberger},
  {Dickinson}, {Giavalisco}, \& {Pettini}}]{Shapley01}
{Shapley}, A.~E., {Steidel}, C.~C., {Adelberger}, K.~L., {et~al.} 2001, \apj,
  562, 95, \dodoi{10.1086/323432}

\bibitem[{{Shen} {et~al.}(2003){Shen}, {Mo}, {White}, {Blanton}, {Kauffmann},
  {Voges}, {Brinkmann}, \& {Csabai}}]{Shen03}
{Shen}, S., {Mo}, H.~J., {White}, S. D.~M., {et~al.} 2003, \mnras, 343, 978,
  \dodoi{10.1046/j.1365-8711.2003.06740.x}

\bibitem[{{Simons} {et~al.}(2021){Simons}, {Papovich}, {Momcheva}, {Trump},
  {Brammer}, {Estrada-Carpenter}, {Backhaus}, {Cleri}, {Finkelstein},
  {Giavalisco}, {Ji}, {Jung}, {Matharu}, \& {Weiner}}]{Simons21}
{Simons}, R.~C., {Papovich}, C., {Momcheva}, I., {et~al.} 2021, \apj, 923, 203,
  \dodoi{10.3847/1538-4357/ac28f4}

\bibitem[{{Springel} {et~al.}(2018){Springel}, {Pakmor}, {Pillepich},
  {Weinberger}, {Nelson}, {Hernquist}, {Vogelsberger}, {Genel}, {Torrey},
  {Marinacci}, \& {Naiman}}]{Springel18}
{Springel}, V., {Pakmor}, R., {Pillepich}, A., {et~al.} 2018, \mnras, 475, 676,
  \dodoi{10.1093/mnras/stx3304}

\bibitem[{{Steidel} {et~al.}(2010){Steidel}, {Erb}, {Shapley}, {Pettini},
  {Reddy}, {Bogosavljevi{\'c}}, {Rudie}, \& {Rakic}}]{Steidel16}
{Steidel}, C.~C., {Erb}, D.~K., {Shapley}, A.~E., {et~al.} 2010, \apj, 717,
  289, \dodoi{10.1088/0004-637X/717/1/289}

\bibitem[{{Steidel} {et~al.}(1996){Steidel}, {Giavalisco}, {Dickinson}, \&
  {Adelberger}}]{Steidel96}
{Steidel}, C.~C., {Giavalisco}, M., {Dickinson}, M., \& {Adelberger}, K.~L.
  1996, \aj, 112, 352, \dodoi{10.1086/118019}

\bibitem[{{Tacchella} {et~al.}(2017){Tacchella}, {Carollo}, {Faber}, {Cibinel},
  {Dekel}, {Koo}, {Renzini}, \& {Woo}}]{Tacchella17}
{Tacchella}, S., {Carollo}, C.~M., {Faber}, S.~M., {et~al.} 2017, \apjl, 844,
  L1, \dodoi{10.3847/2041-8213/aa7cfb}

\bibitem[{{Tacchella} {et~al.}(2016){Tacchella}, {Dekel}, {Carollo},
  {Ceverino}, {DeGraf}, {Lapiner}, {Mandelker}, \& {Primack
  Joel}}]{Tacchella16}
{Tacchella}, S., {Dekel}, A., {Carollo}, C.~M., {et~al.} 2016, \mnras, 457,
  2790, \dodoi{10.1093/mnras/stw131}

\bibitem[{{Tacchella} {et~al.}(2018){Tacchella}, {Carollo}, {F{\"o}rster
  Schreiber}, {Renzini}, {Dekel}, {Genzel}, {Lang}, {Lilly}, {Mancini},
  {Onodera}, {Tacconi}, {Wuyts}, \& {Zamorani}}]{Tacchella18}
{Tacchella}, S., {Carollo}, C.~M., {F{\"o}rster Schreiber}, N.~M., {et~al.}
  2018, \apj, 859, 56, \dodoi{10.3847/1538-4357/aabf8b}

\bibitem[{{Tacchella} {et~al.}(2022){Tacchella}, {Conroy}, {Faber}, {Johnson},
  {Leja}, {Barro}, {Cunningham}, {Deason}, {Guhathakurta}, {Guo}, {Hernquist},
  {Koo}, {McKinnon}, {Rockosi}, {Speagle}, {van Dokkum}, \&
  {Yesuf}}]{Tacchella21}
{Tacchella}, S., {Conroy}, C., {Faber}, S.~M., {et~al.} 2022, \apj, 926, 134,
  \dodoi{10.3847/1538-4357/ac449b}

\bibitem[{{Tadaki} {et~al.}(2017){Tadaki}, {Kodama}, {Nelson}, {Belli},
  {F{\"o}rster Schreiber}, {Genzel}, {Hayashi}, {Herrera-Camus}, {Koyama},
  {Lang}, {Lutz}, {Shimakawa}, {Tacconi}, {{\"U}bler}, {Wisnioski}, {Wuyts},
  {Hatsukade}, {Lippa}, {Nakanishi}, {Ikarashi}, {Kohno}, {Suzuki}, {Tamura},
  \& {Tanaka}}]{Tadaki17}
{Tadaki}, K.-i., {Kodama}, T., {Nelson}, E.~J., {et~al.} 2017, \apjl, 841, L25,
  \dodoi{10.3847/2041-8213/aa7338}

\bibitem[{{Tadaki} {et~al.}(2020){Tadaki}, {Belli}, {Burkert}, {Dekel},
  {F{\"o}rster Schreiber}, {Genzel}, {Hayashi}, {Herrera-Camus}, {Kodama},
  {Kohno}, {Koyama}, {Lee}, {Lutz}, {Mowla}, {Nelson}, {Renzini}, {Suzuki},
  {Tacconi}, {{\"U}bler}, {Wisnioski}, \& {Wuyts}}]{Tadaki20}
{Tadaki}, K.-i., {Belli}, S., {Burkert}, A., {et~al.} 2020, \apj, 901, 74,
  \dodoi{10.3847/1538-4357/abaf4a}

\bibitem[{{van der Walt} {et~al.}(2011){van der Walt}, {Colbert}, \&
  {Varoquaux}}]{numpy2011}
{van der Walt}, S., {Colbert}, S.~C., \& {Varoquaux}, G. 2011, Computing in
  Science Engineering, 13, 22, \dodoi{10.1109/MCSE.2011.37}

\bibitem[{{van der Wel} {et~al.}(2014{\natexlab{a}}){van der Wel}, {Chang},
  {Bell}, {Holden}, {Ferguson}, {Giavalisco}, {Rix}, {Skelton}, {Whitaker},
  {Momcheva}, {Brammer}, {Kassin}, {Martig}, {Dekel}, {Ceverino}, {Koo},
  {Mozena}, {van Dokkum}, {Franx}, {Faber}, \& {Primack}}]{vanderWel14b}
{van der Wel}, A., {Chang}, Y.-Y., {Bell}, E.~F., {et~al.} 2014{\natexlab{a}},
  \apjl, 792, L6, \dodoi{10.1088/2041-8205/792/1/L6}

\bibitem[{{van der Wel} {et~al.}(2014{\natexlab{b}}){van der Wel}, {Franx},
  {van Dokkum}, {Skelton}, {Momcheva}, {Whitaker}, {Brammer}, {Bell}, {Rix},
  {Wuyts}, {Ferguson}, {Holden}, {Barro}, {Koekemoer}, {Chang}, {McGrath},
  {H{\"a}ussler}, {Dekel}, {Behroozi}, {Fumagalli}, {Leja}, {Lundgren},
  {Maseda}, {Nelson}, {Wake}, {Patel}, {Labb{\'e}}, {Faber}, {Grogin}, \&
  {Kocevski}}]{vanderWel14}
{van der Wel}, A., {Franx}, M., {van Dokkum}, P.~G., {et~al.}
  2014{\natexlab{b}}, \apj, 788, 28, \dodoi{10.1088/0004-637X/788/1/28}

\bibitem[{{van Dokkum} {et~al.}(2010){van Dokkum}, {Whitaker}, {Brammer},
  {Franx}, {Kriek}, {Labb{\'e}}, {Marchesini}, {Quadri}, {Bezanson},
  {Illingworth}, {Muzzin}, {Rudnick}, {Tal}, \& {Wake}}]{vanDokkum10}
{van Dokkum}, P.~G., {Whitaker}, K.~E., {Brammer}, G., {et~al.} 2010, \apj,
  709, 1018, \dodoi{10.1088/0004-637X/709/2/1018}

\bibitem[{{van Dokkum} {et~al.}(2013){van Dokkum}, {Leja}, {Nelson}, {Patel},
  {Skelton}, {Momcheva}, {Brammer}, {Whitaker}, {Lundgren}, {Fumagalli},
  {Conroy}, {F{\"o}rster Schreiber}, {Franx}, {Kriek}, {Labb{\'e}},
  {Marchesini}, {Rix}, {van der Wel}, \& {Wuyts}}]{vanDokkum13}
{van Dokkum}, P.~G., {Leja}, J., {Nelson}, E.~J., {et~al.} 2013, \apjl, 771,
  L35, \dodoi{10.1088/2041-8205/771/2/L35}

\bibitem[{{van Dokkum} {et~al.}(2014){van Dokkum}, {Bezanson}, {van der Wel},
  {Nelson}, {Momcheva}, {Skelton}, {Whitaker}, {Brammer}, {Conroy},
  {F{\"o}rster Schreiber}, {Fumagalli}, {Kriek}, {Labb{\'e}}, {Leja},
  {Marchesini}, {Muzzin}, {Oesch}, \& {Wuyts}}]{vanDokkum14}
{van Dokkum}, P.~G., {Bezanson}, R., {van der Wel}, A., {et~al.} 2014, \apj,
  791, 45, \dodoi{10.1088/0004-637X/791/1/45}

\bibitem[{{Webb} {et~al.}(2003){Webb}, {Eales}, {Foucaud}, {Lilly},
  {McCracken}, {Adelberger}, {Steidel}, {Shapley}, {Clements}, {Dunne}, {Le
  F{\`e}vre}, {Brodwin}, \& {Gear}}]{Webb03}
{Webb}, T.~M., {Eales}, S., {Foucaud}, S., {et~al.} 2003, \apj, 582, 6,
  \dodoi{10.1086/344608}

\bibitem[{{Whitaker} {et~al.}(2017){Whitaker}, {Bezanson}, {van Dokkum},
  {Franx}, {van der Wel}, {Brammer}, {F{\"o}rster-Schreiber}, {Giavalisco},
  {Labb{\'e}}, {Momcheva}, {Nelson}, \& {Skelton}}]{Whitaker17}
{Whitaker}, K.~E., {Bezanson}, R., {van Dokkum}, P.~G., {et~al.} 2017, \apj,
  838, 19, \dodoi{10.3847/1538-4357/aa6258}

\bibitem[{{Whitney} {et~al.}(2019){Whitney}, {Conselice}, {Bhatawdekar}, \&
  {Duncan}}]{Whitney19}
{Whitney}, A., {Conselice}, C.~J., {Bhatawdekar}, R., \& {Duncan}, K. 2019,
  \apj, 887, 113, \dodoi{10.3847/1538-4357/ab53d4}

\bibitem[{{Wild} {et~al.}(2009){Wild}, {Walcher}, {Johansson}, {Tresse},
  {Charlot}, {Pollo}, {Le F{\`e}vre}, \& {de Ravel}}]{Wild09}
{Wild}, V., {Walcher}, C.~J., {Johansson}, P.~H., {et~al.} 2009, \mnras, 395,
  144, \dodoi{10.1111/j.1365-2966.2009.14537.x}

\bibitem[{{Williams} {et~al.}(2014){Williams}, {Giavalisco}, {Cassata},
  {Tundo}, {Wiklind}, {Guo}, {Lee}, {Barro}, {Wuyts}, {Bell}, {Conselice},
  {Dekel}, {Faber}, {Ferguson}, {Grogin}, {Hathi}, {Huang}, {Kocevski},
  {Koekemoer}, {Koo}, {Ravindranath}, \& {Salimbeni}}]{Williams14}
{Williams}, C.~C., {Giavalisco}, M., {Cassata}, P., {et~al.} 2014, \apj, 780,
  1, \dodoi{10.1088/0004-637X/780/1/1}

\bibitem[{{Williams} {et~al.}(2017){Williams}, {Giavalisco}, {Bezanson},
  {Cappelluti}, {Cassata}, {Liu}, {Lee}, {Tundo}, \& {Vanzella}}]{Williams17}
{Williams}, C.~C., {Giavalisco}, M., {Bezanson}, R., {et~al.} 2017, \apj, 838,
  94, \dodoi{10.3847/1538-4357/aa662f}

\bibitem[{{Zolotov} {et~al.}(2015){Zolotov}, {Dekel}, {Mandelker}, {Tweed},
  {Inoue}, {DeGraf}, {Ceverino}, {Primack}, {Barro}, \& {Faber}}]{Zolotov15}
{Zolotov}, A., {Dekel}, A., {Mandelker}, N., {et~al.} 2015, \mnras, 450, 2327,
  \dodoi{10.1093/mnras/stv740}

\end{thebibliography}
\end{document}